\documentclass[iop]{emulateapj}

\newcommand{\Msun}{~M_\odot}
\newcommand{\msun}{M_\odot}

\newcommand{\ccm}{\rm ~cm^{-3}}
\newcommand{\kms}{\rm ~km~s^{-1}}
\newcommand{\ergs}{\rm ~erg~s^{-1}}
\newcommand{\ergcms}{\rm ~erg~cm^{-2}~s^{-1}}

\newcommand{\ml}{~\Msun ~\rm yr^{-1}}
\newcommand{\mll}{\Msun ~\rm yr^{-1}}
\newcommand{\wl}{\lambda}
\newcommand{\wll}{\lambda \lambda}

\newcommand{\EE}[1]{\times 10^{#1}}
\newcommand{\angstrom}{\mbox{\normalfont\AA}}
\begin{document}

\title{High Density  Circumstellar Interaction in the Luminous Type IIn SN~2010jl: The first 1100 days}
\author{
Claes Fransson\altaffilmark{1},
Mattias Ergon\altaffilmark{1},
Peter J. Challis\altaffilmark{2},
Roger A. Chevalier\altaffilmark{3},
Kevin France \altaffilmark{4},
Robert P. Kirshner\altaffilmark{2},
G. H. Marion,\altaffilmark{2},
Dan Milisavljevic\altaffilmark{2},
Nathan Smith\altaffilmark{5},
Filomena Bufano\altaffilmark{6},
Andrew S. Friedman,\altaffilmark{2},
Tuomas Kangas\altaffilmark{7,8},
Josefin Larsson\altaffilmark{9},
Seppo Mattila\altaffilmark{10},
Stefano Benetti\altaffilmark{11},
 Ryan Chornock\altaffilmark{2},
Ian  Czekala\altaffilmark{2}, 
Alicia Soderberg\altaffilmark{2},
Jesper Sollerman\altaffilmark{1}
}
\altaffiltext{1}{Department of Astronomy, Oskar Klein Centre, Stockholm University, AlbaNova, 
SE--106~91 Stockholm, Sweden}
\altaffiltext{2}{Harvard-Smithsonian Center for Astrophysics, 60 Garden
Street, Cambridge, MA 02138, USA.}
\altaffiltext{3}{Department of Astronomy, University of Virginia, P.O. Box 400325,  Charlottesville, VA 22904}
\altaffiltext{4}{CASA, University of Colorado, 593UCB Boulder, CO 80309-0593, USA}
\altaffiltext{5}{Steward Observatory, University of Arizona, 933 North Cherry Avenue, Tucson, AZ 85721, USA}
\altaffiltext{6}{Departamento de Ciencias Fisicas, Universidad Andres Bello, Avda. Republica 252, Santiago, Chile}
\altaffiltext{7}{Tuorla Observatory,
University of Turku, V\"ais\"al\"antie 20 FI-21500 Piikki\"o, Finland}
\altaffiltext{8}{Nordic Optical Telescope (NOT), Apartado 474, E-38700 Santa Cruz de La Palma, Spain}
\altaffiltext{9}{KTH, Department of Physics, and the Oskar Klein
Centre, AlbaNova, SE-106 91 Stockholm, Sweden}
\altaffiltext{10}{Finnish Centre for Astronomy with ESO (FINCA),
University of Turku, V\"ais\"al\"antie 20 FI-21500 Piikki\"o, Finland}
\altaffiltext{11}{INAF-Osservatorio Astronomico di Padova, Vicolo dellOsservatorio 5, I-35122 Padova, Italy}

\begin{abstract}
HST and ground based observations of the Type IIn SN
2010jl are analyzed, including photometry,  spectroscopy in the ultraviolet, optical and NIR bands, $26-1128$ days after first detection. At maximum the bolometric luminosity was $\sim3\times10^{43}\ergs$ and even at 850 days exceeds $10^{42}\ergs$.  
A NIR excess, dominating after 400 days, probably originates in dust in the circumstellar medium (CSM). The total radiated energy is $\ga6.5\times10^{50}$ ergs, excluding the dust component.  The spectral lines can be separated into one broad component due to electron scattering, and one narrow with expansion  velocity $\sim100\kms$ from the CSM. 
The broad component is initially symmetric around zero velocity but  becomes blueshifted after $\sim50$ days, while remaining symmetric about a shifted centroid velocity.            
Dust absorption in the ejecta is unlikely to explain the line shifts, and we attribute the shift instead to acceleration by the SN radiation.  
From the optical lines  and the X-ray and dust properties, there is strong evidence for large scale asymmetries in the CSM.  
The ultraviolet lines indicate CNO processing
in the progenitor,  while the optical shows a  number of narrow coronal lines excited by the X-rays. 
The bolometric light curve is consistent with a radiative shock in an $r^{-2}$ CSM with a mass loss rate of $\dot M\sim0.1\ml$.
The total mass lost is $\ga3\Msun$.  
These properties are
consistent with the SN expanding into a CSM characteristic of
an LBV progenitor with a bipolar geometry. 
The apparent absence of nuclear processing is attributed  to a CSM still opaque to electron scattering.

\end{abstract}


\section{INTRODUCTION}
\label{sec-introd}
The progenitors for the narrow-lined Type II supernovae (SN IIn) remain a mystery, and this mystery is even deeper for the rare class of very luminous Type IIn, with $M \la -20$.
While significant evidence points toward an origin from luminous blue variable stars (LBV) \citep[e.g.,][]{Smith2008,Smith2008b,Smith2010}, suggestions for the energy source of the events span a wide range, from pair-instabilities  \citep{Smith2007,GalYam2009}, magnetar power \citep{Kasen2010,Nicholl2013}, or extremely energetic core-collapse SNe  \citep{Umeda2008},  to ongoing shock interactions or shock breakout from an extended progenitor  \citep{Falk1977,Grasberg1987,Smith2007a,Chevalier2011,Ginzburg2012}.

Before SN 2010jl, the most convincing direct detection of a Type IIn progenitor is SN 2005gl at a distance of 66 Mpc, which was a moderately
luminous SN IIn that transitioned into a more normal
SN II. Pre-explosion images of SN 2005gl showed a star at the position of the SN with luminosity consistent with an LBV \citep{Gal-Yam2009}. Recently it has been suggested that SN~1961V was in fact a core-collapse SN IIn, rather than an Eta Carinae-like eruption, in part because the peak absolute magnitude was $-18$ and because there is no surviving star detected in Spitzer images \citep{Kochanek2011,Smith2011b}.  The progenitor of SN~1961V was clearly detected for decades before 1961 as a very luminous blue star, consistent with an extremely massive LBV.
Even more recently, SN 2009ip \citep{Mauerhan2013,Pastorello2012,Fraser2013} and SN 2010mc \citep{Ofek2013b} have become interesting cases of a ``supernova impostor'' becoming a Type IIn SN . 
As we will show, these objects are similar in some ways, but differ from SN 2010jl in the essential feature of total energy release. 

From a Spitzer/IRAC survey of Type IIn SNe, \cite{Fox2011} find mid-IR emission from $\sim 15 \%$ of the SNe. The most likely origin is emission from pre-existing dust heated by the optical emission from circumstellar interaction between the SN and the CSM of the progenitor. Estimates of the dust mass are in the range $10^{-5} - 10^{-1} \Msun$. With a dust to gas ratio of 1:100 this is consistent with the total mass lost in giant  LBV eruptions \citep{SmithOwocki2006}. 

While medium luminosity Type IIn SNe, like SN 1988Z and SN 1995N, often are strong radio and X-ray emitters, the high luminosity IIn SNe are generally very weak. This may be a result of the extended massive envelopes thought to be responsible for the high luminosity of these SNe, as  discussed by \cite{Chevalier2012}.

In this paper we discuss HST and ground based observations of the Type IIn SN 2010jl. 
SN 2010jl was discovered on 2010 Nov.  3.52 by Newton \& Puckett
(2010). 
The first detection of SN 2010jl was, however, from pre-discovery images on 2010 October 9.6 by \cite{Stoll2010}. Based on the presence of narrow Balmer lines it was classified as a Type IIn SN by \cite{Benetti2010}.

From archival HST images \cite{Smith2011} identify a possible
progenitor. Although the exact nature of this star is uncertain, several
facts indicate a progenitor with mass $\ga 30 \Msun$, although it is difficult to distinguish between a single massive star or a young cluster. 
\cite{Stoll2010} find that the host is a low metallicity
galaxy, as found for other luminous Type IIn host galaxies, which typically have metallicities $\la 0.4 \  Z_\odot$ \citep{Neill2011,Chatzopoulos2011}.

The peak V  and I magnitudes were 13.7 and 13.0, respectively, corresponding  to  absolute magnitudes  M$_V \approx -19.9$ and M$_I \approx -20.5$  \citep{Stoll2010}. This is considerably more luminous than the average Type IIn SN, M$_V = -18.4$ \citep{Kiewe2012}, and places
it among the very luminous Type IIn SNe, and close to the super-luminous SNe, defined as $M  \la -21$ \citep{Gal-Yam2012}.  The light curve showed a slow evolution with a decline by $\sim 1$ mag during the first 200 days \citep{Andrews2011}. 

\cite{Smith2012} report optical spectroscopy of the SN from 2 -- 236 days after discovery. The spectrum shows a number of strong emission
lines, with a narrow and a broad component. The H$\alpha$ line is
well fitted with a `Lorentzian' profile with full width half maximum FWHM
$\sim  1800 \kms$ that extends to $\sim 6000 \kms$.
The narrow H$\alpha$ line was found to be double-peaked, and can be
approximated by a Gaussian emission component with FWHM =
120 $\kms$. The  shift of the line peak of the broad H$\alpha$ line was seen as evidence for dust formation in the post-shock gas. However,  Smith et. al. discussed also other scenarios based on a combination of electron scattering and continuum absorption. In this paper we show that electron scattering can explain all aspects of the line shapes.

\cite{Zhang} presented an extensive data set, including UBVRI  photometry and low resolution spectroscopy  during the first $\sim 525$ days after first detection.  Although the light curves and the optical low resolution spectra agree well up to $\sim 200$ days, there are some important differences in the later observations and interpretation between Zhang et al. and this paper (see Section \ref{sec_phot_results}).

From  Spitzer 3.6 and 4.5 $\mu$m and JHK photometry at an age of 90 -- 108 days \cite{Andrews2011}  find evidence for pre-existing dust with a mass of $0.03-0.35$ $\Msun$  at a distance of $\sim 6\EE{17}$ cm  from the SN.
Here, we complement these observations with more extensive coverage of the very late phases.  An important clue to the geometry is that \cite{Patat2011} find strong, wavelength independent polarization at a level of $\sim 2 \%$, which indicates an asphericity with axial ratio $\sim 0.7$. 

SN 2010jl was discovered as an X-ray source 
by \cite{Immler2010}. Subsequent Chandra observations by Chandra et al. (2012) on 2010 Dec. $7-8$, and 2011 Oct.
17, i.e., 59 days and 373 days after first detection, indicate a hard spectrum with $kT \ga 10$ keV and a very large column density at the first epoch, $N_H \sim 10^{24}$ cm$^{-2}$, which decreased to $N_H \sim 3 \times 10^{23}$ cm$^{-2}$ at 373 days. This absorption is most likely intrinsic to the SN or its CSM.
NuSTAR and XMM observations by \cite{Ofek2013} in 2012 Oct. -- Nov. (days $728-754$) were able to provide a tighter constraint on the X-ray temperature of $\sim 12$ keV. In addition, they also present UV and X-ray observations with the SWIFT satellite.  No radio detections have been reported.

\cite{Ofek2013} also discussed the optical light curve and found from a mass loss rate of $\sim 0.8 \ml$, based on their assumed wind velocity of $300 \kms$, and a total mass lost of $\ga10  \ \msun$. Because of their limited optical photometry they assumed a constant bolometric correction to the photometry. They also obtained spectroscopy at only a few epochs, and these data are most important at very late phases. While most of the qualitative conclusions in their paper are similar to ours, there are some important differences which we discuss in the paper. They do not analyze the emission line spectrum in any detail, partly due to the lower spectral resolution of their data which prevents separation of the narrow component from the broad. 

Recently \cite{Borish2014} discuss near-IR spectroscopic observations where they find evidence for high velocity gas from the He I $\wl 10,830$ line with velocities up to 5500 $\kms$. This gas may be related to the X-ray emitting gas, which originates from shocks with similar velocities.  

Based on a single spectrum at 513 days \cite{Maeda2013}  proposed that the line shifts of the Balmer lines, discussed by  \cite{Smith2012} and in this paper, result from dust formation in the ejecta. This is also the subject of a paper by \cite{Gall2014}, which makes a detailed analysis of the dust properties based on this interpretation. In this paper, based on  extensive observations, we show that this interpretation, however, has severe problems. 

In this paper we discuss an extensive set of optical, UV, and IR photometric and spectroscopic observations of SN 2010jl, together with interpretation of these observations. Compared to other papers, our data set includes broader wavelength coverage, more extensive time coverage, and higher spectral resolution. We present a unique set of UV observations obtained with HST.  X-ray observations and information about the progenitor from the literature enable us to provide a more complete picture of the event. In particular, we provide strong constraints on the structure of the progenitor and its environment. 

Our paper is organized as follows. In Sect. \ref{sec-obs} we report the observations and give our results in Sect. \ref{sec-results}.   In Sect. \ref{sec-discussion} we discuss these in relation to  the properties of the CSM, dynamics, energetics and  different progenitor scenarios. We also put SN 2010jl in the context of other IIn SNe. Our main conclusions are summarized in Sect. \ref{sec-conculsions}.

Based on the first detection we adopt 2010 Oct 9 (Julian Date  2,455,479) as the explosion date of the SN.  
With  a recession velocity of the host galaxy   3207 $\pm 30 \kms  $  \citep{Stoll2010},  we adopt a distance to UGC 5189A of $49 \pm 4$ Mpc in agreement with \cite{Smith2011} .

\section{OBSERVATIONS}
\label{sec-obs}

\subsection{Photometry}

Most of the optical imaging was obtained with the 1.2-m telescope at the F. L. Whipple Observatory (FLWO) using the KeplerCam instrument. KeplerCam data were reduced using IRAF and IDL procedures described in \citet{Hicken07}. A few additional late time epochs were obtained with the 6.5-m Baade telescope at the Magellan Observatory using the IMACS instrument and the 2.5-m Nordic Optical Telescope (NOT) using the ALFOSC instrument. These were reduced with IRAF using standard methods. High quality NOT imaging was used to construct $B$-, $V$-, $r$- and $i$-band templates by PSF subtraction of the SN using the IRAF DAOPHOT package. These templates were then subtracted from the FLWO, Magellan Observatory and NOT imaging using the HOTPANTS package. Template subtraction is necessary to obtain accurate photometry after $\sim$100 days as the SN is located close to the center of the galaxy.

The optical photometry was calibrated to the Johnson-Cousins (JC) and Sloan Digital Sky Survey (SDSS) standard systems using local reference stars in the SN field, in turn calibrated using standard field observations on five photometric nights. As the SN Spectral Energy Distribution (SED), in particular at late phases, is line-dominated, we have applied the technique of S-corrections \citep{Stritzinger02} to transform from the natural system of each instrument to the standard systems. This was done for all bands except the $u$-band. Color-terms and filter response functions were adopted from \citet{Hicken12} for the FLWO and from Ergon et al. (2013) for the NOT. Filter response functions for the JC and SDSS standard systems were adopted from \citet{Bessel12} and \citet{Doi10}.

Most of the near-infrared (NIR) imaging was obtained with the 1.3-m Peters Automated Infrared Imaging Telescope (PAIRITEL) at the FLWO. The data were processed into mosaics using the PAIRITEL Mosaic Pipeline version 3.6 implemented in python. Details of PAIRITEL observations and reduction of SN data can be found in \citet{Friedman12}. Two additional late time epochs were obtained with NOT using the NOTCAM instrument. This high quality imaging was used to construct $J$-, $H$-, $K$-band templates by PSF subtraction of the SN using the IRAF DAOPHOT package. These templates were then subtracted from the PAIRITEL imaging using the HOTPANTS package. As in the optical, template subtraction is necessary to obtain accurate photometry after $\sim$100 days and is particularly important for the late $J$-band photometry.

The NIR photometry was calibrated to the 2MASS standard system using reference stars from the 2MASS Point Source catalogue \citep{Skrutskie06} within the SN field. As the PAIRITEL was one of the two telescopes used for the 2MASS survey the natural system photometry is already on the 2MASS standard system and no transformation is needed. The NOT photometry was transformed to the 2MASS standard system using linear color-terms from Ergon et al. (2013). The absence of strong lines in the X-shooter NIR spectrum indicates that this is sufficient and that S-corrections is not needed.

As a check we can compare with the J, H and K$_s$ magnitudes from  \cite{Andrews2011} at 104 days. Our J and H magnitudes agree within $\sim 0.1 \ $  mag with those of Andrews et al. However, our K$_s$ magnitude close to this epoch is $\sim 12.7$, while that of Andrews et al. is $\sim 13.75$. Judging from the SED plot in Andrews et al. it, however, seems that there is a typographical error and that the magnitude should be $12.75\pm 0.1$. In this case there is excellent agreement.

In Table \ref{table_phot_opt} and Table \ref{table_phot_ir}. we give the photometry for all epochs, including errors. 
\begin{deluxetable*}{lccccccl}
\tabletypesize{\footnotesize}
\tablecaption{Optical photometry.\label{table_phot_opt}
}
\tablewidth{0pt}
\tablehead{
\colhead{JD} & \colhead{epoch$^a$} &\colhead{u'}& \colhead{B}&\colhead{V}& \colhead{r'} &\colhead{i'} &\colhead{Telescope(Instrument)}\\ 
}
\startdata
55508.98 & 29.98 & 13.85 (0.12) & 14.08 (0.03) & 13.79 (0.03) & 13.49 (0.02) & 13.54 (0.03) &  FLWO (KeplerCam) \\
55510.98 & 31.98 & 13.91 (0.12) & 14.10 (0.03) & 13.80 (0.03) & 13.49 (0.02) & 13.55 (0.03) &  FLWO (KeplerCam) \\
55513.96 & 34.96 & 13.94 (0.12) & 14.13 (0.03) & 13.84 (0.03) & 13.52 (0.02) & 13.58 (0.03) &  FLWO (KeplerCam) \\
55516.00 & 37.00 & 14.00 (0.12) & 14.16 (0.03) & 13.85 (0.03) & 13.53 (0.02) & 13.60 (0.03) &  FLWO (KeplerCam) \\
55522.91 & 43.91 & ... & ... & 13.88 (0.03) & 13.55 (0.02) & 13.63 (0.03) &  FLWO (KeplerCam) \\
55527.02 & 48.02 & 14.12 (0.12) & 14.30 (0.03) & 13.95 (0.03) & 13.64 (0.02) & 13.68 (0.03) &  FLWO (KeplerCam) \\
55529.95 & 50.95 & 14.19 (0.12) & 14.35 (0.03) & 14.02 (0.03) & 13.66 (0.02) & 13.74 (0.03) &  FLWO (KeplerCam) \\
55530.93 & 51.93 & 14.24 (0.12) & 14.31 (0.03) & 13.99 (0.03) & 13.61 (0.02) & 13.74 (0.03) &  FLWO (KeplerCam) \\
55533.98 & 54.98 & 14.22 (0.12) & 14.38 (0.03) & 14.07 (0.03) & 13.71 (0.02) & 13.79 (0.03) &  FLWO (KeplerCam) \\
55537.01 & 58.01 & 14.25 (0.12) & 14.40 (0.03) & 14.11 (0.03) & 13.71 (0.02) & 13.83 (0.03) &  FLWO (KeplerCam) \\
55540.95 & 61.95 & 14.30 (0.12) & 14.39 (0.03) & 14.09 (0.03) & 13.69 (0.02) & 13.82 (0.03) &  FLWO (KeplerCam) \\
55558.99 & 79.99 & 14.57 (0.12) & 14.62 (0.03) & 14.32 (0.03) & 13.87 (0.02) & 14.03 (0.03) &  FLWO (KeplerCam) \\
55564.02 & 85.02 & 14.54 (0.12) & 14.77 (0.03) & 14.46 (0.03) & 13.89 (0.02) & 14.13 (0.03) &  FLWO (KeplerCam) \\
55565.02 & 86.02 & 14.52 (0.12) & 14.84 (0.03) & 14.47 (0.03) & 13.89 (0.02) & 14.14 (0.03) &  FLWO (KeplerCam) \\
55566.03 & 87.03 & 14.61 (0.12) & 14.82 (0.03) & 14.51 (0.03) & 13.91 (0.02) & 14.18 (0.03) &  FLWO (KeplerCam) \\
55566.91 & 87.91 & 14.61 (0.12) & 14.80 (0.03) & 14.48 (0.03) & 13.88 (0.02) & 14.15 (0.03) &  FLWO (KeplerCam) \\
55568.98 & 89.98 & 14.64 (0.12) & ... & 14.51 (0.03) & 13.90 (0.02) & 14.20 (0.03) &  FLWO (KeplerCam) \\
55570.00 & 91.00 & 14.62 (0.12) & 14.81 (0.03) & 14.50 (0.03) & 13.91 (0.02) & 14.20 (0.03) &  FLWO (KeplerCam) \\
55571.98 & 92.98 & 14.62 (0.12) & 14.86 (0.03) & 14.51 (0.03) & 13.91 (0.02) & 14.21 (0.03) &  FLWO (KeplerCam) \\
55572.86 & 93.86 & 14.63 (0.12) & 14.84 (0.03) & 14.54 (0.03) & 13.92 (0.02) & 14.21 (0.03) &  FLWO (KeplerCam) \\
55575.89 & 96.89 & 14.62 (0.12) & 14.91 (0.03) & 14.55 (0.03) & 13.94 (0.02) & 14.24 (0.03) &  FLWO (KeplerCam) \\
55576.76 & 97.76 & ... & ... & 14.65 (0.03) & 13.94 (0.02) & 14.28 (0.03) &  FLWO (KeplerCam) \\
55578.00 & 99.00 & 14.76 (0.12) & 14.92 (0.03) & 14.57 (0.03) & 13.93 (0.02) & 14.26 (0.03) &  FLWO (KeplerCam) \\
55580.88 & 101.88 & 14.66 (0.12) & ... & 14.62 (0.03) & 13.94 (0.02) & 14.27 (0.03) &  FLWO (KeplerCam) \\
55587.89 & 108.89 & 14.80 (0.12) & ... & 14.62 (0.03) & 13.95 (0.02) & 14.34 (0.03) &  FLWO (KeplerCam) \\
55588.88 & 109.88 & ... & 14.94 (0.03) & 14.63 (0.03) & 13.96 (0.02) & 14.33 (0.03) &  FLWO (KeplerCam) \\
55589.87 & 110.87 & 14.72 (0.12) & 14.95 (0.03) & 14.63 (0.03) & 13.98 (0.02) & 14.34 (0.03) &  FLWO (KeplerCam) \\
55594.91 & 115.91 & 14.71 (0.12) & 15.00 (0.03) & 14.70 (0.03) & 13.97 (0.02) & 14.37 (0.03) &  FLWO (KeplerCam) \\
55597.77 & 118.77 & ... & 14.98 (0.03) & 14.69 (0.03) & 13.99 (0.02) & 14.39 (0.03) &  FLWO (KeplerCam) \\
55600.86 & 121.86 & 14.94 (0.12) & 15.00 (0.03) & 14.71 (0.03) & 13.98 (0.02) & 14.40 (0.03) &  FLWO (KeplerCam) \\
55602.70 & 123.70 & 14.76 (0.12) & ... & 14.71 (0.03) & 13.99 (0.02) & 14.41 (0.03) &  FLWO (KeplerCam) \\
55605.94 & 126.94 & 14.79 (0.12) & 15.09 (0.03) & 14.75 (0.03) & 14.01 (0.02) & 14.44 (0.03) &  FLWO (KeplerCam) \\
55606.82 & 127.82 & 14.85 (0.12) & ... & ... & ... & ... &  FLWO (KeplerCam) \\
55607.93 & 128.93 & 14.88 (0.12) & 15.08 (0.03) & 14.77 (0.03) & 14.02 (0.02) & 14.44 (0.03) &  FLWO (KeplerCam) \\
55608.86 & 129.86 & 14.83 (0.12) & ... & 14.75 (0.03) & 14.00 (0.02) & 14.45 (0.03) &  FLWO (KeplerCam) \\
55615.95 & 136.95 & 14.81 (0.12) & ... & 14.79 (0.03) & 14.00 (0.02) & 14.51 (0.03) &  FLWO (KeplerCam) \\
55623.65 & 144.65 & 14.82 (0.12) & 15.13 (0.03) & 14.81 (0.03) & 14.02 (0.02) & 14.53 (0.03) &  FLWO (KeplerCam) \\
55628.78 & 149.78 & 14.87 (0.12) & 15.12 (0.03) & 14.83 (0.03) & ... & 14.56 (0.03) &  FLWO (KeplerCam) \\
55631.66 & 152.66 & 14.98 (0.12) & 15.14 (0.03) & 14.84 (0.03) & 14.02 (0.02) & 14.55 (0.03) &  FLWO (KeplerCam) \\
55647.73 & 168.73 & 15.02 (0.12) & 15.17 (0.03) & 14.90 (0.03) & 14.02 (0.02) & 14.63 (0.03) &  FLWO (KeplerCam) \\
55667.74 & 188.74 & ... & ... & 14.96 (0.03) & 14.03 (0.02) & 14.69 (0.03) &  FLWO (KeplerCam) \\
55668.81 & 189.81 & ... & ... & 14.98 (0.03) & 14.03 (0.02) & 14.69 (0.03) &  FLWO (KeplerCam) \\
55669.74 & 190.74 & ... & ... & 14.97 (0.03) & 14.03 (0.02) & 14.69 (0.03) &  FLWO (KeplerCam) \\
55673.70 & 194.70 & ... & 15.21 (0.03) & 14.97 (0.03) & 14.04 (0.02) & 14.69 (0.03) &  FLWO (KeplerCam) \\
55682.68 & 203.68 & ... & 15.23 (0.03) & 15.00 (0.03) & 14.04 (0.02) & 14.72 (0.03) &  FLWO (KeplerCam) \\
55686.68 & 207.68 & ... & 15.22 (0.03) & ... & 14.03 (0.02) & 14.73 (0.03) &  FLWO (KeplerCam) \\
55688.67 & 209.67 & ... & 15.24 (0.03) & 14.98 (0.03) & 14.02 (0.02) & 14.72 (0.03) &  FLWO (KeplerCam) \\
55691.70 & 212.70 & ... & 15.21 (0.03) & 15.02 (0.03) & 14.04 (0.02) & 14.76 (0.03) &  FLWO (KeplerCam) \\
55696.63 & 217.63 & ... & ... & 15.01 (0.03) & 14.04 (0.02) & 14.74 (0.03) &  FLWO (KeplerCam) \\
55698.67 & 219.67 & ... & ... & 15.04 (0.03) & 14.04 (0.02) & 14.76 (0.03) &  FLWO (KeplerCam) \\
55704.68 & 225.68 & ... & ... & 15.01 (0.03) & 14.04 (0.02) & 14.78 (0.03) &  FLWO (KeplerCam) \\
55717.68 & 238.68 & ... & 15.27 (0.03) & 15.05 (0.03) & 14.05 (0.02) & 14.83 (0.03) &  FLWO (KeplerCam) \\
55855.99 & 376.99 & ... & 16.38 (0.03) & ... & 14.72 (0.02) & 15.72 (0.03) &  FLWO (KeplerCam) \\
55866.97 & 387.97 & ... & 16.39 (0.03) & 16.15 (0.03) & 14.79 (0.02) & 15.79 (0.03) &  FLWO (KeplerCam) \\
55888.01 & 409.01 & ... & 16.71 (0.03) & 16.40 (0.03) & 15.01 (0.02) & 15.98 (0.03) &  FLWO (KeplerCam) \\
55888.98 & 409.98 & ... & 16.70 (0.03) & 16.43 (0.03) & 15.02 (0.02) & 15.98 (0.03) &  FLWO (KeplerCam) \\
55916.98 & 437.98 & ... & 17.08 (0.03) & 16.77 (0.03) & 15.25 (0.02) & 16.27 (0.03) &  FLWO (KeplerCam) \\
55918.82 & 439.82 & ... & 17.20 (0.03) & 16.82 (0.03) & 15.29 (0.02) & 16.27 (0.03) &  FLWO (KeplerCam) \\
55948.84 & 469.84 & ... & 17.47 (0.03) & 17.06 (0.03) & 15.52 (0.02) & 16.52 (0.03) &  FLWO (KeplerCam) \\
55952.80 & 473.80 & ... & 17.41 (0.03) & 17.03 (0.03) & 15.50 (0.02) & 16.57 (0.03) &  FLWO (KeplerCam) \\
55972.87 & 493.87 & ... & 17.54 (0.03) & 17.33 (0.03) & 15.62 (0.02) & 16.73 (0.03) &  FLWO (KeplerCam) \\
55978.77 & 499.77 & ... & 17.62 (0.03) & 17.26 (0.03) & 15.65 (0.02) & 16.71 (0.03) &  FLWO (KeplerCam) \\
56009.73 & 530.73 & ... & 17.80 (0.03) & 17.62 (0.03) & 15.81 (0.02) & 16.97 (0.03) &  FLWO (KeplerCam) \\
56037.66 & 558.66 & ... & 18.16 (0.03) & 17.83 (0.03) & 15.95 (0.02) & 17.23 (0.03) &  FLWO (KeplerCam) \\
56040.65 & 561.65 & ... & 18.01 (0.03) & 17.84 (0.03) & 15.99 (0.02) & ... &  FLWO (KeplerCam) \\
56060.63 & 581.63 & ... & 18.36 (0.03) & 17.89 (0.03) & 16.06 (0.02) & 17.33 (0.03) &  FLWO (KeplerCam) \\
56065.63 & 586.63 & ... & 18.29 (0.03) & 18.02 (0.03) & 16.12 (0.02) & 17.48 (0.03) &  FLWO (KeplerCam) \\
56241.73 & 762.73 & ... & 19.43 (0.03) & 19.19 (0.03) & 17.04 (0.02) & 18.76 (0.03) &  NOT (ALFOSC) \\
56248.83 & 769.83 & ... & 19.61 (0.03) & 19.31 (0.03) & 17.10 (0.02) & 18.52 (0.03) &  MAG (IMACS) \\
56327.54 & 848.54 & ... & 19.91 (0.03) & 19.73 (0.03) & 17.59 (0.02) & ... &  NOT (ALFOSC) \\
56396.48 & 917.48 & 19.16 (0.12) & 20.43 (0.03) & 20.40 (0.03) & 18.19 (0.02) & 20.14 (0.03) &  NOT (ALFOSC) \\
 \enddata
 \tablenotetext{a}{Relative to first detection date, JD  2,455,479.0.}
\end{deluxetable*}

\begin{deluxetable*}{lccccl}
\tabletypesize{\footnotesize}
\tablecaption{NIR Photometry. \label{table_phot_ir}
}
\tablewidth{0pt}
\tablehead{
\colhead{JD} &\colhead{Epoch}&\colhead{J}& \colhead{H}&\colhead{K}&\colhead{Telescope(Instrument) }\\ 
}
\startdata
55505.94 & 26.94 & 12.84 (0.06) & 12.56 (0.07) & ... &  PAIRITEL (2MASS) \\
55507.99 & 28.99 & ... & ... & 12.32 (0.08) &  PAIRITEL (2MASS) \\
55508.98 & 29.98 & ... & 12.57 (0.07) & ... &  PAIRITEL (2MASS) \\
55511.96 & 32.96 & 12.88 (0.06) & ... & ... &  PAIRITEL (2MASS) \\
55514.90 & 35.90 & 12.89 (0.06) & 12.56 (0.07) & ... &  PAIRITEL (2MASS) \\
55517.97 & 38.97 & ... & 12.68 (0.07) & ... &  PAIRITEL (2MASS) \\
55518.95 & 39.95 & 12.88 (0.06) & ... & 12.36 (0.08) &  PAIRITEL (2MASS) \\
55524.00 & 45.00 & 12.92 (0.06) & 12.61 (0.07) & ... &  PAIRITEL (2MASS) \\
55525.03 & 46.03 & 12.87 (0.06) & ... & ... &  PAIRITEL (2MASS) \\
55533.85 & 54.85 & 13.03 (0.06) & ... & ... &  PAIRITEL (2MASS) \\
55536.95 & 57.95 & 13.04 (0.06) & ... & 12.34 (0.08) &  PAIRITEL (2MASS) \\
55537.95 & 58.95 & 13.04 (0.06) & 12.69 (0.07) & 12.41 (0.08) &  PAIRITEL (2MASS) \\
55540.87 & 61.87 & 13.05 (0.06) & 12.82 (0.07) & 12.43 (0.08) &  PAIRITEL (2MASS) \\
55569.92 & 90.92 & 13.27 (0.06) & 13.04 (0.07) & 12.54 (0.08) &  PAIRITEL (2MASS) \\
55572.84 & 93.84 & 13.26 (0.06) & ... & ... &  PAIRITEL (2MASS) \\
55575.83 & 96.83 & 13.28 (0.06) & 13.07 (0.07) & 12.69 (0.08) &  PAIRITEL (2MASS) \\
55578.79 & 99.79 & 13.30 (0.06) & 13.03 (0.07) & 12.66 (0.08) &  PAIRITEL (2MASS) \\
55599.77 & 120.77 & 13.42 (0.06) & 13.19 (0.07) & 12.76 (0.08) &  PAIRITEL (2MASS) \\
55602.70 & 123.70 & 13.44 (0.06) & 13.13 (0.07) & 12.81 (0.08) &  PAIRITEL (2MASS) \\
55607.80 & 128.80 & 13.47 (0.06) & ... & ... &  PAIRITEL (2MASS) \\
55617.74 & 138.74 & ... & 13.25 (0.07) & 12.86 (0.08) &  PAIRITEL (2MASS) \\
55620.76 & 141.76 & 13.53 (0.06) & ... & 12.86 (0.08) &  PAIRITEL (2MASS) \\
55623.77 & 144.77 & ... & 13.29 (0.07) & ... &  PAIRITEL (2MASS) \\
55629.73 & 150.73 & 13.55 (0.06) & ... & 12.84 (0.08) &  PAIRITEL (2MASS) \\
55640.73 & 161.73 & 13.63 (0.06) & 13.31 (0.07) & ... &  PAIRITEL (2MASS) \\
55643.73 & 164.73 & 13.59 (0.06) & ... & 12.91 (0.08) &  PAIRITEL (2MASS) \\
55644.71 & 165.71 & 13.57 (0.06) & 13.32 (0.07) & ... &  PAIRITEL (2MASS) \\
55647.79 & 168.79 & 13.61 (0.06) & ... & 12.99 (0.08) &  PAIRITEL (2MASS) \\
55651.72 & 172.72 & 13.63 (0.06) & 13.36 (0.07) & 12.96 (0.08) &  PAIRITEL (2MASS) \\
55670.66 & 191.66 & 13.61 (0.06) & ... & ... &  PAIRITEL (2MASS) \\
55676.68 & 197.68 & 13.65 (0.06) & 13.49 (0.07) & 13.06 (0.08) &  PAIRITEL (2MASS) \\
55688.63 & 209.63 & 13.63 (0.06) & 13.51 (0.07) & 12.95 (0.08) &  PAIRITEL (2MASS) \\
55705.68 & 226.68 & 13.72 (0.06) & 13.51 (0.07) & ... &  PAIRITEL (2MASS) \\
55708.67 & 229.67 & 13.66 (0.06) & 13.52 (0.07) & ... &  PAIRITEL (2MASS) \\
55947.89 & 468.89 & 13.76 (0.06) & 12.76 (0.07) & 11.88 (0.08) &  PAIRITEL (2MASS) \\
55949.87 & 470.87 & ... & ... & 11.79 (0.08) &  PAIRITEL (2MASS) \\
55999.65 & 520.65 & 14.01 (0.06) & ... & 11.86 (0.08) &  PAIRITEL (2MASS) \\
56002.59 & 523.59 & ... & 12.81 (0.07) & ... &  PAIRITEL (2MASS) \\
56010.63 & 531.63 & ... & 12.80 (0.07) & 11.96 (0.08) &  PAIRITEL (2MASS) \\
56011.63 & 532.63 & 14.06 (0.06) & ... & ... &  PAIRITEL (2MASS) \\
56013.71 & 534.71 & ... & 12.84 (0.07) & 11.94 (0.08) &  PAIRITEL (2MASS) \\
56014.63 & 535.63 & 14.06 (0.06) & ... & ... &  PAIRITEL (2MASS) \\
56015.69 & 536.69 & ... & 12.82 (0.07) & 11.86 (0.08) &  PAIRITEL (2MASS) \\
56017.69 & 538.69 & 14.06 (0.06) & 12.84 (0.07) & ... &  PAIRITEL (2MASS) \\
56019.70 & 540.70 & ... & ... & 11.88 (0.08) &  PAIRITEL (2MASS) \\
56025.70 & 546.70 & 14.14 (0.06) & ... & ... &  PAIRITEL (2MASS) \\
56026.68 & 547.68 & 14.15 (0.06) & 12.88 (0.07) & ... &  PAIRITEL (2MASS) \\
56027.69 & 548.69 & 14.10 (0.06) & ... & ... &  PAIRITEL (2MASS) \\
56028.68 & 549.68 & 14.16 (0.06) & ... & ... &  PAIRITEL (2MASS) \\
56029.68 & 550.68 & 14.16 (0.06) & ... & 11.95 (0.08) &  PAIRITEL (2MASS) \\
56031.68 & 552.68 & 14.16 (0.06) & ... & 12.03 (0.08) &  PAIRITEL (2MASS) \\
56035.61 & 556.61 & ... & 12.88 (0.07) & 12.02 (0.08) &  PAIRITEL (2MASS) \\
56036.68 & 557.68 & 14.17 (0.06) & ... & ... &  PAIRITEL (2MASS) \\
56055.71 & 576.71 & 14.22 (0.06) & ... & 11.98 (0.08) &  PAIRITEL (2MASS) \\
56058.71 & 579.71 & ... & 12.97 (0.07) & ... &  PAIRITEL (2MASS) \\
56060.68 & 581.68 & 14.26 (0.06) & ... & ... &  PAIRITEL (2MASS) \\
56061.67 & 582.67 & 14.30 (0.06) & 12.99 (0.07) & ... &  PAIRITEL (2MASS) \\
56066.65 & 587.65 & 14.23 (0.06) & ... & ... &  PAIRITEL (2MASS) \\
56217.02 & 738.02 & 15.03 (0.06) & ... & ... &  PAIRITEL (2MASS) \\
56226.02 & 747.02 & 14.99 (0.06) & 13.50 (0.07) & 12.37 (0.08) &  PAIRITEL (2MASS) \\
56228.02 & 749.02 & 15.09 (0.06) & ... & ... &  PAIRITEL (2MASS) \\
56229.00 & 750.00 & ... & 13.58 (0.07) & ... &  PAIRITEL (2MASS) \\
56231.97 & 752.97 & 15.12 (0.06) & ... & 12.29 (0.08) &  PAIRITEL (2MASS) \\
56344.39 & 865.39 & 15.80 (0.06) & 13.77 (0.07) & 12.60 (0.08) &  NOT (NOTCAM) \\
56374.44 & 895.44 & 15.90 (0.06) & 14.02 (0.07) & 12.73 (0.08) &  NOT (NOTCAM) \\
56621.69	& 1142.7&17.28 (0.25)&14.40 (0.11)&12.99 (0.12)&	NOT	(NOTCAM)\\
 \enddata
\end{deluxetable*}

\subsection{Spectroscopy}
\label{sec-spec}

\subsubsection{$HST$-COS and STIS Observations and Data Reduction}

HST observations  with the Space Telescope Imaging Spectrograph
(STIS) of SN2010jl were obtained at four epochs, 23, 33, 107 and 573 days after first
detection. The observations were carried out as part of program GO-12242
`UV Studies of a Core Collapse Supernova'.
The G230LB and G430L gratings were used at each epoch with
the $52\times 0.2\arcsec$ slit. The data were reduced using the standard HST Space
Telescope Science Data Analysis System (STSDAS) routines
to bias subtract, flat-field, extract, wavelength calibrate,
and flux-calibrate each SN spectrum.  The spectral resolution of these
gratings corresponds to $500-600 \kms$ at the short wavelength limit of the gratings
and $250-300 \kms$ at the long limit. Table \ref{hsttable}  summarizes the STIS
observations including exposure time and spectral resolution.
\ifx \apjloutput \undefined
\begin{deluxetable*}{lccclccr}
\footnotesize 
\fi
\tabletypesize{\scriptsize}
\tablewidth{0in}
\tablecaption{Log of HST observations for SN 2010jl\label{hsttable}}
\tablehead{
\colhead{Date} & \colhead{J.D.}&\colhead{Epoch}&\colhead{Instrument}&\colhead{Grism/Grating}&\colhead{Wavelength}&\colhead{Spectral}&\colhead{Exposure}  \\
\colhead{}  &\colhead{2,450,000+} &\colhead{(days)$^a$}&\colhead{}&\colhead{} &\colhead{ range (\AA)}&\colhead{resolution} &\colhead{s}   \\
}
\startdata
2010-11-11&5512.37&33.9&STIS&G230LB&1685-3065&615-1135&3550\\
          &5512.45&&STIS&G430L&2900-5700&530-1040&700\\
2010-11-22&5522.69&44.2&COS&G130M&1150-1450&16,000-21,000&1000\\
          &5522.70&&COS&G160M&1405-1775&16,000-21,000&1000\\
          &5522.56&&STIS&G230LB&1685-3065&615-1135&3575\\
          &5522.64&&STIS&G430L&2900-5700&530-1040&700\\
2011-01-23&5585.33&106.8&COS&G130M&1150-1450&16,000-21,000&2380\\
          &5585.39&&COS&G160M&1405-1775&16,000-21,000&2950\\
          &5585.13&&STIS&G230LB&1685-3065&615-1135&2300\\
          &5585.27&&STIS&G430L&2900-5700&530-1040&1200\\
2012-05-03&6051.45&573.0&STIS&G230L&1568-3184&615-1135&4400\\
          &6051.39&&STIS&G430L&2900-5700&530-1040&600\\
          2012-06-20&6098.75&620.6&COS&G130M&1150-1450&16,000-21,000&7400\\
          &6099.41&&COS&G160M&1405-1775&16,000-21,000&7400
          \enddata


\tablenotetext{a}{Relative to first detection date, 2010 Oct. 9, JD 2,455,479} 
\ifx \apjloutput \undefined
\end{deluxetable*}
\fi

The SN was also observed with the medium resolution far-UV modes of $HST$-COS (G130M and G160M) at 44 days, 107 days and at 621 days.  A description of the COS instrument and on-orbit performance characteristics can be found in \cite{Osterman2011} and \cite{Green2012}.  All observations were centered on SN 2010jl (R.A. = 09$^{\mathrm h}$ 42$^{\mathrm m}$ 53.33$^{\mathrm s}$, Dec. = +09\arcdeg 29\arcmin 41.8\arcsec ; J2000) and COS performed NUV imaging target acquisitions with the PSA/MIRRORB mode. The MIRRORB configuration introduces optical distortions into the target acquisition image, but a first-order analysis indicates that the observations were centered on the point-like SN. It is  important to realize that COS has only limited
spatial resolution, which means that objects as far as $2 \arcsec$ \  from
the center contribute to the spectrum. This is  important
for objects with nearby H II regions. 
Our last COS spectrum of SN 2010jl, at 621 days, has strong galaxy contamination from this source.

The G130M data were processed with the COS calibration pipeline, CALCOS\footnote{We refer the reader to the Cycle 18 COS Instrument Handbook for more details: {\tt http://www.stsci.edu/hst/cos/documents/handbooks/current/cos\_cover.html}} v2.12, and combined with the custom IDL coaddition procedure described by~\citet{Danforth2010} and~\citet{Shull2010}. 
The coaddition routine interpolates all detector segments and grating settings onto a common wavelength grid, and makes a correction for the detector QE-enhancement grid.  No correction for the detector hex pattern is performed.  
The data cover the 1155~--~1773~\AA\ bandpass, with two breaks at the COS detector segment gaps,  [$\lambda_{gap}$,$\Delta\lambda_{gap}$] = [1302~\AA, 16~\AA] and [1592~\AA, 20~\AA] for the G130M and G160M modes, respectively.  The resolving power of the medium resolution COS modes is $R$~$\equiv$~$\lambda / \Delta\lambda$~$\approx$~18,000 ($\Delta$$v$~=~17 km s$^{-1}$). 

\subsubsection{Optical spectroscopy}

Complementing the HST observations, an extensive spectroscopic campaign involving several ground based telescopes was launched, and a log of all spectroscopic  observations is available in Table~\ref{alfosctable}.
\begin{deluxetable*}{lccccccl}
\tabletypesize{\footnotesize}
\tablecaption{Journal of spectroscopic observations \label{alfosctable}}
\tablehead{
\colhead{UT date } & \colhead{ J.D.M} &\colhead{Epoch$^a$}&\colhead{Range}& \colhead{FWHM res.}&\colhead{Exposure}&\colhead{airmass}&\colhead{Instrument}   \\ 
 &&\colhead{days}&\colhead{\AA} &\colhead{\AA}&\colhead{s}&&\\
}
\startdata
2010-11-07  &  5507.53& 28.5 &$3475-7415$&6.2 &  240 &   1.13&FAST/300GPM\\
2010-11-09& 5509.71 &  30.7 &  $6350-6850$ &  1.2 & 600x4 &   1.30 & ALFOSC/gr17 \\
2010-11-10 &   5510.51&31.5&$3475-7415$&6.2     &600&    1.16&FAST/300GPM\\
2010-11-10&   5510.72 &  31.7 &  $3500-5060$ & 1.9 & 400x3 &   1.25 & ALFOSC/gr16 \\
2010-11-14& 5515.4\phantom{0}& 36.4&$3177-8527$&6.5&120&1.72&MMT/300GPM\\
2010-11-15& 5516.4\phantom{0}& 37.4&$3188-8534$&6.5&120&1.14&MMT/300GPM\\
2010-11-16& 5517.4\phantom{0}& 38.4&$3186-4516$&1.45&1200&1.40&MMT/1200GPM\\
2010-11-16& 5517.5\phantom{0}& 38.5&$4395-5731$&1.45&750&1.35&MMT/1200GPM\\
2010-11-16& 5517.5\phantom{0}& 38.5&$5587-6931$&1.45&600&1.30&MMT/1200GPM\\
2010-11-16& 5517.5\phantom{0}& 38.5&$3177-8399$&6.5&120&1.15&MMT/300GPM\\
2010-11-18& 5518.68 &   39.7 &  $6350-6850$ &  1.2 & 600x2 &  1.37 & ALFOSC/gr17 \\
2010-11-18&  5518.71 &   39.7 &   $3500-5060$ & 1.9 & 500x3 &  1.21 & ALFOSC/gr16 \\
2010-11-24& 5524.76 &   45.8 &  $6350-6850$ & 1.2 & 600x2 &  1.07 & ALFOSC/gr17 \\
2010-11-24&  5524.72 & 45.7 &  $3500-5060$ & 2.9 & 500x3 & 1.13 & ALFOSC/gr16 \\
2010-11-28& 5529.4\phantom{0}& 50.4 &$5653-7538$&2.0&900&1.4&MMT/832GPM\\
2010-12-06  &  5536.54& 57.5&$3475-7415$&6.2&900&1.11&FAST/300GPM\\
2010-12-07 &   5537.49& 58.5&$3475-7415$&6.2&1500&1.08&FAST/300GPM\\
2010-12-09& 5540.40& 61.4&$5575-6908$&1.45&450&1.1&MMT/1200GPM\\
2010-12-11  &  5541.50& 62.5&$3475-7415$&6.2&900&1.08&FAST/300GPM\\
2010-12-12  &  5542.47& 63.5&$3475-7415$&6.2&1020&1.09&FAST/300GPM\\
2010-12-16  & 5546.41  & 67.4 & $3850-9000$ & 0.56 & 3600 & 1.14 & TRES\\
2010-12-19  & 5549.47  & 70.5 & $3850-9000$ & 0.56 & 3000 & 1.08 & TRES\\\
2010-12-28  & 5558.37  & 79.4 & $3850-9000$ & 0.56 & 3600 & 1.17 & TRES\\
2010-12-28& 5558.63 &   79.6 &  $6350-6850$ & 1.2 & 600x3 &   1.13 & ALFOSC/gr17 \\
2010-12-28& 5558.65 &   79.7 &   $3500-5060$ & 2.9 & 600x3 &  1.08 & ALFOSC/gr16 \\
2011-01-02  &  5563.36&84.4&$3475-7415$&6.2&900&1.19&FAST/300GPM\\
2011-01-05  & 5566.43& 87.4&$3475-7415$&6.2&1020&1.08&FAST/300GPM\\
2011-01-08   & 5569.39& 90.4&$3475-7415$&6.2&900&1.09&FAST/300GPM\\
2011-01-13  & 5574.45  & 95.5 & $3850-9000$ & 0.56 & 3600 & 1.15 & TRES\\
2011-01-28   & 5589.29& 110.3&$3475-7415$&6.2&300&1.20&FAST/300GPM\\
2011-01-31&  5592.55 &  113.6 &   $6350-6850$ & 1.2 & 600x4 &  1.10 & ALFOSC/gr17 \\
2011-01-31& 5592.59 & 113.6 &   $3500-5060$ & 1.9 & 600x2 & 1.06 & ALFOSC/gr16 \\
2011-02-05   & 5597.41& 118.4&$3475-7415$&6.2&1200&1.18&FAST/300GPM\\
2011-02-26   & 5618.14 & 139.1&$3475-7415$&6.2&1200&1.56&FAST/300GPM\\
2011-03-01  &  5621.27& 142.3&$3475-7415$&6.2&1500 &1.08&FAST/300GPM\\
2011-03-01& 5621.5\phantom{0} &142.5 &   $6350-6850$ & 1.2 & 1200x3 & 1.18 &ALFOSC/gr17 \\
 2011-03-03   & 5623.23 & 144.2&$3475-7415$&6.2&   1544   & 1.10&FAST/300GPM\\
2011-03-09   & 5629.21     & 150.2&$3475-7415$&6.2&900  &  1.11&FAST/300GPM\\
2011-03-10&5630.56 &151.6  &   $3500-5060$ & 1.9 & 900x1 &  1.17 &ALFOSC/gr16 \\
2011-03-16  &  5636.23     & 157.2&$3475-7415$&6.2&900  &  1.08&FAST/300GPM\\
2011-03-29&5650.41 & 171.4 &   $6350-6850$ & 1.2 & 900x3 & 1.07 &ALFOSC/gr17\\
2011-04-03  &  5654.18    & 175.2&$3475-7415$&6.2& 819  &  1.08&FAST/300GPM\\
2011-04-08 & 5660.41& 181.4 & $3500-5060$ & 1.9 & 900x3 & 1.06 &ALFOSC/gr16 \\
2011-05-19&5701.40 & 222.4 &   $6350-6850$ & 1.2 & 900x3 & 1.31 &ALFOSC/gr17\\
2011-05-21& 5703.39&224.4  &   $3500-5060$ & 1.9 & 900x2 & 1.29 &ALFOSC/gr16 \\
2011-10-28  &  5862.52   & 383.5&$3475-7415$&6.2& 1010   & 1.28&FAST/300GPM\\
2011-10-30  &  5864.48 & 385.5&$3475-7415$&6.2&   1200  &  1.44&FAST/300GPM\\
2011-11-03   & 5868.49    &389.5&$3475-7415$&6.2&1800  & 1.33&FAST/300GPM\\
2011-11-28    & 5894.5\phantom{0}    &415.5&$5653-7538$&0.71/px&900&1.2& MMT/832GPM \\
2011-12-24  &  5919.51   &440.5&$3475-7415$&6.2& 1800& 1.15&FAST/300GPM\\
2011-12-27  &  5922.36  &443.4&$3475-7415$&6.2&  1800 & 1.26&FAST/300GPM\\
2011-12-31 &   5926.31  &447.3&$3475-7415$&6.2&  1800  &  1.47&FAST/300GPM\\
2012-01-14  &  5940.23  & 461.2&$3000-10200$&0.1&5376&1.22-1.41&X-SHOOTER\\
2012-01-14  &  5940.23  & 461.2&$10200-24800$&0.3&5928&1.22-1.41&X-SHOOTER\\
2012-01-18  &  5944.32  & 465.3&$3475-7415$&6.2&1800&1.16&FAST/300GPM\\
2012-03-24   & 6010.15   &531.2&$3475-7415$&6.2& 1800  & 1.14&FAST/300GPM\\
2012-05-12  &  6059.15   &580.2&$3475-7415$&6.2& 1800 &  1.22&FAST/300GPM\\
2012-05-15  &  6062.18    &583.2 &$3475-7415$&6.2&1800  & 1.43&FAST/300GPM\\
2012-05-29& 6076.18  &  597.2  &$4000-6900$&  3.5  &  900x3  & 1.9  &  MDM/OSMOS \\
2012-06-15  &  6093.16   &614.2&$3475-7415$&6.2& 1200 &  2.40&FAST/300GPM\\
2012-10-20  &  6221.00   & 742.0  &$3475-7415$& 6.2 & 1800 & 1.38 & FAST/300GPM \\
2012-11-12  &  6244.00   & 765.0  &$3475-7415$& 6.2 & 1200 & 1.10 & FAST/300GPM \\
2012-11-14  &  6246.00   & 767.0  &$3475-7415$& 6.2 & 1800 & 1.20 & FAST/300GPM \\
2012-12-21&6282.8\phantom{0}&803.8&$5587-6931$&1.45&1200&1.1&MMT/1200GPM\\
2013-02-02&6325.8\phantom{0}&846.8&$3177-8527$&6.5&1800&1.2&MMT/300GPM\\
2013-02-03&6326.8\phantom{0}&847.8&$5587-6931$&1.45&1800&1.3&MMT/1200GPM\\
2013-11-10&6606.7\phantom{0}&1127.7&$6400-6900$&1.45&1800&1.3&MMT/1200GPM\\
 \enddata
\tablenotetext{a}{Relative to first detection date, JD  2,455,479.0.}
\end{deluxetable*}

Optical spectroscopy of SN~2010jl was obtained on 16  epochs with 
the 2.5 m Nordic Optical Telescope (NOT)  on La Palma, Spain with the ALFOSC spectrograph. 
To optimize the resolution for this narrow line SN we used two set-ups. 
Grism 17 covers basically only the H$\alpha$ region, while grism 16 was used for the bluer part, covering the rest of the Balmer series and connecting with the HST NUV data. 
All observations were performed at the parallactic angle, and the airmass was always less than 1.4 at start of the observations. To further enhance  the resolution we used narrow slits, of width 0.9 arcsec for grism 17 and 0.5-0.75 arcsec for grism 16.
The spectra were reduced in a standard manner using IRAF
scripts as implemented in the QUBA pipeline.
Wavelength calibrations were determined from exposures of HeNe arc lamps, and were checked against bright night sky emission lines.
Flux calibration was performed by means of
spectrophotometric standard stars. 

A large number of optical spectra (3400--7300~\AA) were also obtained at the the FLWO 1.5 m Tillinghast telescope using the FAST spectrograph \citep{Fabricant98} from day 28 to day 767.   FLWO/FAST data are reduced using a combination of standard IRAF and custom IDL procedures \citep{Matheson05}.  High resolution spectra were also obtained with TRES (Tillinghast Reflector Echelle Spectrograph), which is a fiber-fed echelle spectrograph on the 1.5-meter Tillinghast telescope at FLWO.


Moderate-resolution optical spectra were obtained with the 2.4 m Hiltner telescope at MDM Observatory, on Kitt Peak, AZ on day 597. The Ohio State Multi-Objet Spectrograph (OSMOS; \citealt{Martini2011}) was used with the VPH grism and $1.4\arcsec$ inner slit in combination with the MDM4K CCD detector. These spectra were reduced and calibrated employing standard techniques in IRAF. Cosmic rays and obvious cosmetic defects have been removed. Wavelengths were checked against night sky emission lines and flux calibrations were applied using observations of \cite{Stone1977} and \cite{Massey1990} standard stars.

The 6.5 m MMT spectra were all taken
by the Blue Channel spectrograph. All observation
used the $1.0 \arcsec$ slit oriented at the parallactic
angle. The reduction procedure was the same as for the FLWO/FAST spectra.

One medium resolution  spectrum with X-shooter at the Very Large Telescope at ESO (VLT) on day 461 (2012 Jan. 14)  was obtained, including both the optical and NIR ranges. The slit used was 1.0\arcsec \ in the UV and 0.9\arcsec \ in the optical and NIR. The resolutions were 4350, 7450 and 5300 in the UV, optical and NIR, respectively, corresponding to 69, 40 and 56 $\kms$, respectively. The X-shooter spectrum was pre-reduced using version 1.1.0 of the dedicated ESO pipeline \citep{Goldoni}, with calibration frames (biases, darks, flat fields and arc lamps) taken during daytime.
Spectrum extraction and flux calibration were done using standard IRAF tasks. For the latter we used  a spectrophotometric standard taken from the ESO list (http://www.eso.org/sci/facilities/paranal/instruments/xshooter/tools/specphot\_list.html) and observed during the same night. Telluric bands were removed using a telluric standard spectrum taken at the same airmass as the SN. 

Absolute fluxes of all our spectra were determined from r' band photometry. With the strong H$\alpha$ line from the supernova, this is the band least affected by the galaxy background. For the last epoch MMT spectrum at 1128 days we do not have any simultaneous optical photometry and we estimate that the absolute flux is only accurate to $\sim 25$ \%. 

\section{RESULTS}
\label{sec-results}

\subsection{Reddening}
\label{sec-redd}

From \cite{Schlegel1998}, \citet{Smith2011} estimate  a Milky Way reddening corresponding to  E$_{\rm (B-V)} = 0.027$ mag.  Based on the weak Na I lines, \cite{Smith2011} assume negligible reddening from the host galaxy. From the damping wings of the Ly$\alpha$ absorption in our COS spectra we can make an independent estimate of this for both the host and for the Milky Way. 

\begin{figure}[t!]
\begin{center}
\resizebox{85mm}{!}{\includegraphics[angle=0]{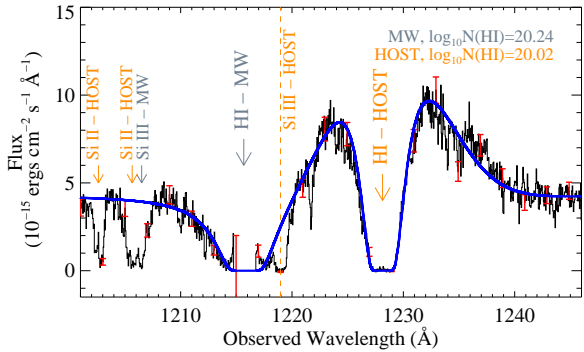}}
\caption{Fit to the Ly$\alpha$ absorptions on 23 January 2011 from the Milky Way and from the host galaxy.  The blue line shows the fit, while the black line shows the observations and the red error bars the one sigma error. 
}
\label{fig_nh}
\end{center}
\end{figure}

We discuss the COS spectrum in detail later. Here we use the damping wings of the Ly$\alpha$ absorption to derive a value of the column density, $N_{\rm H}$  \citep[e.g.,][Chap. 9.4]{Draine2011}.  The Milky Way absorption is affected by the Si III $\lambda 1206.5$ absorption on the red side, so we use the blue side for the fit. For the host galaxy we instead use the red side to avoid the  slight overlap with the Milky Way absorption. From fits of these, shown in Figure   \ref{fig_nh},  we derive a value of $N_{\rm H} = (1.05\pm 0.3)\EE{20}$ cm$^{-2}$ for $b = 10 \kms$ for the host galaxy, while the corresponding value is N(HI)$_{\rm MW} = 1.75 (\pm 0.25 ) \times 10^{20} \ {\rm cm}^{-2}$ from the Milky Way.  There is some uncertainty on the Milky Way column density, because the Si III 1206.5 \AA \  absorption line from the host galaxy falls on the red edge of the Milky Way Ly$\alpha$ profile. 

Using the  $N_{\rm H I}$ vs. E$_{\rm (B-V)}$ relation in \cite{Bohlin1978}, and assuming the same $N(H_2)/N(H I)$ ratio
for the host galaxy of 2010jl as in the Milky Way,  we find E$_{\rm (B-V)} = 0.036$ mag  from the Milky Way and E$_{\rm (B-V)} = 0.022$ mag  for the host galaxy.  The value for the Milky Way agrees well with that derived from the FIR emission, but we also note that there is a non-neglible absorption from the host galaxy. 
For the remainder of the paper, where needed, we will adopt a value of  E$_{\rm (B-V)} = 0.058$ mag for the total reddening.

\subsection{Light curves and total energy output}
\label{sec_phot_results}
\begin{figure}
\begin{center}
\resizebox{85mm}{!}{\includegraphics[scale=.80,angle=0]{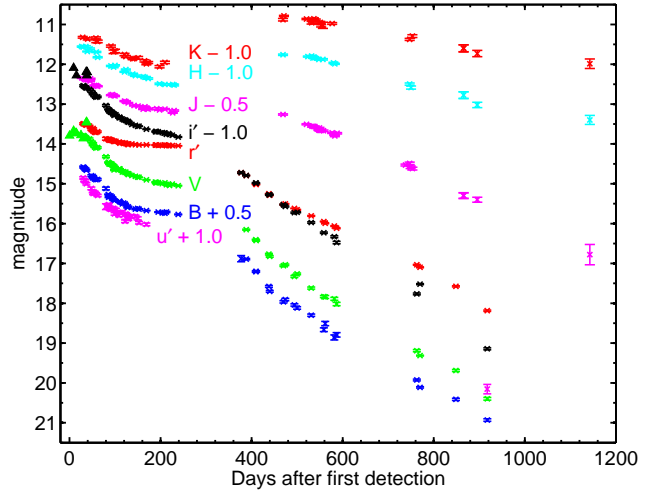}}
\caption{Light curves of SN 2010jl in different bands. Triangles are early, pre-discovery I- and V band observations from  \cite{Stoll2010}.}
\label{fig_photometry}
\end{center}
\end{figure}

Figure \ref{fig_photometry} shows the light curve from our photometry, complemented with early measurements from \cite{Stoll2010}.
The B, V and i' bands  all show nearly the same slow decline by $\sim 0.8$ mag/100 days during the first $\sim 175$ days. After this the light curves become almost constant up to $\sim 240$ days.  
However, the r' band already shows signs of flattening at age 100 days.
We discuss this band, dominated by H$\alpha$, in more detail in Sect. \ref{sec-profiles}.  Being most sensitive to the decreasing temperature, the u' band shows the steepest decline. 

After the first gap produced by conjunction with the sun, all optical bands show a considerably faster decline at  $\ga 300$ days. From 400 -- 850 days the decline in the optical bands is roughly linear. The decline rate is $8.0 \times 10^{-3}$ mag/day in i',  $4.9 \times 10^{-3}$ mag/day in r', $8.0 \times 10^{-3}$ mag/day in V and $7.1\times 10^{-3}$ mag/day in B.  
We attribute the slow r band decline to emission in H$\alpha$:  as seen in the spectra, H$\alpha$ is stronger with respect to the continuum as time goes by  (Sect. \ref{sec_spec_results}). 

 Compared to the R-band photometry of \cite{Ofek2013} we agree within less than 0.1 mag over the whole period. 
Up to day 200 our optical photometry agrees within $\sim 0.1-0.2$ mag in the r', B and V bands also with the light curves by \cite{Zhang}, in spite of the difference in filter profiles between our r' band and their R band. This is probably because the r'  and R bands are both dominated by H$\alpha$. Our i' magnitudes are $\sim 0.5$ mag fainter than the I magnitudes by Zhang et al. and  $\sim 0.3$ mag fainter in u' compared to their U band. Most likely this difference can be explained by the different filter responses between the SDSS i' and u' bands and the standard I and U bands. 

At epochs later than 350 days our magnitudes are 0.5 -- 1.5 mag fainter in all bands, a difference which increases with time. The origin of this is not clear, but we note that Zhang et al.  do not give any errors in their table of the photometry for these epochs. Also, our photometry showed a similar flattening of the light curve until we subtracted the background from the host galaxy.

The NIR light curves show a similar decline to the optical up to $\sim 200$ days. After this the light curves flatten considerably in J and H,  while the K-band shows an increase compared to the flux at earlier times. The difference between the NIR and the optical behavior is real and has important consequences for our physical picture for SN~2010jl.  As we discuss in Sect. \ref{sec_dust}, we interpret this emission as coming from dust  in the CSM of the progenitor. 

The pseudo-bolometric light curve for the flux in the wavelength range
3500-25000 \AA \ was calculated using the method described in Ergon
et al. (2013). In the optical region where we have a well sampled
spectral sequence, we made use of both photometry and spectroscopy to
accurately calculate the bolometric luminosity, whereas in the NIR region
we used photometry to interpolate the shape of the SED. 
The SED has then been de-reddened using E$_{\rm (B-V)} = 0.036$ mag  from the Milky Way and E$_{\rm (B-V)} = 0.022$ mag  for the host galaxy, as derived in Sect. \ref{sec-redd}. For the Milky Way reddening we use the extinction law by  \cite{Cardelli1989} and being a star forming galaxy with WR features \citep{Shirazi2012} we use the star burst extinction from \cite{Calzetti1994} for the host galaxy.

\begin{figure}
\begin{center}
\resizebox{85mm}{!}{\includegraphics[scale=0.1,angle=0]{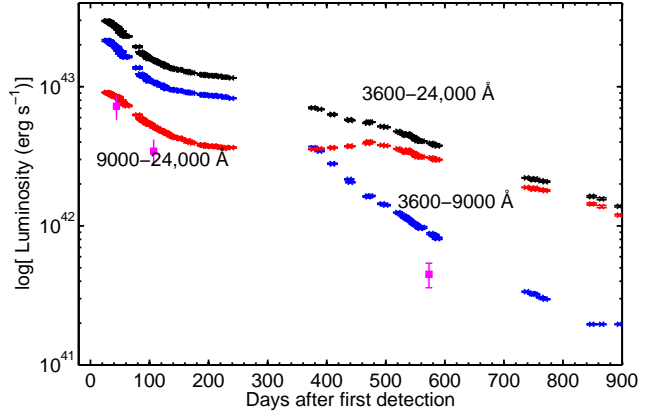}}
\caption{Pseudo-bolometric light curves of SN 2010jl. The contributions from the $3600-9000$ \AA \ and $9000-24,000$ \AA \ ranges are shown separately. Also shown as magenta squares are the UV luminosities at the time of our HST observations. The day 44 and day 107 luminosities include the $1100 - 3600$ \AA \ range, while  day 573 only includes the STIS $1700 - 3600$ \AA \ range. }
\label{fig_bollc}
\end{center}
\end{figure}

As is clear from
Figure \ref{fig_photometry}, the contribution from the NIR bands becomes
increasingly important. The luminosity in the NIR region (9000-24000
\AA ) is shown in Figure  \ref{fig_bollc} as red points, whereas the total
(3500-25000 \AA ) luminosity is shown as black points. While the optical bolometric contribution starts to drop at $\sim 350$ days the NIR part has a long plateau from 200 to $\sim 450$ days.  At age 100 days, the IR represents 33 \% of the flux, equaling the optical output at day 400; by age 500 days, it is 73\% and by day 770, 85\%.
This era is dominated by a dust contribution (Sect. \ref{sec_dust}), and reflects both the instantaneous energy output and  a contribution from a light echo. In Sect.  \ref{sec_dust} we show that the SN itself dominates the NIR at early epochs, while later the NIR comes from an echo. 

The  pseudo-bolometric light curve in Figure  \ref{fig_bollc} only includes the optical and NIR  bands and ignores both the UV and midIR. Especially at the early phases, the comparatively hot spectrum has a large UV contribution. We will discuss the spectra in more detail later; here we only estimate this contribution from an integration of the flux from COS and STIS for days 44, 107 and 573. We do not include the COS spectrum for day 621 because this is severely contaminated by the host galaxy. 

On day 44 we find a luminosity in the $1100-3600$ \AA  \ range equal to $7.2 \times 10^{42} \ergs$ and for day 107 $3.4 \times 10^{42} \ergs$. For day 573 we find a  luminosity of  $4.5 \times 10^{41} \ergs$ in the STIS $1700-3600$ \AA \ range. We estimate the uncertainty in these fluxes as $\sim 25 \%$. There is therefore  a substantial UV contribution to the bolometric luminosity especially at late epochs, up to $\sim 50 \%$ of the optical luminosity at the last epoch. There may also be a substantial EUV and X-ray luminosity that is not included in our observations.

Up to day $\sim 200$  our total bolometric luminosity agrees with that of \cite{Zhang}. Zhang et al. have, however, no NIR (or UV) observations. These account for an increasing fraction of the luminosity as the event ages, so their luminosities are increasingly incomplete. In addition, the background contribution from the host is very important after day $\sim 350$, as was discussed above.  At early epochs there is some discrepancy between our bolometric luminosities and those of Zhang et al.  It is not clear what produces this difference: it may be due to  Zhang et al. using photometry alone, while we use a combination of photometry and spectroscopy to combine the bands. 

Assuming that the luminosity is the same as at our first determinations at day 27 (see below),  
we find a total integrated energy from day 0 to day 920 in the $3600 - 9000$ \AA \ range of $3.3\times10^{50} \ \rm ergs$ and in the $9000 - 24,000$ \AA \ range $2.6\times10^{50} \  \rm ergs$. The total in the $3600 - 24,000$ \AA \ range is therefore $5.8\times10^{50} \  \rm ergs$.  
The correct way to use the IR measurements depends on the interpretation (Sect. \ref{sec_dust}) and on the relative contributions between the photosphere emission and dust: NIR light from an echo reflects the optical - UV output, while NIR light from  shock heated dust should be added to the energy budget.

 To disentangle the dust and direct, photospheric luminosity from the SN we have made fits to the Spectral Energy Distribution (SED) from the photometry for most epochs. We have here included the Spitzer 3.6 and 4.5 $\mu$m fluxes from  \cite{Andrews2011} at  day 87 (JD 2,455,565) and from  \cite{Fox2013} for day 254 (JD 2,455,733), as well as unpublished Spitzer observations from the Spitzer archive from days 465, 621 and 844. For these epochs we find fluxes of 8.55,  8.63, 7.73 mJy  at 3.6 $\mu$m and 8.30, 8.66, 8.24 mJy at 4.5 $\mu$m. 

\begin{figure}[t!]
\begin{center}
\resizebox{85mm}{!}{\includegraphics[angle=0]{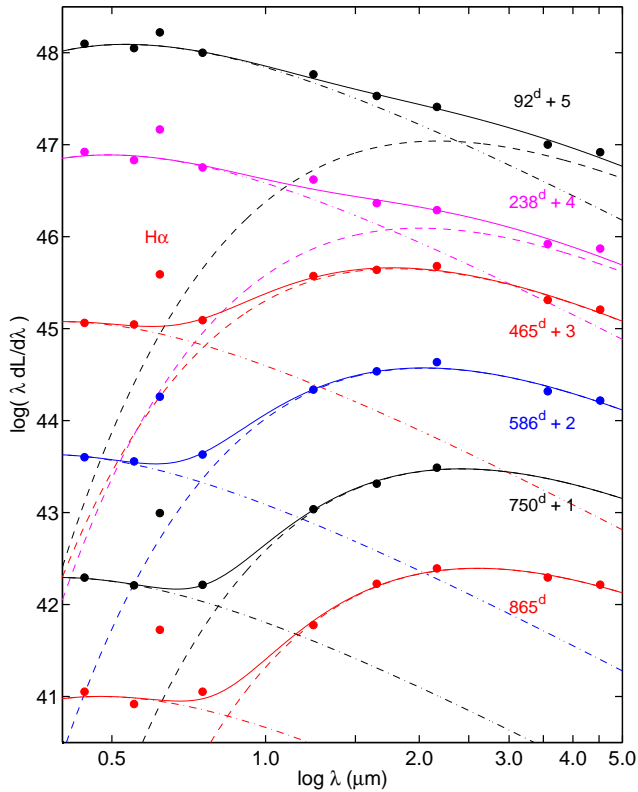}}
\caption{Spectral energy distributions at different epochs, together with blackbody fits from a dust component and a photospheric component. Because of the dominance of the H$\alpha$ line in the r-band this band is not included in the fits. For clarity each spectrum has been shifted by one decade in luminosity relative to the previous. }
\label{fig_sed}
\end{center}
\end{figure}

In Figure \ref{fig_sed} we show the SEDs for these dates, as well as for one epoch (750 days) where we only have optical and NIR data.  
These SEDs  are modeled  with two blackbodies with different temperatures and we minimize the chi square with the photospheric effective temperature and radius  $T_{\rm eff}$ and $R_{\rm phot}$, and the corresponding parameters for the dust shell, $T_{\rm dust}$, $R_{\rm dust}$, as free parameters. We discuss the motivation for this choice of spectrum further in Sect. \ref{sec_dust}. In these fits we exclude the r-band, because it is dominated by the H$\alpha$ line. For the Spitzer photometry at day 254, 621, and 844, where no other photometry exists, we interpolate the Spitzer fluxes and the  NIR and optical photometry to days 238, 586 and 865, respectively. Given the slow evolution of the Spitzer fluxes at these epochs, this should only introduce a minor error. 

\begin{figure}[t!]
\begin{center}
\resizebox{85mm}{!}{\includegraphics[angle=0]{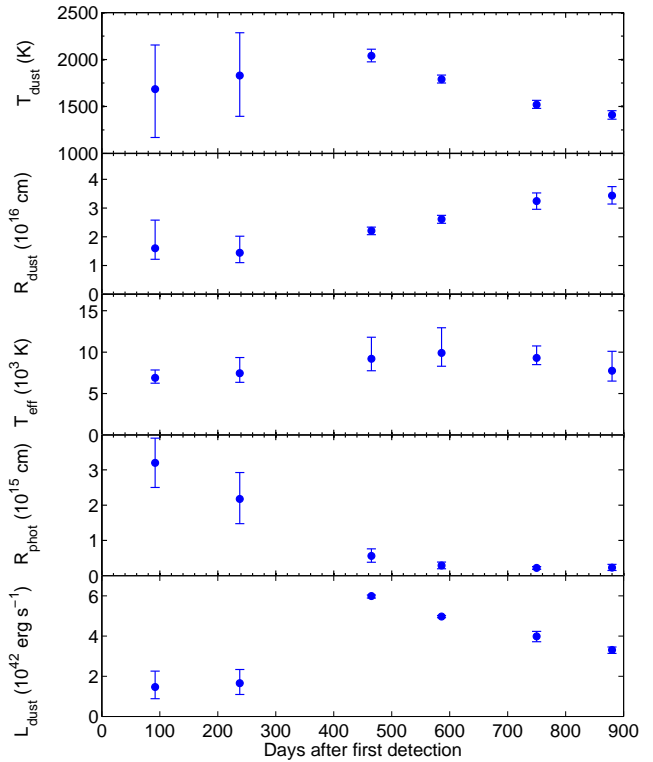}}
\caption{Blackbody temperature, radius and luminosity for the dust component  and radius and effective temperature for the SN component for the epochs in Figure \ref{fig_sed}.}
\label{fig_dust_tbb_rbb}
\end{center}
\end{figure}

From these models we find that up to day $\sim 400$ the photospheric contribution dominates also the NIR, while at later stages the dust component takes over.  
The best fit parameters are for these models given in Table \ref{table_dust_param} for the different dates, and in Figure \ref{fig_dust_tbb_rbb} we show the blackbody temperature, radius and total luminosity of the dust and photospheric components from these fits, together with the standard deviations. The large errors in $T_{\rm dust}$ and $R_{\rm dust}$ at the first two epochs is a result of the dominance of the photospheric component also for the NIR, while the errors in $T_{\rm eff}$ is mainly a result of the relatively few bands (mainly B, V, I) where there is a large contribution from this component.
\begin{deluxetable*}{lcccc}
\tabletypesize{\footnotesize}
\tablecaption{Parameters for the SED fits assuming two blackbody components. \label{table_dust_param}}
\tablewidth{0pt}
\tablehead{
\colhead{Epoch}& \colhead{$T_{\rm eff}$}&\colhead{$R_{\rm  phot}$}& \colhead{$T_{\rm dust}$}&\colhead{$R_{\rm  dust}$}\\
days&K&$10^{15}$ cm&K&$10^{16}$  cm\\
}
\startdata
\phantom{0}92& \phantom{0}6900     &3.20&1685& 1.60\\
238& \phantom{0}7450    &2.18&1830& 1.44\\
465& \phantom{0}9200    &0.56&2040& 2.21\\
586&  \phantom{0}9900    &0.29&1790& 2.61\\
750& \phantom{0}9300    &0.22&1520& 3.24\\
880& \phantom{0}7750  &0.23&1410& 3.43\\
\tableline
 \enddata
\end{deluxetable*}

At epochs earlier than  $\sim 400$ days the dust temperature is constant within errors at $\sim 1850 \pm 200$ K, and then  slowly decays to $\sim 1400$ K at 850 days. The blackbody radius is $\sim (1-2)\times10^{16}$ cm for the first $\sim 300$ days, and then slowly  increases to $\sim 3\times 10^{16}$ cm at the last observation. The dust luminosities we obtain for the first epochs are lower than the NIR luminosities in Figure \ref{fig_bollc}. The reason for this, as can be seen in Figure \ref{fig_sed}, is that the photospheric contribution dominates the J, H and K bands for these epochs. At epochs later than the 465 day observation the opposite is true, which is a result of including the total dust emission from the blackbody fit and not only the NIR bands.  

Already at $\sim 90$ days  \cite{Andrews2011} found from NIR and Spitzer observations an IR excess due to warm dust, but with  a  lower temperature of $\sim 750$ K  than we find. Andrews et al., however, only include the Spitzer fluxes to the dust component, while we  include also the J, H and K bands in this component, explaining our higher dust temperatures. We note that \cite{Andrews2011} underestimate the K-band flux 
in their SED fit.



\begin{figure}[t!]
\begin{center}
\resizebox{85mm}{!}{\includegraphics[angle=0]{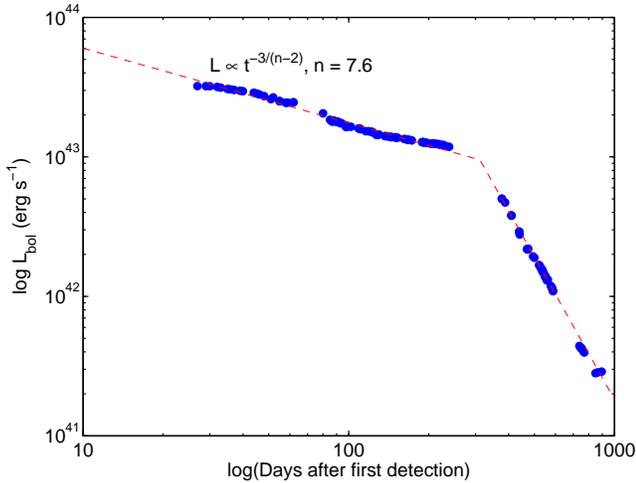}}
\caption{Bolometric light curve for the SN ejecta, excluding the dust echo. The dashed lines show power law fits to the early and late light curve used to construct the density distribution of the explosion. Note the pronounced break in the light curve at $\sim 320$ days.  The dashed lines give power law fits to the luminosity before and after the break  (see Sect. \ref{sec_energy} for a discussion).  }
\label{bol_log_log}
\end{center}
\end{figure}

Using the SED fitting  we can improve on the bolometric light curve by separating the SN and dust  contributions of the  IR flux to the bolometric luminosity and add this to the BVri contribution in Figure \ref{fig_bollc}. Based on the UV flux at the epochs with HST observations we multiply this by a factor 1.25 (Sect. \ref{sec_phot_results}). In this way we arrive at the bolometric light curve from the SN ejecta alone in Figure \ref{bol_log_log}, now shown in a log -- log plot. From this we see that the bolometric light curve from the ejecta  can   be accurately characterized by a power law decay  from  $\sim 20 -   320$ days, given by   $L(t) \sim 1.75\times 10^{43}  (t/100 \ {\rm days})^{-0.536} \ergs$ and a final steep decay $L(t) = 8.71 \times 10^{42} (t/320 \ {\rm days})^{-3.39} \ergs$ after day 320.  

\cite{Ofek2013}  estimate the bolometric light curve by assuming a constant bolometric correction of $-0.27$ mag to the R-band photometry. With this assumption they  find a  flatter light curve with $L(t) \propto  t^{-0.36}$ for the same explosion date as we use here. The reason for this difference is that the R-band decays slower than most of the other bands, as can be seen from Fig. \ref{fig_photometry}. The bolometric light curve will therefore be steeper than the R-band light curve. 

The slope depends on the assumed shock breakout date. 
\cite{Ofek2013}  discuss this based on the light curve and and find a likely range of 15-25 days before I-band maximum, corresponding to JD 2,455,469 - 2,455,479. Using 2,455,469 instead of our 2,455,479 would change the best fit luminosity decline to $L(t) \sim 1.9\times 10^{43}  (t/100 \ {\rm days})^{-0.61} \ergs$.  

To estimate the total energy output from the SN we assume that the bolometric luminosity before our first epoch at 26 days was constant at the level at  26 days, which is supported by the early observations by  \cite{Stoll2010}, shown in Figure  \ref{fig_photometry}. The total energy from the SN (excluding the echo) is then $6.5 \times 10^{50}$ ergs. In addition, there is a contribution from the EUV as well as X-rays and mid-IR (Sect. \ref{sec-big}). Even ignoring these, we note that the total radiated energy is a large fraction of the energy of a `normal' core collapse SN (see Sect. \ref{sec-big}). 

\subsection{Spectroscopic evolution}
\label{sec_spec_results}

\begin{figure*}
\begin{center}
\resizebox{150mm}{!}{\includegraphics[scale=.80,angle=0]{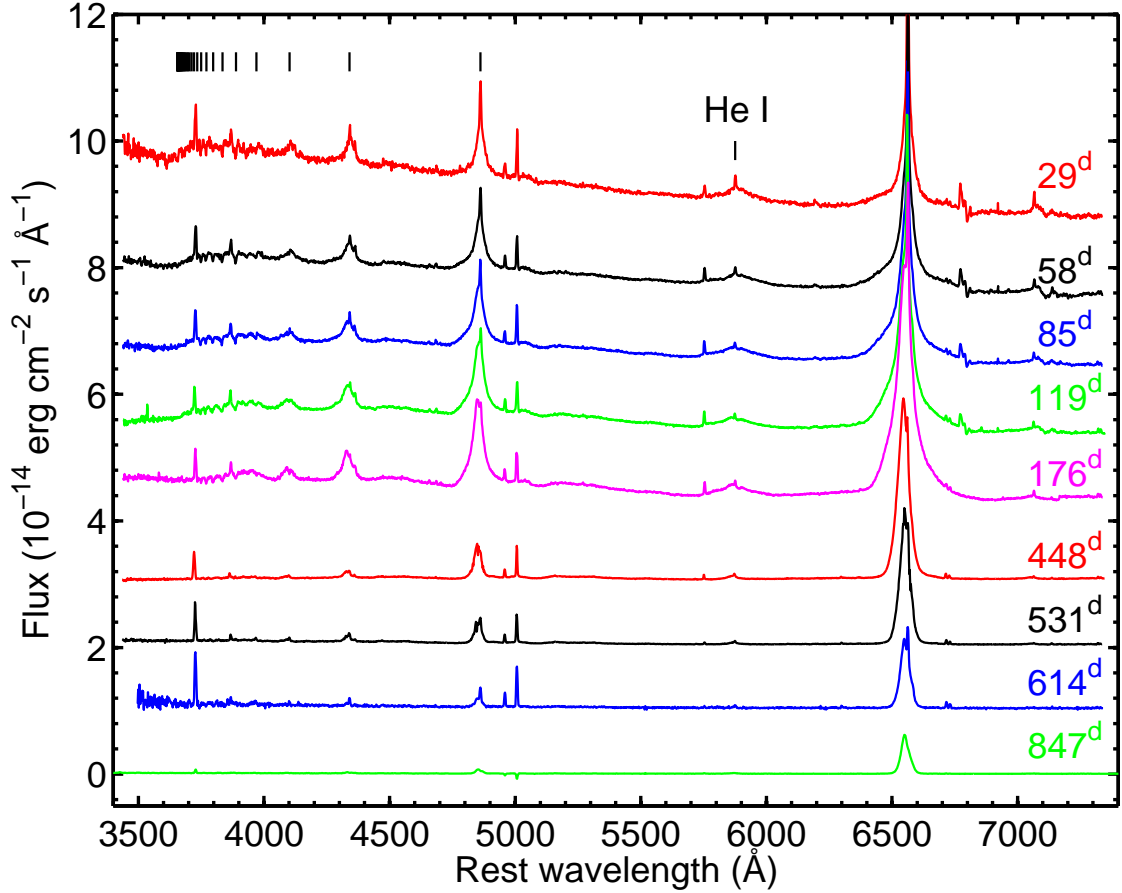}}
\caption{Spectral sequence in the optical from observations with FAST and MMT. Each spectrum has been shifted upwards by $10^{-14}  \  \mathrm{ erg ~s^{-1} ~cm^{-2}~ } \angstrom^{-1}$ relative to the one below. The wavelengths of the Balmer lines are shown, as well as the broad He I $\wl 5876$ line.}
\label{fig_fast_spec}
\end{center}
\end{figure*}

\begin{figure*}
\begin{center}
\resizebox{150mm}{!}{\includegraphics[scale=.80,angle=0]{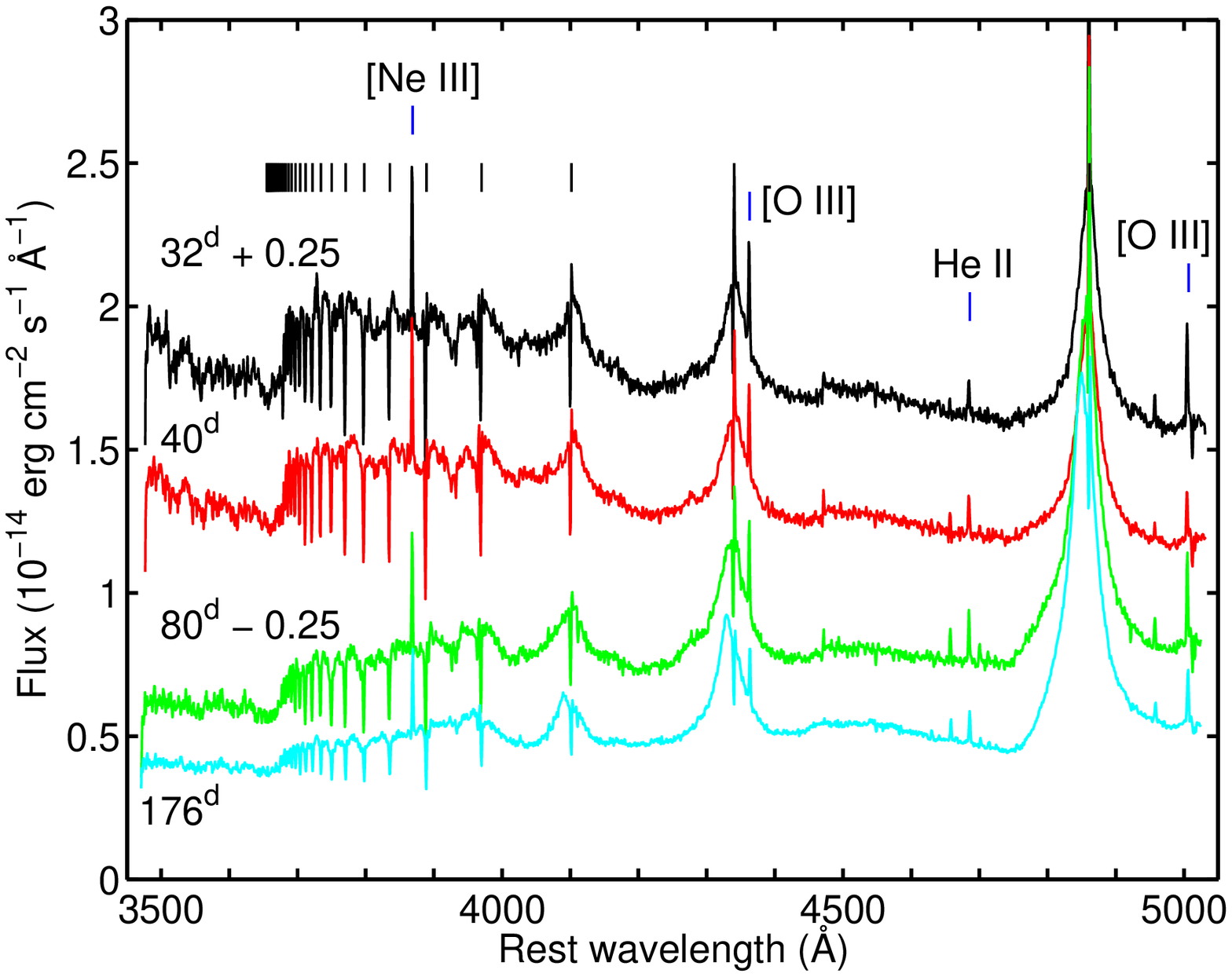}}
\caption{The grism 16 spectral series from NOT showing the Balmer series bluewards of 
H$\alpha$ as well as the Balmer decrement. Note the narrow lines due to [Ne III] $\wl 3868.8$, [O III] $\wll 4959, 5007$, and He II $\wl 4686$, and the unusually strong [O III] $\wl 4363$ line. The day 32 and day 80 spectra have been shifted upward and downward, respectively, by $2.5 \times 10^{-15}  \ergs cm^{-2} \angstrom^{-1}$.}
\label{fig1}
\end{center}
\end{figure*}

Figures \ref{fig_fast_spec} and  \ref{fig1} show the SN~2010jl spectral sequence for days 29 -- 847 obtained with FAST at FLWO and
with grism 16 at NOT, respectively. The former have the advantage of showing the full spectral interval between 3500 -- 7200 \AA, while the NOT spectra below 5100 \AA \  have a higher dispersion, showing the narrow line profiles better.  

From our first optical spectra at 29 days to the last at  848 days we see surprisingly little change in the lines present (Figure \ref{fig_fast_spec}). The main difference  is that the continuum is getting substantially redder with time. This is also apparent from the steep light curve of the u' band in Figure  \ref{fig_photometry}. At the same time the  Balmer discontinuity at 3640 \AA \ becomes weaker.  

The most conspicuous features of the spectra are the strong, symmetric Balmer emission lines, from H$\alpha$ up to H$\delta$. To the blue of H$\delta$, the broad emission lines blend into a continuum. In addition to these blended broad emissions in Figure \ref{fig1}, each Balmer line up to at least H16 shows a narrow blueshifted P-Cygni absorption. 
In addition to the Balmer lines
and several He I lines, some of which show prominent P-Cygni profiles (Sect. \ref{sec-narrow}), we also see
narrow emission lines from [Ne III] $\lambda3868$, [O III]
$\lambda\lambda4363, 4959, 5007$, several [Fe III] lines, as well as He II $\lambda4686$ (for more details see Sect. \ref{sec-narrow}). Our high resolution X-shooter spectrum reveals an additional large number of  weaker narrow lines (Sect. \ref{xshooter}).

\begin{figure*}
\begin{center}
\resizebox{150mm}{!}{\includegraphics[angle=0]{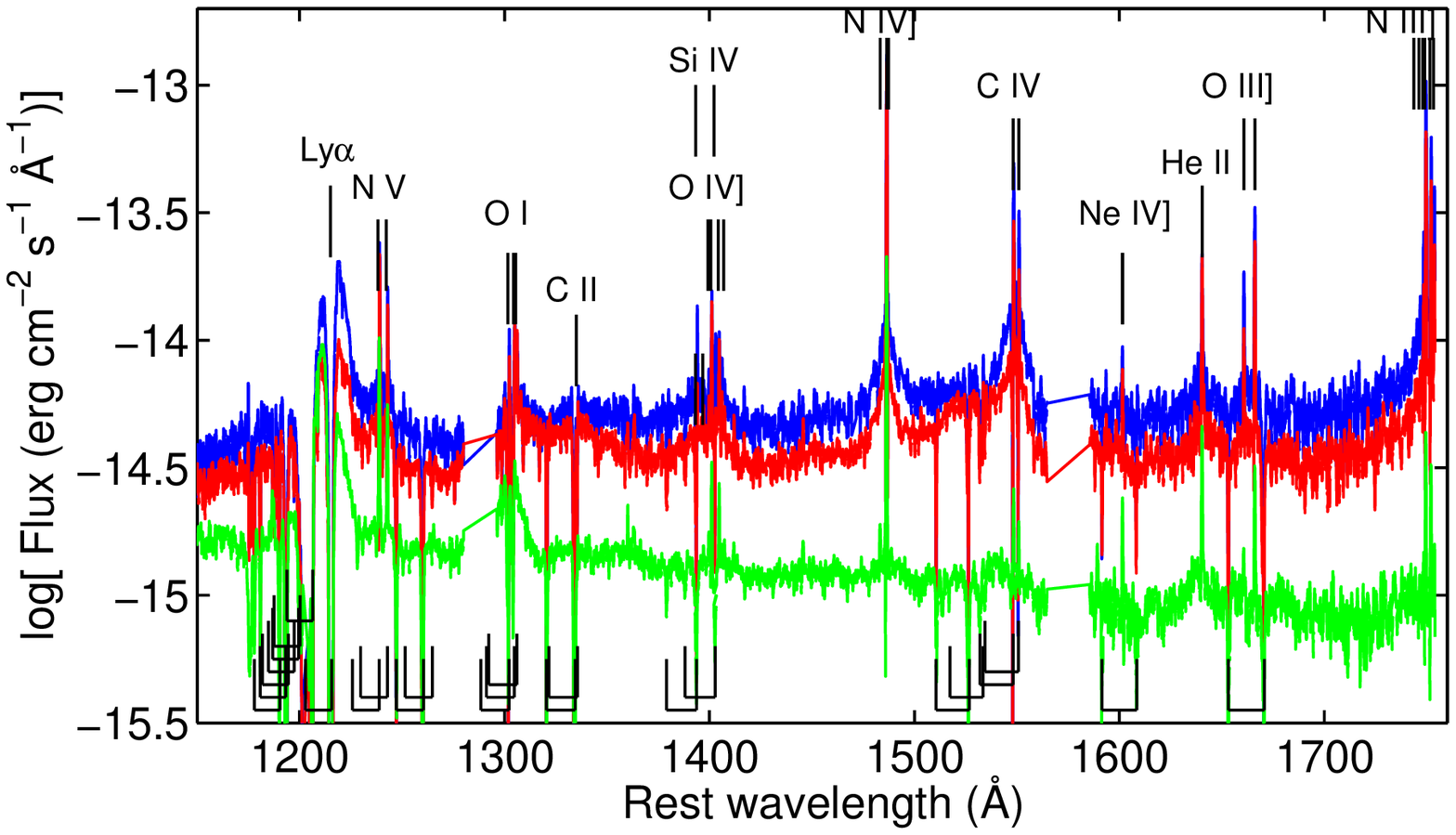}}
\caption{COS spectra from day 44 (blue), day 107 (red) and day 620 (green) with the most important lines marked. We also show the positions of the strongest absorptions from the Milky Way and host galaxy in the lower part of the figure. The continuum in the day 620 spectrum is dominated by the continuum emission from the host galaxy background.
}
\label{fig_full_spec}
\end{center}
\end{figure*}

\begin{figure*}
\begin{center}
\resizebox{125mm}{!}{\includegraphics[angle=0]{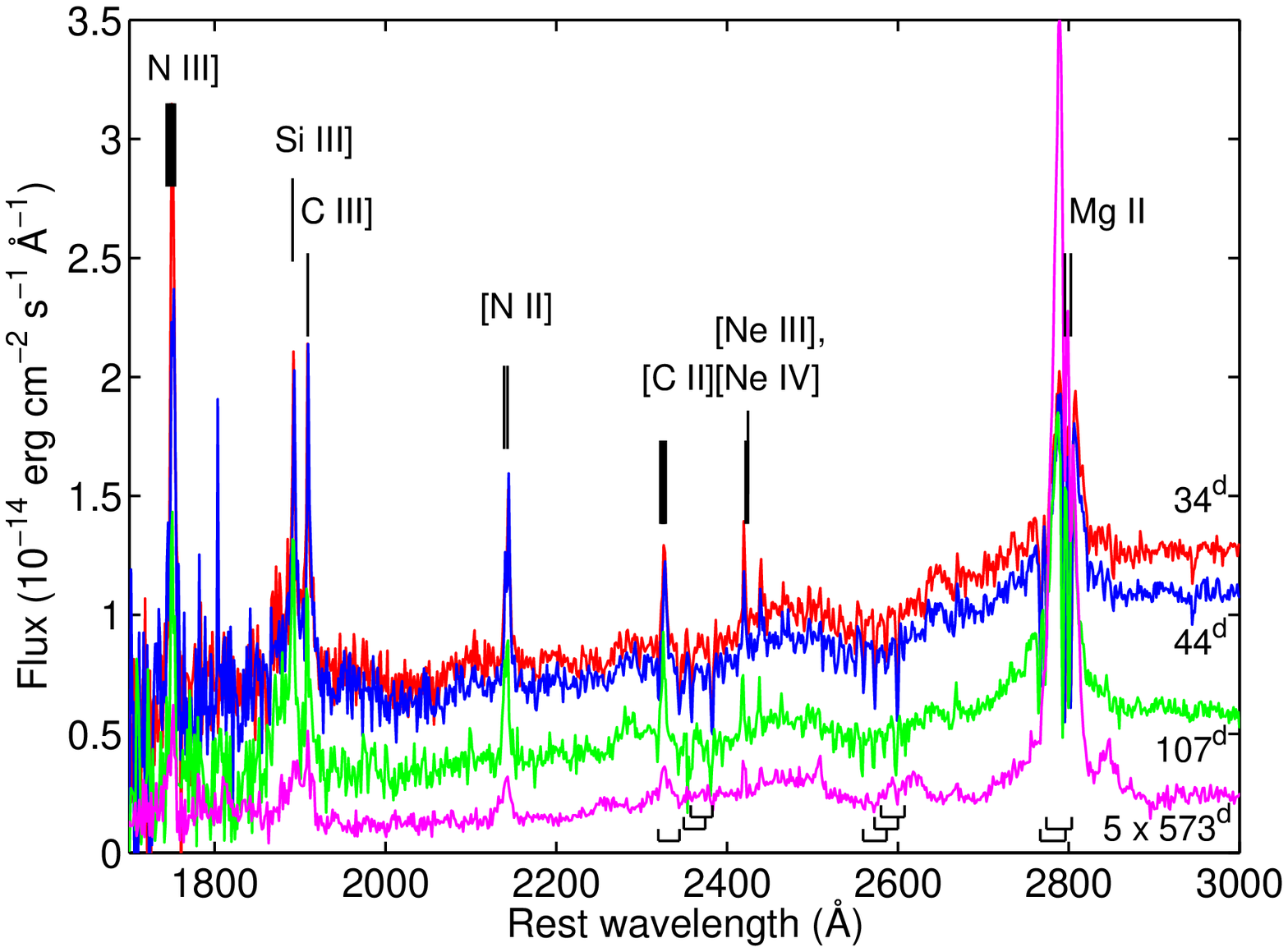}}
\resizebox{125mm}{!}{\includegraphics[angle=0]{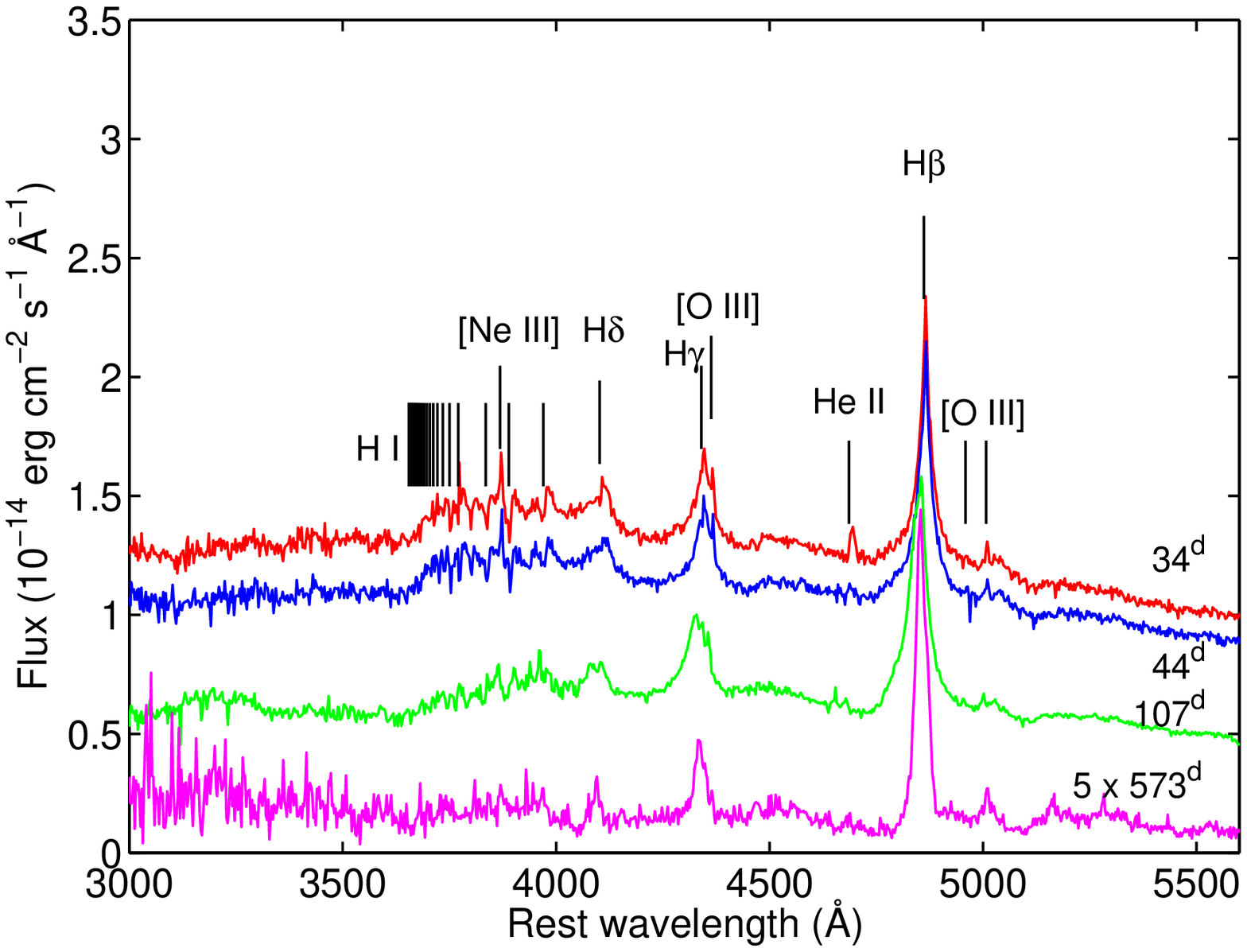}}
\caption{UV and optical STIS spectra from days 34 (red), 44 (blue), 107 (green) and 573 (magenta). The last spectrum has been multiplied by a factor of 5.0 for clarity. Note the strong [N II] and N III] lines
in the UV spectrum, as well as the broad, symmetric line profiles for the
strongest lines.  }
\label{fig_full_stis_spec}
\end{center}
\end{figure*}

Figure \ref{fig_full_spec} shows the far-UV spectrum with COS from
days 44, 107, and 621,  while Figure \ref{fig_full_stis_spec}
shows the STIS spectra from days 34, 44, 107 and 573. 
Again, we see a number of prominent, broad lines with strong, narrow emission components, but also in many cases P-Cygni absorptions. The main difference from the optical is that in the UV we find mainly  lines of highly ionized elements, like C III-IV, N III-V, O III-IV and Si III-IV, although there are also neutral and low ionization lines of C II, O I and Mg II. 
Comparison of the STIS spectra reveals little evolution
of the spectrum between the first two epochs. The COS spectra show a similar slow evolution. 
 At the time of the third epoch the continuum has faded by a factor of $\sim 2$, while the narrow lines have decreased by a modest amount. 

In addition to the lines from the SN, the spectrum also shows a number of interstellar absorption lines from the Milky Way and the host galaxy, shown in the lower part of Figure \ref{fig_full_spec}. In wavelength order  we identify these as Si II $\wll$ 1190.4, 1193.3, 1194.5, 1197.4, N I $\wll$ 1199.6, 1200.2, Si III $\wl$ 1206.5, 
N V $\wll$ 1238.8, 1242.8, Si II $\wll$ 1260.4, 1264.7, O I $\wll$ 1302.2, 1304.9, 1306.0, 
C II $\wll$ 1334.5, 1335.7, Si IV $\wll$ 1393.8, 1402.8, Si II $\wll$ 1526.7, 1533.4, 
C IV $\wll$ 1548.2, 1550.8, Fe II $\wl$ 1608.5, and Al II $\wl$ 1670.8. The lower S/N and spectral resolution of the STIS spectra makes ISM line identifications more difficult. We do, however, detect a number of strong lines, Fe II $\wll$ 2344.2, 2374.5, 2382.8, 2586.6, 2600.2, 2607.9, Mg II $\wll$ 2796.4, 2803.5. All these lines are commonly seen in different directions of the Milky Way as well as in different galaxies. 

\begin{figure}
\begin{center}
\resizebox{85mm}{!}{\includegraphics[angle=0]{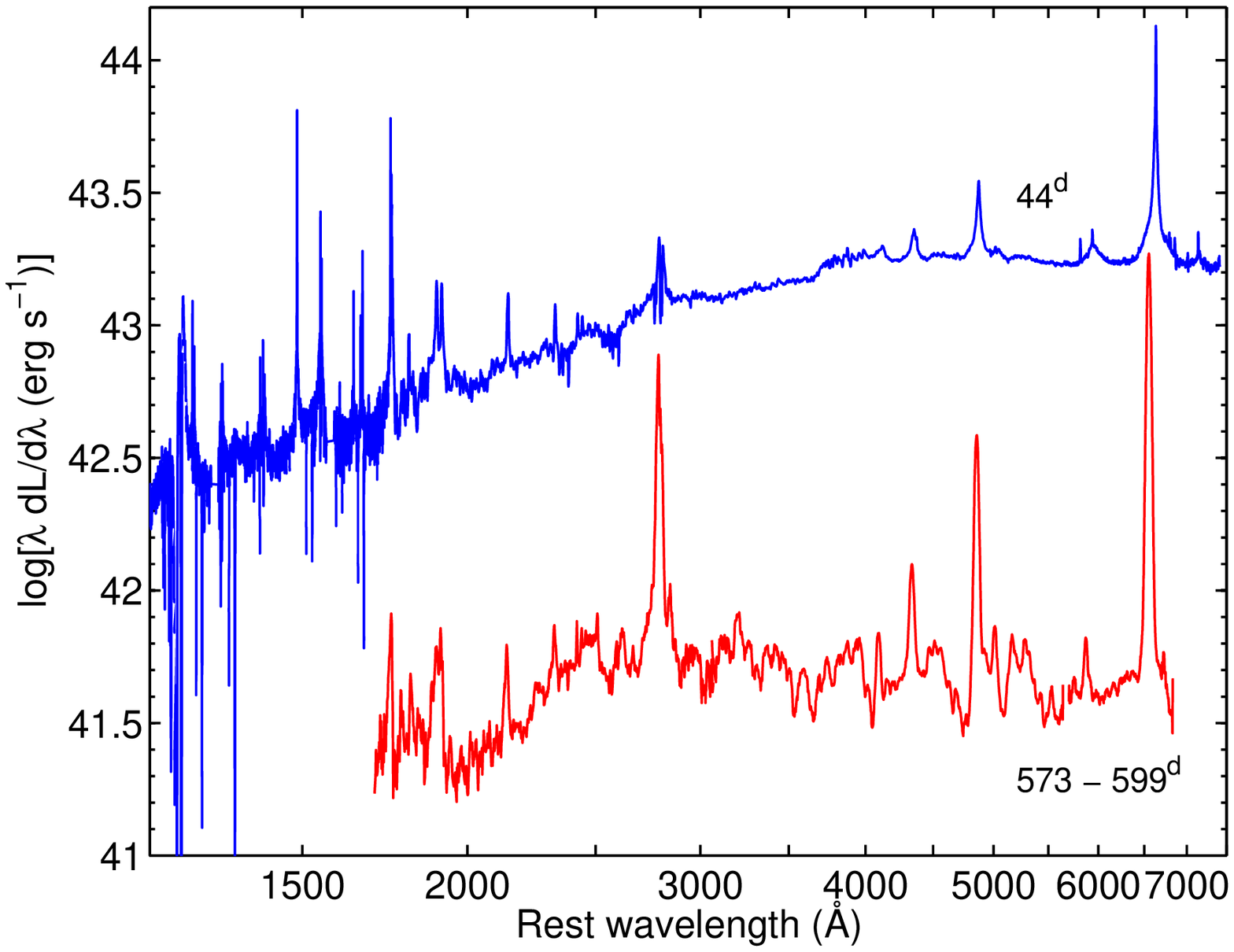}}
\caption{Combined UV and optical spectra from day 44 and day 573 -- 599 from  COS, STIS and ground based observations. For $\lambda > 5700$ \AA \ on day 44, we use  an interpolated spectrum from the FAST day 32 and day 58 observations. Both spectra are dereddened. We  plot $\log \lambda dL/d\lambda$ against $\log \lambda$, which gives the luminosity distribution per wavelength decade.}
\label{fig_full_opt_uv}
\end{center}
\end{figure}

To show the relative contributions from the UV and optical ranges we   show in Figure \ref{fig_full_opt_uv} the complete UV/optical spectrum at two epochs, one early and one very late. For the first epoch this is a combination of the COS and STIS spectra from day 44 and an interpolation of the FAST spectra from  day 32 and day 58 to this date. For the late spectrum we exclude the COS spectrum because of the strong contamination of the continuum by the host galaxy by the large COS aperture. The emission lines in this spectrum are well-defined and, with a few exceptions, free from contamination. 

From Figure \ref{fig_full_opt_uv} we note the increasing importance of the UV at late phases, as was also concluded from Figure \ref{fig_bollc}. This is similar to other Type IIn SNe, like SN 1995N \citep{Fransson2002} and SN 1998S \citep{Fransson2005}, but in contrast to Type IIP and Ibc SNe, which rapidly become faint in the UV. We also see a clear change in  character of the spectrum, from one dominated by a continuum at early epochs to one dominated by line emission at the last epochs. 

\subsection{Line profiles and flux evolution}
\label{sec-profiles}

The medium resolution optical spectra in
Figure \ref{fig1}, as well as the high S/N COS spectrum in
Figure \ref{fig_full_spec},  clearly show that there are two
distinct velocity components in the strongest emission lines. This is
apparent for the H I and Mg II lines, but also the high ionization UV
lines like C IV, N III-V, O III-IV, Si IV and N V. The broad
components extend  to $\sim 2400-10,000 \kms$, while the narrow components
have an order of magnitude lower velocities.  The maximum velocity to where the broad component can be traced depends mainly on the flux of the line and S/N of the spectra. For both of these reasons the H$\alpha$ line displays the most extended wings. We now discuss the properties of these components in more detail.

\subsubsection{The broad component}
\label{sec-broad}

\begin{figure}
\begin{center}
\resizebox{85mm}{!}{\includegraphics[scale=.60,angle=0]{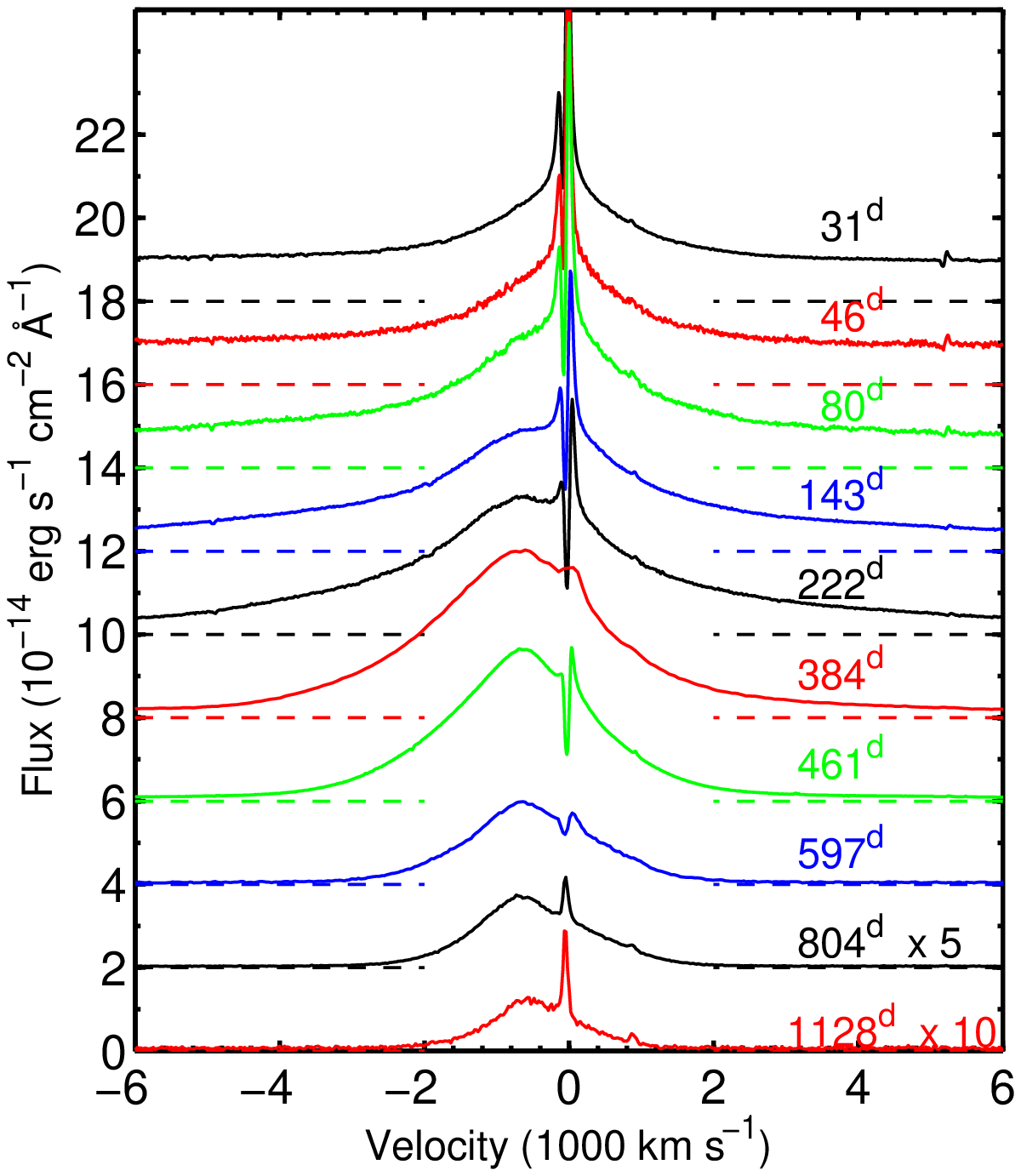}}
\caption{Evolution of the broad component of H$\alpha$ from day  31 to day 1128 from NOT grism 17  (upper five panels), FAST (day 384), X-shooter (day 461), MDM (day 597) and MMT (days 804 and 1128). Note the evolution from a symmetric profile centered at $v= 0$ to the late  blueshifted line profile. The narrow P-Cygni component is smoothed out in the lower resolution day 384 FAST spectrum and partly also in the day 597 MDM spectrum. Each spectrum is shifted by $2 \times 10^{-14} \ergs cm^{-2} \angstrom^{-1}$. The dashed lines show the zero level for each date. For clarity, the day 804 spectrum has been multiplied by a factor 5 and the day 1128 spectrum by a factor 10.}
\label{fig3b}
\end{center}
\end{figure}

In Figure \ref{fig3b} we show the SN~2010jl
spectral sequence of the H$\alpha$ line from day 31 to day 1128. The broad lines display a  smooth, peaked profile, characteristic of
electron scattering \citep{Munch1948,Auer1972,Hillier1991,Chugai2001}. At early times the
lines are symmetric between the red and blue wings, while  after
$\sim 50$ days they display a pronounced bump to the blue.   
As we discuss in detail in Sect. \ref{sec_broad}, we interpret this as a result of a macroscopic velocity. We discuss (and reject) the alternative, in which dust absorption determine the line shape, in Sect. \ref{sec_broad}.

\begin{figure}
\begin{center}
\resizebox{85mm}{!}{\includegraphics[scale=.60,angle=0]{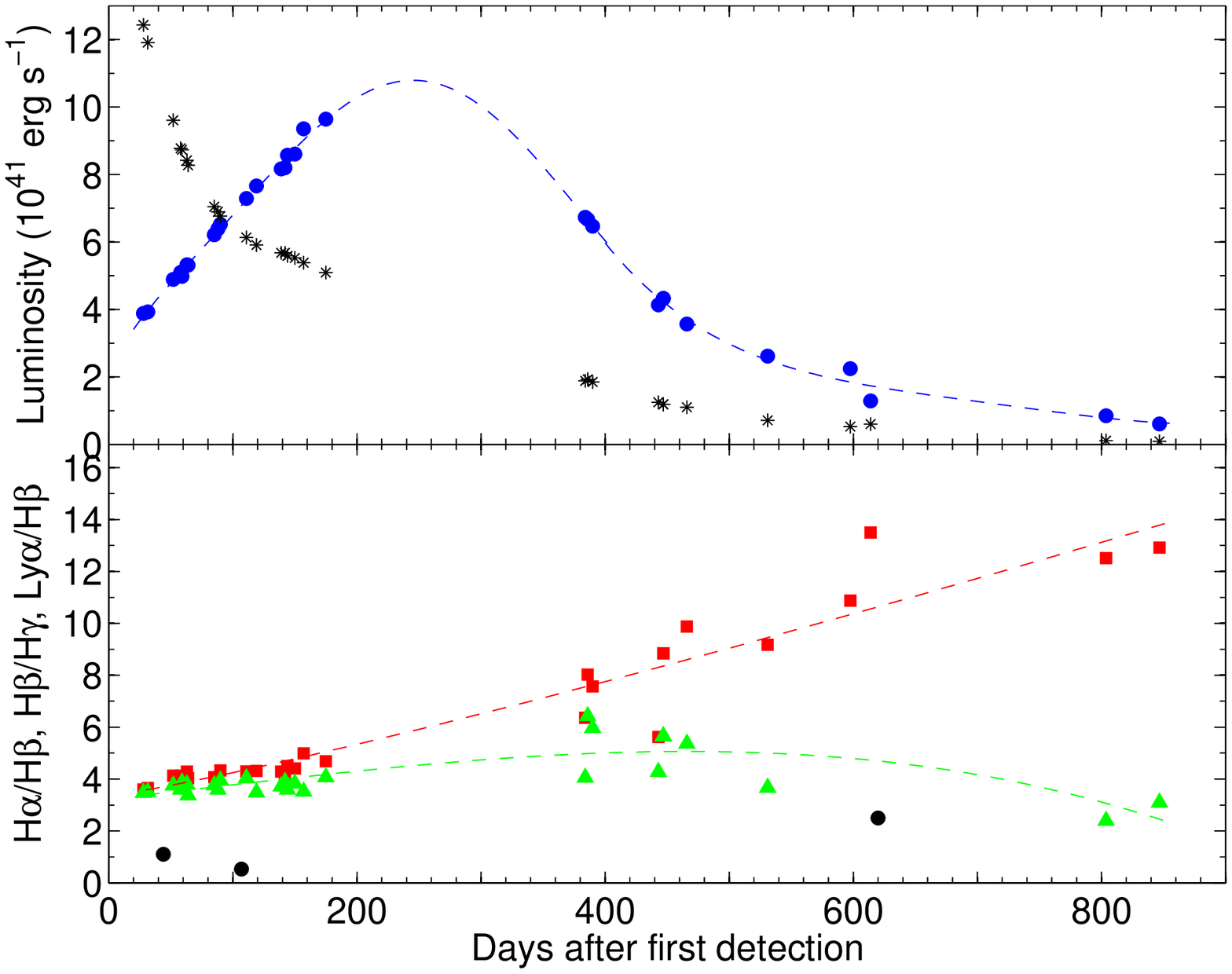}}
\caption{Upper panel: Evolution of the continuum subtracted flux of the broad H$\alpha$ line within $\pm 10,000 \kms$ (blue dots) and the continuum flux in the same region (black asterisks). Lower panel: The H$\alpha$/H$\beta$ ratio (red squares) and the H$\beta$/H$\gamma$ ratio (green triangles) for the same period. The dashed lines are polynomial least square fits to the respective data points. We have also added the Ly$\alpha$/H$\beta$ ratios for the three epochs where we have COS-spectra  (black dots). Luminosities and line ratios are corrected for reddening. }
\label{fig_haflux} 
\end{center}
\end{figure}

To calculate the flux of H$\alpha$,  H$\beta$, and H$\gamma$ we determine the continuum level on either side of the line  between $-15,000$ to $-12,000 \kms$ and between $12,000$ to $15,000 \kms$. A continuum level is then determined as a linear interpolation between these velocities, and subtracted from the total flux. Figure \ref{fig3b} shows that this should be an accurate  estimate.

In Figure \ref{fig_haflux} we show the H$\alpha$ luminosity together with the H$\alpha$/H$\beta$ and H$\beta$/H$\gamma$ ratios. The dashed lines give  polynomial least square fits to the data. In addition, we show the continuum flux in the same velocity interval,  $\pm 10,000 \kms$, as the lines. From the figure we see that  H$\alpha$  increases by a factor $\sim 2.5$ from day 29 to day 175.
After the first observational gap between $\sim 200$ days and $\sim 400$ days the flux has dropped somewhat, and then drops by another factor of $\sim 10$ from 400 to 850 days. From the polynomial fit it is likely that the maximum H$\alpha$ luminosity occurred at $\sim 260$ days when the SN was close to the sun. The bell shaped light curve of  the H$\alpha$ luminosity is in contrast to the continuum light curve in the same velocity region, shown as black dots, which decreases monotonically, and by a factor of $\sim 100$ from the first observations up  to 800 days. The total luminosity (continuum plus H$\alpha$) remains nearly constant up to $\sim 400$ days, after which it too drops. The increasing flux in H$\alpha$ during the first $\sim 300$ days explains the flatter light curve of the r' band shown in Figure \ref{fig_photometry}. 

For the first $\sim 500$ days our H$\alpha$ luminosity agrees well with the total luminosity from the `narrow' and `intermediate' components of \cite{Zhang} Their terminology is different from that used in this paper. Their `narrow' and `intermediate' components are both part of our `broad', while their data do not resolve what we refer to as the `narrow' component. Our light curve, however, covers approximately one more year, showing a continued decay of H$\alpha$. 

The H$\alpha$/H$\beta$ flux ratio increases steadily from $\sim 3.6$ initially to $10-15$ at $\sim 800$ days. The H$\beta$/H$\gamma$ ratio similarly increases  from $\sim 3.4$ initially to $\sim 5.0$ at $\sim 400$  days, after which it decreases to $\sim 2.8$ at 800 days (Figure  \ref{fig_haflux}).  For 
Balmer lines higher than H$\delta$ the broad components blend together into a  quasi-continuum,  and the Balmer decrement for these is therefore difficult to determine. 

\begin{figure}
\begin{center}
\resizebox{82mm}{!}{\includegraphics[angle=0]{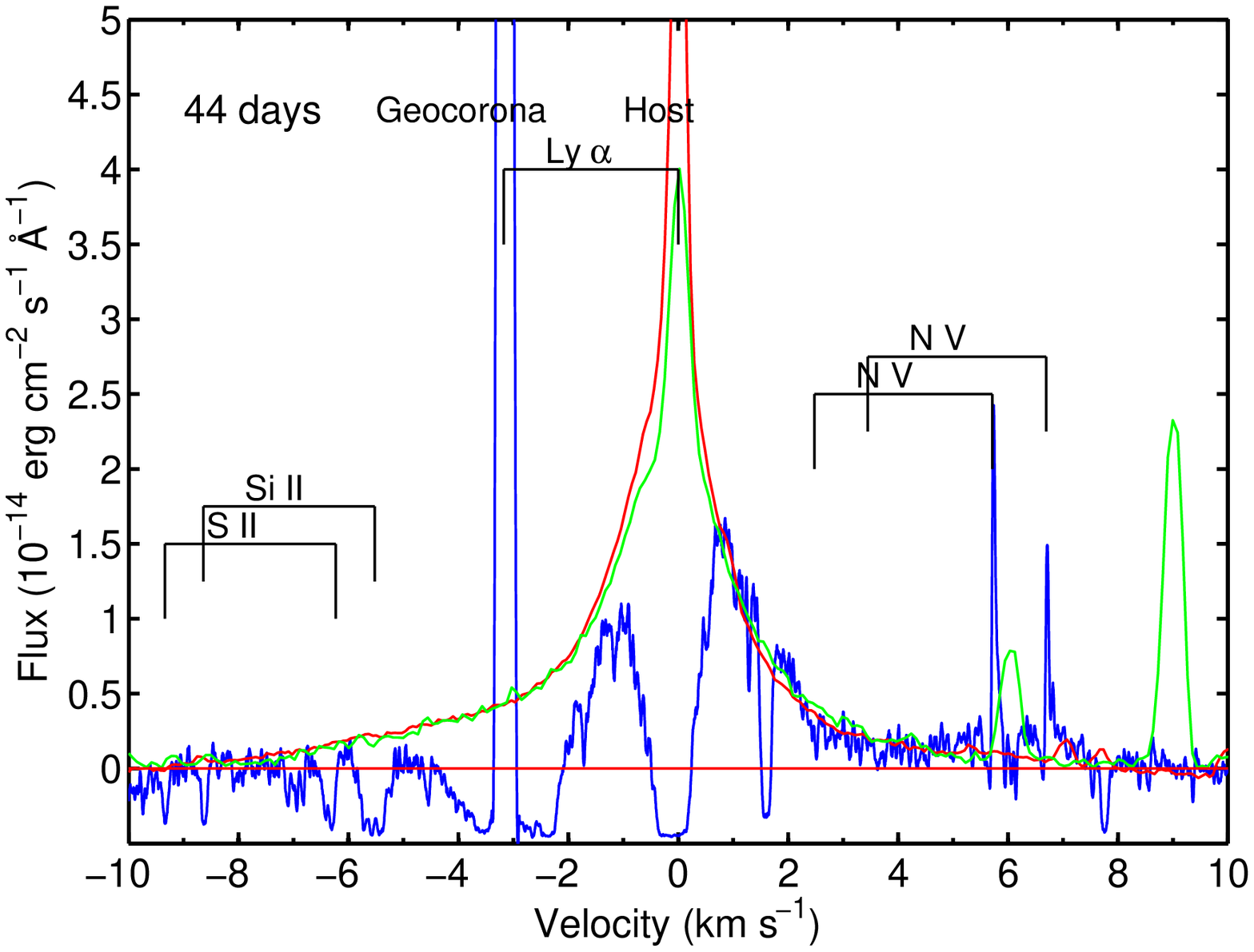}}
\resizebox{82mm}{!}{\includegraphics[angle=0]{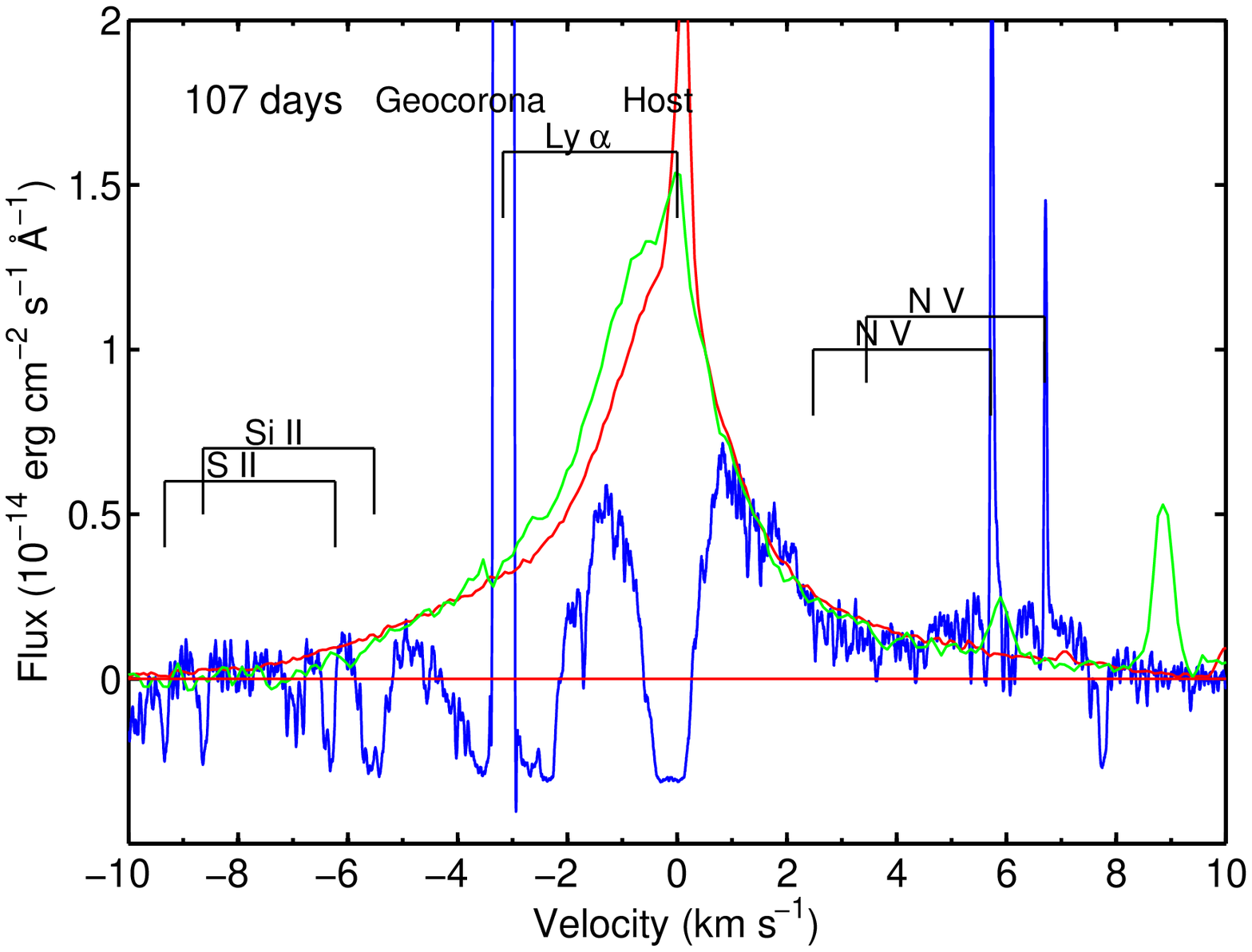}}
\resizebox{82mm}{!}{\includegraphics[angle=0]{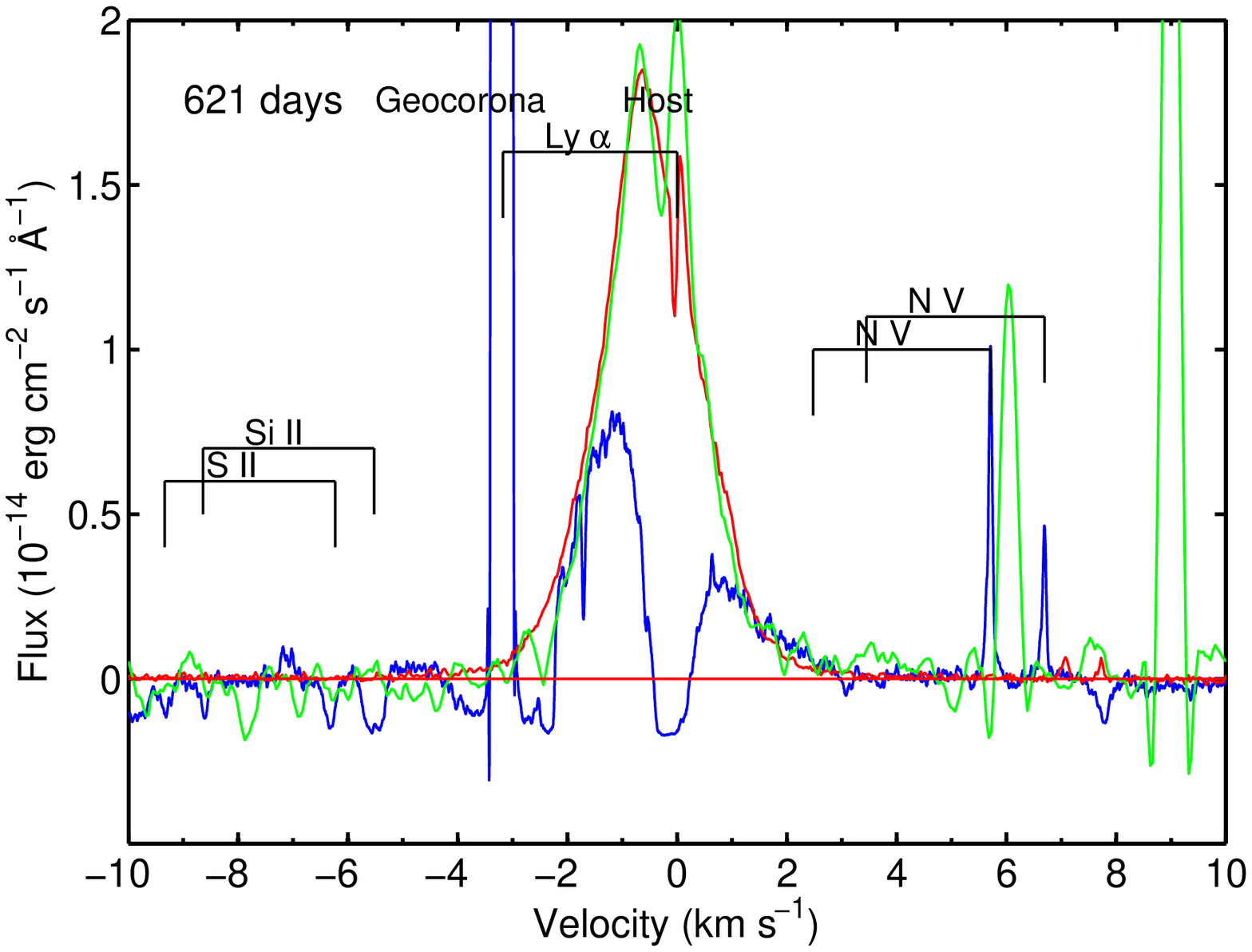}}
\caption{The Ly$\alpha$ line profile (blue) from the COS spectra of days 44,  107 and 621 compared to the H$\alpha$ (red) and H$\beta$ (green) lines. The scale and continuum level of H$\alpha$  and H$\beta$ lines have been adjusted to agree with those of Ly$\alpha$. The
  positions of the N V emission lines are marked, as well as the
  strongest interstellar lines from the host galaxy and our Galaxy in the Ly$\alpha$ spectrum.}
\label{fig_lprof_lya}
\end{center}
\end{figure}

The Ly$\alpha$ line, shown in Figure \ref{fig_lprof_lya}, is
severely distorted by interstellar absorption lines both from the
host galaxy and from the Milky Way. The broad Ly$\alpha$ absorption
from the host galaxy makes it impossible to determine if any narrow
emission is present. The red side of the broad component is relatively
unaffected by absorptions and extends to $\sim 2460 \kms$. The blue
side reaches into the damping profile of the host galaxy absorption. However, a
lower limit of $\sim 2300 \kms$ to the blue extent of the line can
be determined. As is shown by the comparison with the
H$\beta$ line from the same epoch STIS spectrum, the two lines have
similar profiles in the velocity ranges where Ly$\alpha$ is
unaffected by absorptions.  The lower flux of the Ly$\alpha$ line between $-2000 \kms$ and $- 1000 \kms$ compared the H$\alpha$ and H$\beta$ lines is caused by the overlapping ISM absorptions, as well as the Si III $\lambda 1206.5$ line (Sect. \ref{sec-redd}). A similar reasoning applies to the other epochs, although we also observe a change in the ratio of the blue and red segments of the line from the first two epochs to the last. This is caused by the blueshift of the line, so that more of the red side is absorbed by the host galaxy absorption, while the blue region is increasing for the same reason.

In Sect. \ref{sec-redd} we found the damping wings of the Ly$\alpha$ absorption imply a value of N$_H \sim (1.05\pm 0.3)\EE{20}$ cm$^{-2}$ for $b = 10 \kms$ from the host galaxy. This means that the large column density derived from the Chandra observations, $\sim 10^{24} \ \rm cm^{-2}$ \citep{Chandra2012}, must be intrinsic to the SN itself or its CSM.

  The ratios between the Ly$\alpha$, H$\beta$, and H$\alpha$ lines are useful as indicators of the conditions in the broad line region of AGNs \citep[e.g., ][]{Krolik1978,Fraser2013}. Because of the broad absorptions the total flux of the Ly$\alpha$ line can not be directly determined. Instead we estimate the line ratio of this line to H$\alpha$ and H$\beta$ by first subtracting the continuum and then scaling the line profiles so that these fit each other as well as possible for the parts of the lines unaffected by interstellar absorptions (see Figure \ref{fig_lprof_lya}). The scaling factor then gives the line ratio. 

Including reddening with $E_{\rm B-V}=0.058$ (Sect. \ref{sec-redd}), we find for the first COS observation at day 44  $F({\rm Ly}\alpha)/F({\rm H}\beta) = 1.1$ and   $F({\rm H}\alpha)/F({\rm H}\beta) = 2.8$, for day 107 $F({\rm Ly\alpha})/F({\rm H}\beta) = 0.53$ and $F({\rm H}\alpha)/F({\rm H}\beta)= 4.1$ and for day 620 $F({\rm Ly}\alpha)/F({\rm H}\beta) = 2.5$ and $F({\rm H}\alpha)/F({\rm H}\beta) = 9.0$. Because of the difficulty in the  calibration of the COS-spectra relative to the optical spectra the uncertainty in the Ly$\alpha/$H$\beta$ ratio is large. We have plotted the Ly$\alpha/$H$\beta$ ratio in Figure \ref{fig_haflux} as black dots. 

Although the uncertainties in the Ly$\alpha$ flux are large, we see from these ratios that the broad line region has a Balmer decrement, as well as a Ly$\alpha$/H$\alpha$ ratio, very different from the recombination values. This is similar to what is found in the dense broad line regions in AGNs, and indicates a combination of high optical depths in the lines and high densities. Models show that  column densities of $\ga 10^{24} \ \rm cm^{-2}$ in combination with high densities, $\ga 5 \times 10^8 \ \rm cm^{-3}$, are needed to get these ratios \citep{Krolik1978,Kwan1984,Fraser2013}. A flat X-ray spectrum also lowers the Ly$\alpha$/H$\alpha$ ratio and increases the  H$\alpha$/H$\beta$ ratio  \citep{Kwan1986}. There is, however, a substantial degeneracy between the density, column density and X-ray flux. In Sect. \ref{sec-big} we discuss this further based on real models. 
 
Besides the H I lines, also N V $\wll 1238.8, 1242.8$, O I $\wll
1302.2-1306.0$, O IV] $\wll 1397.2-1407.3$ + SI IV $\wll 1393.8, 1402.8$,
  N IV] $\wl 1486.5$, C IV $\wll 1548.2-1550.8$  have  broad
    components. This is probably also the case for He II $\wl 1640.5$, O
    III] $\wll 1660.8-1666.2$, N III] $\wll 1746.8-1754.0$, and Si III]
          $\wl 1892.0$ / C III] $\wl 1908.7$, although the flux of the
            broad component is comparable to the S/N for these lines.

\begin{figure}
\begin{center}
\resizebox{85mm}{!}{\includegraphics[angle=0]{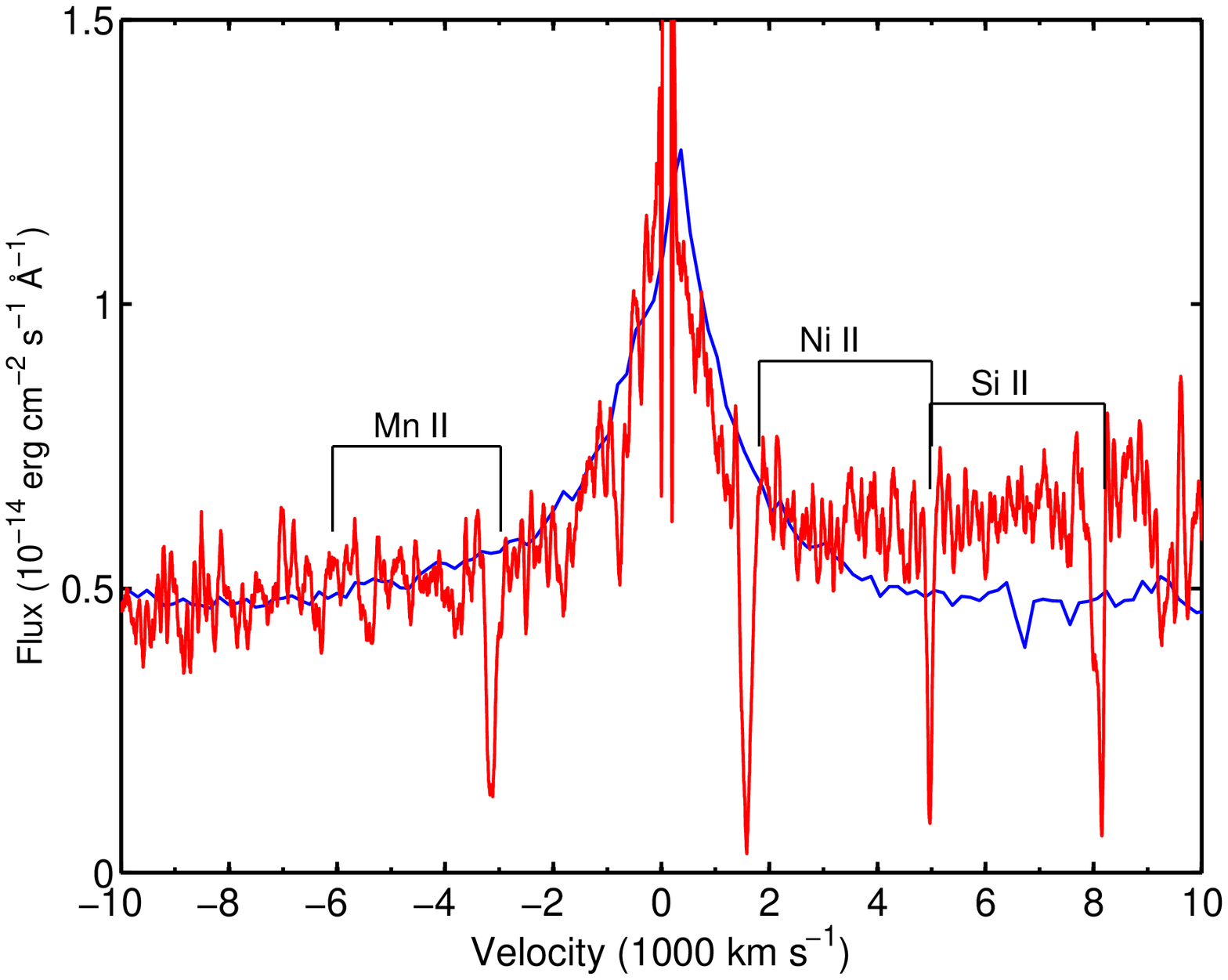}}
\resizebox{85mm}{!}{\includegraphics[angle=0]{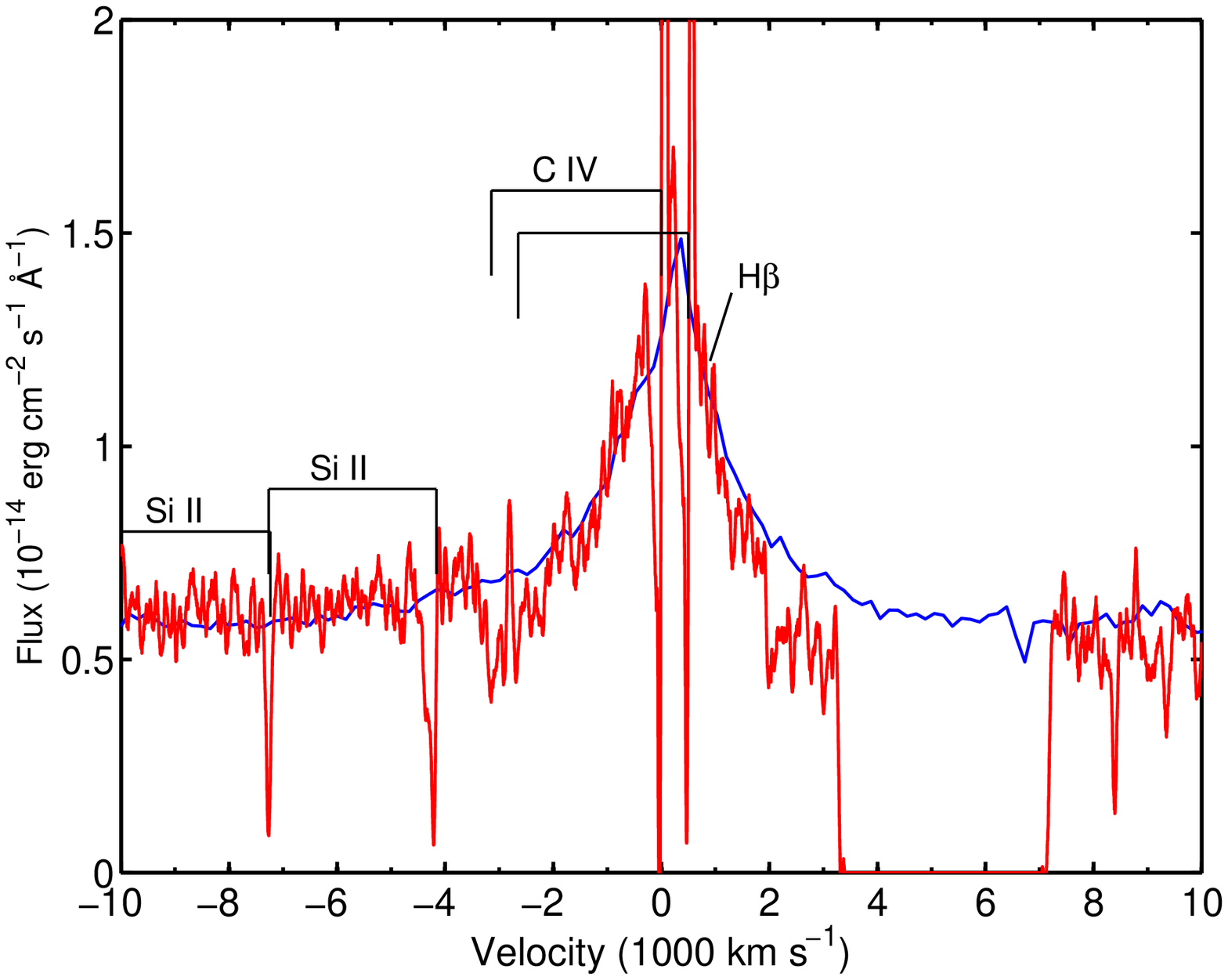}}
\caption{Comparison of the day 44 N IV]  (left) and C IV (right) lines with the H$\beta$ line profile (blue). The positions of some of the interstellar absorptions lines from the host and our Galaxy are marked. 
}
\label{fig_lprof_niv_civ}
\end{center}
\end{figure}

In Figure \ref{fig_lprof_niv_civ} we compare the broad
components from the strong C IV and N IV] lines with the profile of H$\beta$. Although the S/N in the COS observation is lower, 
 we see that the profiles are basically identical, as expected for
  electron scattering from lines originating at similar depths.

\begin{figure}[ht!]
\begin{center}
\resizebox{85mm}{!}{\includegraphics[angle=0]{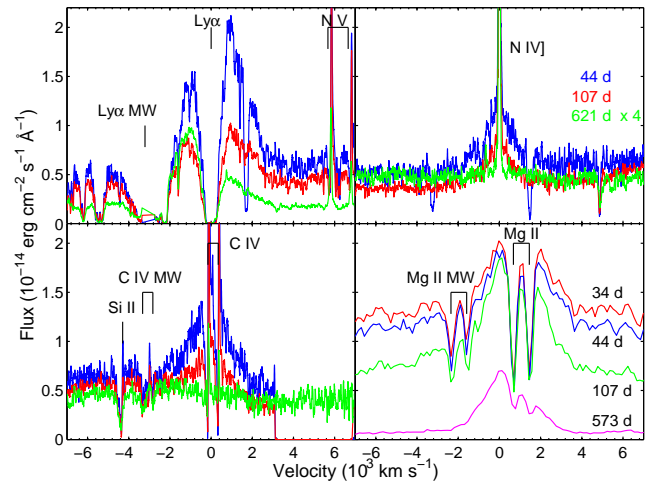}}
\caption{Evolution of the  profiles of the Ly$\alpha$, N IV], C IV and Mg II lines. The Ly$\alpha$, N IV], C IV spectra are from COS, while the Mg II profile is from the lower resolution STIS spectra. Note that for clarity the flux of the COS spectrum for day 621 has been multiplied by a factor of 4.0.
}\label{fig_lprof_lya_mgii}
\end{center}
\end{figure}

Figure \ref{fig_lprof_lya_mgii} shows the evolution of the most important broad UV lines from the COS and STIS observations. The scattering wings of both  the N IV] and C IV lines get fainter between the first two observations at 44 and 107 days, opposite to the evolution of the H$\alpha$ line.  At 621 days the wings of the high ionization lines have completely disappeared, in contrast to the Ly$\alpha$, O I $\wll
1302.2-1306.0$  and Mg II lines, which all still have strong wings in the last HST observations. The evolution of the latter lines agree more with H$\alpha$. 

\subsubsection{The narrow component}
\label{sec-narrow}

\begin{figure}[t!]
\begin{center}
\resizebox{85mm}{!}{\includegraphics[scale=.60,angle=0]{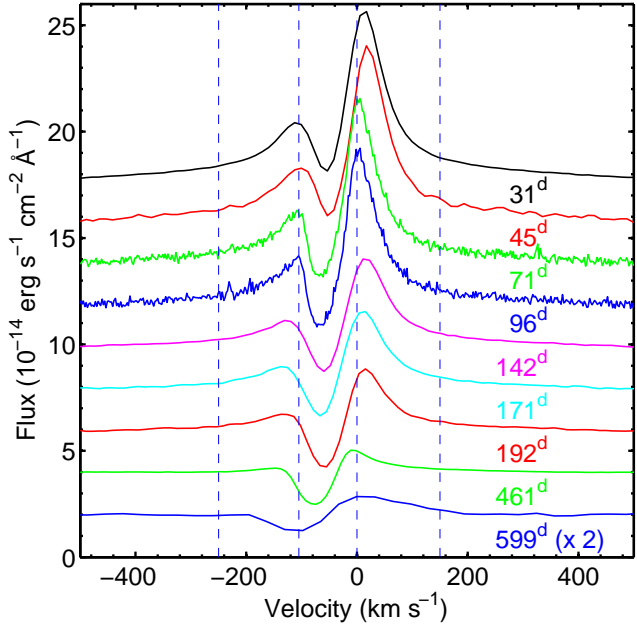}}
\caption{The narrow component of H$\alpha$ from day 31 to day 599. The `continuum' level has been set as the average flux between 300 -- 500 $\kms$ and each spectrum has then been shifted by $2\times 10^{-14} \ergs$ cm$^{-2}$ s$^{-1}$ \AA$^{-1}$ relative to the previous. The vertical dashed lines are at $-250$, $-105$, $0$ and  $+150 \kms$. The day 96 and day 142 spectra are from TRES, the day 461 from X-shooter and the day 599 spectrum from MDM, while the others come from NOT/grism 17. The flux of the day 599 spectrum has been multiplied by a factor 2. 
}
\label{fig3}, 
\end{center}
\end{figure}

As  noted by  \citet{Smith2011}, many of the optical
lines display a narrow component, in most cases with  a P-Cygni
profile, as shown in Figure \ref{fig1}. 
In Figure \ref{fig3} we show the narrow H$\alpha$ line on a larger velocity scale from NOT/Grism17, TRES, X-shooter and MDM spectra. The peak of the line is close to zero velocity, while the red wing reaches $\sim 150 \kms$, as marked by the vertical lines. The blue wing is more complex, consisting of both an absorption, reaching  $\sim -105 \kms$, and a high velocity emission reaching $\sim -250 \kms$.  These velocities do not change appreciably with time although the emission components get weaker. 

The emission wing on the blue side of the P-Cygni absorption is  likely to be caused by electron scattering in the CSM of the SN. 
Examples of this are known for  Wolf-Rayet stars, where  \citet[][see  his Figure 8]{Hillier1991} calculated the P-Cygni profile including electron scattering in the line forming zone and obtained a line profile similar to that seen for H$\alpha$. The expansion velocity of the CSM should therefore be $\sim -105 \kms$, rather than the higher velocity characteristic of the blue wing. This also shows that the electron scattering optical depth is not negligible in the CSM of the SN. 

\begin{figure}[t!]
\begin{center}
\resizebox{85mm}{!}{\includegraphics[scale=.60,angle=0]{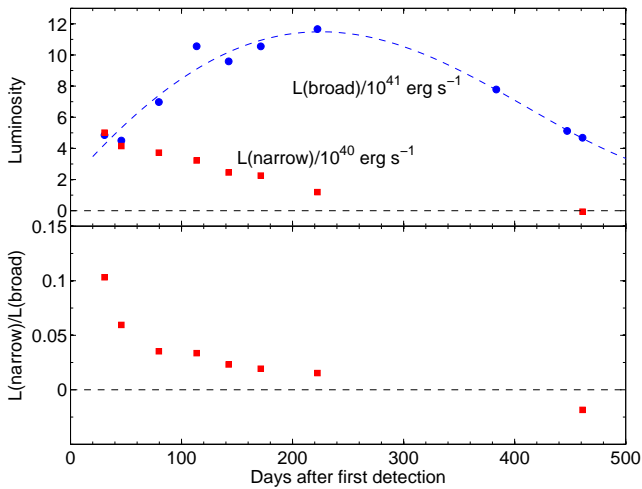}}
\caption{Upper panel: The luminosity of the broad and narrow components of H$\alpha$ from day 31 to day 461. Only spectra with resolution higher than FWHM 1.2 \AA \  have been included for the narrow component.  The day 96 and day 142 spectra are from TRES, the day 461 from X-shooter, while the others come from NOT/grism 17. For the narrow line we give the net luminosity in the P-Cygni line. The last measurement at 461 days has a stronger absorption component than emission component and the luminosity is therefore negative. Note that the luminosity scale of the narrow and broad lines are different by a factor 10. Lower panel: Ratio of these luminosities. }
\label{fig_narrowHa}, 
\end{center}
\end{figure}

In Figure \ref{fig_narrowHa} we show the net flux of the narrow H$\alpha$ line from our highest dispersion spectra. For the narrow component we have interpolated the broad line flux on each side of the line ($-250 \kms$ and $200 \kms$, respectively), and then calculated the net emission above this level. In the day 461 spectrum the P-Cygni absorption is stronger than the emission and the luminosity of the line is consequently negative. 

The main change with time of the narrow component is  a decrease of the flux (and equivalent width) of the line, from a net emission line to a line with  zero or even negative equivalent width. When we compare the broad and narrow emission we see that they evolve independently from each other. The narrow component is in fact more correlated to the continuum luminosity than to the broad H$\alpha$. 

Of the He I lines the following lines are detected: $\wl 3889$ (blended with H 8-2), $\wl 3965$ (P-Cygni), $\wl 4471$ (P-Cygni),  $\wl4713$  (only emission),  $\wl4922$ (probable), $\wl5016$ (P-Cygni), $\wl5876$ (P-Cygni ), $\wl6678$ (P-Cygni), $\wl7065$ (pure emission). Although some of these are seen in H II regions, the fact that P-Cygni profiles are observed for most of them shows that they originate in the SN environment. Additional He I lines are present in the NIR (Sect. \ref{xshooter}).

\begin{figure}
\begin{center}
\resizebox{85mm}{!}{\includegraphics[angle=0]{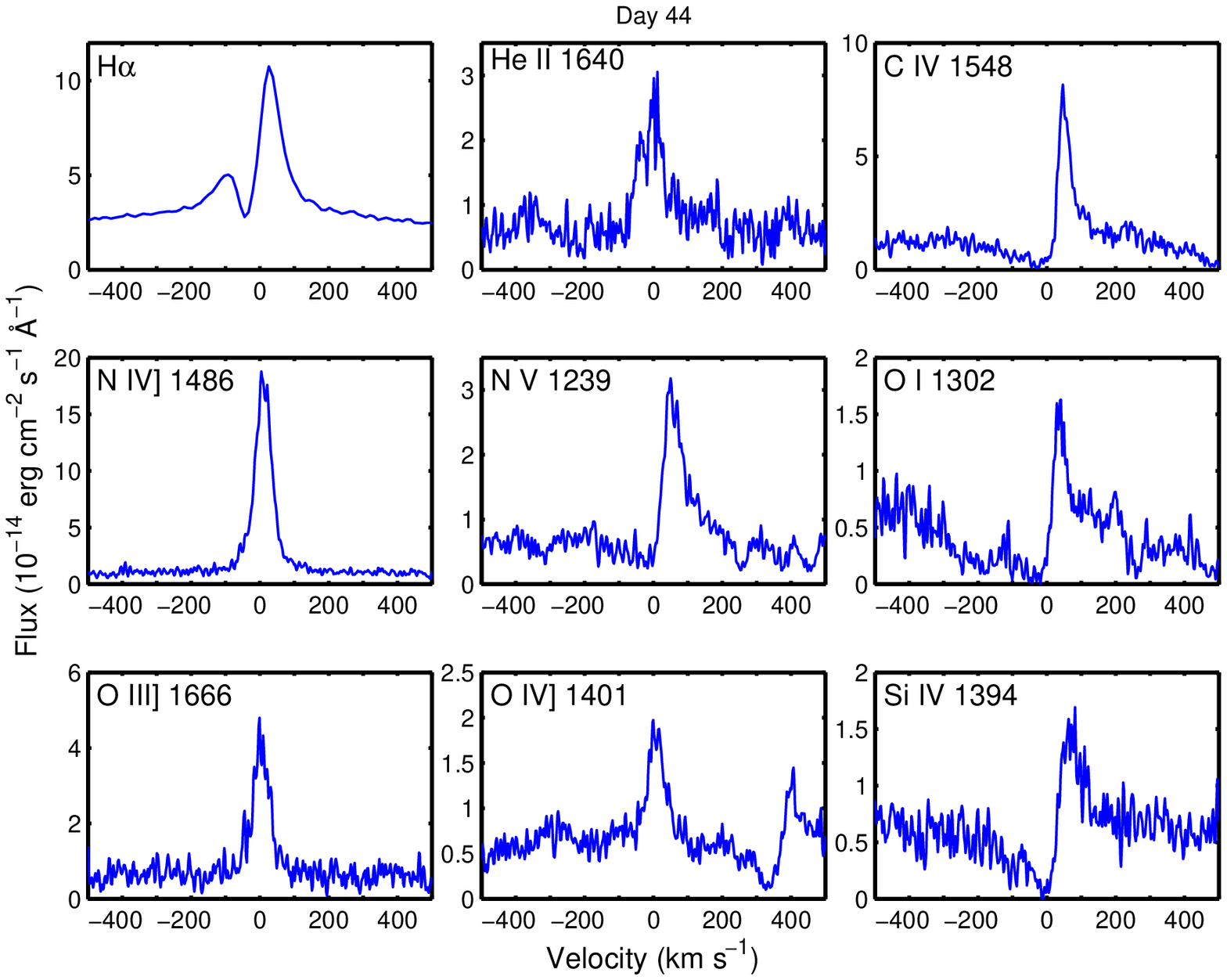}}
\resizebox{85mm}{!}{\includegraphics[angle=0]{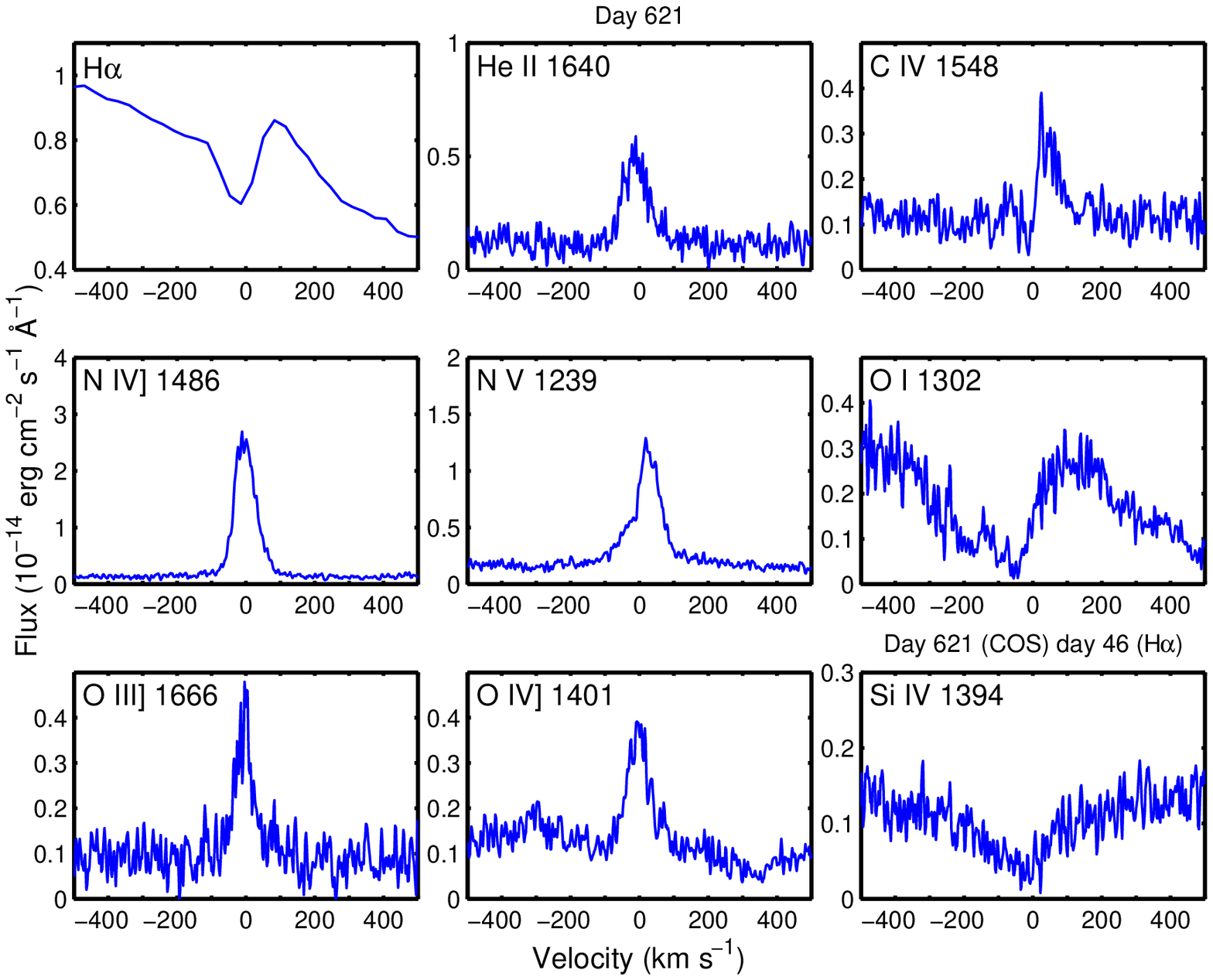}}
\caption{The central, low velocity region of the line profiles of the H$\alpha$, He II, C IV, N IV], N V, O I, O III], O IV] and Si IV lines on a velocity scale from day 44  (upper three rows) and day 621 (lower three rows). The H$\alpha$ spectra are from days 46 and 597, respectively.}
\label{fig_lprof1}
\end{center}
\end{figure}

Figure \ref{fig_lprof1} shows a sample of the most important other narrow emission lines in the UV and optical ranges
on a velocity scale for the day 44 and day 621 spectra. When comparing the line profiles of these lines one has to take into account the absorption of the UV resonance lines from the host galaxy at redshift 3207 $\kms$ and the Milky Way.  An important case is  the C IV  $\wll 1548.2, 1550.8$ doublet. 
The velocity separation of the $\wl$1550.8 component relative  to the $\lambda 1548.2$ component corresponds to 500 $\kms$. The absorption from the Milky Way does not interfere with the $\lambda 1548.2$ line below 2700 $\kms$. Absorption from the host may, however, be important. Typical rotational velocities are $100 - 200 \kms$, and this may well interfere with both the absorption and emission components, making a determination of the expansion velocity difficult for these lines. 

When we compare the different lines in Figure \ref{fig_lprof1} we see that the absorptions of  the O I $\wl$ 1302.2, Si IV $\wl 1393.8$ and the C IV $\wl 1548.2$ lines all are considerably wider than the intercombination lines. 
The strong N IV] $\wl 1486.5$ line has a FWHM $\sim 50 \kms$
  and extends to $\sim 100 \kms$ to the red side and $\sim 80 \kms$ to the
  blue. These velocities are similar to the red emission component of
  the C IV lines, but considerably less than the
  C IV absorption. Similar widths are found for the O III] $\wl 1666.2$ and He II $\wl 1640.4$ lines, although the S/N are lower for these lines. In contrast, the velocity of the blue edge of the C IV $\lambda 1548.2$ absorption
corresponds to $\sim -280 \kms$.  The extent of the strong emission
component to the red is, however, only $\sim 100 \kms$, although it is difficult to determine this
accurately.

The higher velocity of the absorption of the C IV  $\wll 1548.2, 1550.8$ lines probably originate in the host galaxy ISM, or possibly in higher velocity CSM gas of low density whose emission is too faint to be seen. This is also consistent with the other resonance lines. For the O I $\wl$ 1302.2 line one can even distinguish a separation between the P-Cygni absorption at $\sim 100 \kms$ and the higher absorption component. 
 
 When we compare the line profiles of the day 44 and day 621
spectra we do not see any clear evolution in velocity widths.
The velocities of the high ionization narrow UV lines are similar the optical lines, i.e., $\sim 100 \kms$. 

The fact that the Balmer and the He I lines have P-Cygni absorptions
show that the population of these  excited levels are high even at $\sim 600$ days. This
in turn argues for a high density, $\ga 10^8 \ccm$, in the line
forming region (Sect. \ref{sec-big}). In Sect. \ref{sec-fluxes} we discuss other independent density
determinations which indicate  high density circumstellar gas.

Tables \ref{table_stis_1} and
\ref{table_cos} give fluxes of the identified narrow UV lines from the STIS and COS spectra. 
To measure the fluxes of the different lines we first determine the
continuum level on both sides of the line and fit this to a
polynomial, which is subtracted from the total flux in the line. This introduces 
uncertainties in both the continuum level and the extent of the
line, which is most serious for the lower resolution STIS spectrum
which also has a lower S/N. 
In addition, the low spectral resolution of STIS also has
the consequence that multiplets, like N III] $\wll 1746.8 - 1754.0$, are not resolved. To complicate things further, several lines, like the N III], Si III] and C III] lines, sit on top of broad electron scattering profiles (see below). The higher spectral resolution COS
spectra are less sensitive to these effects.  Unfortunately, this setting of COS does not cover the C III] lines.  
\begin{deluxetable*}{llccccl}
\tabletypesize{\footnotesize}
\tablecaption{STIS Line Catalog. \label{table_stis_1}
}
\tablewidth{0pt}
\tablehead{
\colhead{Species} & \colhead{$\lambda_{vac}$} && \colhead{Line Flux}  & && \colhead{Notes}   \\ 
    &  \colhead{(\AA)} & & (10$^{-14}$ ergs cm$^{-2}$ s$^{-1}$) && \\
    &&day 34&day 44&day 107&day 573&
}
\startdata
O III]&1660.81-1666.15&2.8:&2.4:&--&&at wavelength limit \\
N III]&1746.82-1754.00&18.4\phantom{0}$\ \pm 2.0$\phantom{0}&13.5\phantom{0}$\ \pm 2.0$\phantom{0}&5.88$\ \pm 0.8$&0.60$\ \pm 0.10$&\\
Si III]&1882.71-1896.64&8.80$\ \pm 0.6$&9.12$\ \pm 1.5$&7.64$\ \pm 0.5$&0.51$\ \pm 0.15$&\\
C III]&1906.68-1909.60&8.50$\ \pm 1.2$&8.40$\ \pm 1.2$&5.20$\ \pm 1.0$&0.56$\ \pm 0.15$&\\
N II]&2139.68-2143.45&4.37 $\ \pm 0.1$&5.01$\ \pm 0.2$&3.36$\ \pm 0.2$&0.45$\ \pm 0.10$& \\
C II]&2324.21-2328.83&2.46$\ \pm 0.3$&2.75 $\ \pm 0.2$&2.12$\ \pm 0.2$&0.37$\ \pm 0.10$ &\\
Mg II&2796.34-2803.52&14.5\phantom{0}$\ \pm 0.5$\phantom{0}&25.6\phantom{0}$\ \pm 1.0$\phantom{0}&38.4\phantom{0}$\ \pm2.0$\phantom{0}&12.8$\ \pm 2.00$&broad component\\
$[$Ne III$]$&3869.85&1.44$\ \pm 0.1$&1.37$\ \pm 0.2$&0.77$\ \pm 0.3$&-& \\
H I&4102.89&7.73$\ \pm 0.3$&6.72$\ \pm 3.0$&5.43$\ \pm 1.8$&0.57$\ \pm 0.20$&broad component \\
H I&4341.68&18.6\phantom{0}$\ \pm 0.9$\phantom{0}&16.2\phantom{0}$\ \pm 2.5$\phantom{0}&19.1\phantom{0}$\ \pm1.5$\phantom{0}&2.39$\ \pm 0.30$&broad component \\
$[$O III$]$&4364.45&0.85$\ \pm 0.2$&0.56$\ \pm 0.3$&0.50$\ \pm 0.15$&0.12$\ \pm 0.03$& \\
He II&4687.02&1.45$\ \pm 0.2$&0.57$\ \pm 0.3$&$- $&-&marginal detection\\
H I&4862.68&46.9\phantom{0}$\ \pm 8.0$\phantom{0}&46.5\phantom{0}$\ \pm 7.0$\phantom{0}&58.7\phantom{0}$\ \pm 3.0$\phantom{0}&9.03$\ \pm 0.50$&broad component \\
$[$O III$]$&5008.24&0.79$\ \pm 0.2$&0.70$\ \pm 0.3$&\phantom{0}0.30$\ \pm 0.15$&0.23$\ \pm 0.10$& \\
 \enddata
\end{deluxetable*}
\begin{deluxetable*}{lcccccccl}
\tabletypesize{\footnotesize}
\tablecaption{SN2010jl COS Line Catalog. \label{table_cos} }
\tablewidth{0pt}
\tablehead{
\colhead{Species} & \colhead{$\lambda_{vac}$} &&& \colhead{Line Flux}&&&Notes   \\ 
    & (\AA)  &&& (10$^{-15}$ ergs cm$^{-2}$ s$^{-1}$)&&&&\\
    &&day 44&&day 107&&day 620&&\\
}
\startdata
N V      &   1238.82&    -0.76\phantom{0} &$\pm 0.3$ &   -0.44\phantom{0}&$\pm 0.2$&0.08&$\pm 0.05$&P-Cygni abs \\
N V      &   1238.82&     7.38 & $\pm 0.5$&    6.64&$\pm 0.5$&4.24&$\pm 0.10$& \\
N V      &   1242.80&    -0.68\phantom{0}&$\pm 0.2$&    -0.35\phantom{0}&$\pm 0.2$&-0.64&$\pm 0.20$&P-Cygni abs \\
N V      &   1242.80&     3.25&$\pm 0.5$&     2.96 &$\pm 0.4$&1.68&$\pm 0.08$&\\
O I      &   1302.17&    -2.40\phantom{0} &$\pm1.0 $&    -2.05\phantom{0} &$\pm 0.9$&-2.01&$\pm$&P-Cygni abs\\
O I      &   1302.17&     1.71&$\pm 0.6$&     1.33 &$\pm 0.6$&-&&\\
O I      &   1304.86&    -&-&    -&-&-1.42&$\pm$&complex P-Cygni abs \\
O I      &   1304.86&     2.64 &$\pm 0.3$&    2.05 &$\pm 0.3$&-&&\\
O I      &   1306.03&     1.55&$\pm0.7$&     1.84 &$\pm 0.3$&-&&\\
Si II     &   1309.28&    -0.61\phantom{0} &$\pm 0.3$&    -0.57\phantom{0} &$\pm 0.2$&&&P-Cygni abs\\
Si II     &   1309.28&     0.37 &$\pm 0.5$&     0.07 &$\pm 0.4$&&&\\
Si IV    &   1393.76&    -2.21\phantom{0} &$\pm 0.9$&    -1.66\phantom{0} &$\pm 0.5$&-1.23&$\pm 0.15$&P-Cygni abs \\
Si IV    &   1393.76&     3.36&$\pm 1.6$&     1.04 &$\pm 0.4$&-&&\\
O IV]    &   1397.20&     0.53&$\pm 0.2$&     0.39 &$\pm 0.2$&-&& \\
O IV]    &   1399.78&     0.92&$\pm 0.2$&     0.77 &$\pm 0.2$&0.22&$\pm 0.10$&\\
O IV     &   1401.16&     4.04&$\pm 0.2$&     3.59 &$\pm 0.3$&0.59&$\pm 0.20$&\\
Si IV    &   1402.77&    -1.53\phantom{0} &$\pm 0.5$&    -0.99\phantom{0} &$\pm 0.3$&-1.03&$\pm$&P-Cygni abs\\
Si IV    &   1402.77&     2.16 &$\pm 0.9$&     0.48 &$\pm 0.3$&-&&\\
O IV]    &   1404.81&     2.20 &$\pm 0.5$&     2.52 &$\pm 0.3$&0.56&$\pm 0.15$&\\
S IV]    &   1406.00&     1.06 &$\pm 0.2$&     0.72 &$\pm 0.2$&-&&\\
O IV]    &   1407.38&     0.40 &$\pm 0.2$&     0.57 &$\pm 0.2$&-&&\\
N IV]    &   1483.32&    -&&-&&0.25&$\pm0.05$&\\
N IV]    &   1486.50&    54.7\phantom{00} &$\pm 1.8$&    51.7\phantom{00} &$\pm 1.5$&9.61&$\pm0.10$&\\
C IV     &   1548.19&    -4.25\phantom{0} &$\pm 1.7$&    -2.85\phantom{0} &$\pm 0.5$&-&&P-Cygni abs\\
C IV     &   1548.19&    15.9\phantom{00} &$\pm 1.0$&     8.98 &$\pm 0.8$&0.58&$\pm0.07$&\\
C IV     &   1550.77&    -4.36\phantom{0} &$\pm 0.4$&    -2.54\phantom{0} &$\pm 0.5$&-&&P-Cygni abs\\
C IV     &   1550.77&     8.90&$\pm 2.0$&     4.90 &$\pm 0.7$&0.31&$\pm0.05$&\\
Ne IV]    &   1601.5?&     -&& -&&0.65&$\pm0.07$&\\
He II    &   1640.40&     8.55 &$\pm 0.3$&     8.64 &$\pm 0.3$&1.81&$\pm0.10$&\\
O III]   &   1660.81&     4.96&$\pm 0.2$&     2.88&$\pm 0.2$&0.53&$\pm0.10$&\\
O III]   &   1666.15&    12.6\phantom{00} &$\pm 0.3$&     9.64 &$\pm 0.3$&1.48&$\pm0.10$&\\
N III]   &   1746.82&     1.53 &$\pm 0.5$&    0.73 &$\pm 0.3$&-&&\\
N III]   &   1748.65&     5.51 &$\pm 0.7$&     3.89 &$\pm 0.3$&-&&\\
N III]   &   1749.67&    42.9\phantom{00} &$\pm 0.8$&    27.1\phantom{00} &$\pm 0.7$&1.62&$\pm0.07$&\\
N III]   &   1752.16&    20.8\phantom{00} &$\pm 0.7$&    15.1\phantom{00} &$\pm 0.4$&0.86&$\pm0.05$&\\
N III]   &   1754.00&     7.44&$\pm 0.7$&     6.47 &$\pm 0.3$&0.23&$\pm0.05$&\\
 \enddata
\end{deluxetable*}

A further complication, when using these lines for diagnostic purposes, is that all the resonance lines in the UV have P-Cygni components. The photons scattered out of the line of sight (LOS) to the observer, which give rise to the absorption component, are partly compensated for by those scattered into the LOS from the area of the wind outside of the projected photosphere. The net emission in a line from the CSM comes from the emission component plus the scattered photons. The fraction of the photons from the back of the SN, and which are in the line of sight of the optically thick ejecta, do not contribute to the emission. An upper limit to this contribution is the absorbed flux relative to the continuum in the absorption component. The exact fraction depends on the emissivity as function of the radius from the photosphere. For this reason we list in Table \ref{table_cos} both the flux in the emission component and the flux deficit in the absorption component for the COS spectra. 

An inspection of Table \ref{table_stis_1}, as well as Figure \ref{fig_full_stis_spec}, shows that between the first two epochs there is relatively little evolution  of the flux of the narrow lines in the STIS range, although the continuum has decreased somewhat. At 107 days the fluxes of both continuum and lines have, however, decreased by a factor of $\sim 2$. The same trends can be seen in the COS spectra (Table \ref{table_cos} and Figure \ref{fig_full_spec}). The decrease in the narrow line fluxes correlates  with the broad component of the high ionization UV lines (Figure \ref{fig_lprof_lya_mgii}).
In the last STIS spectrum at 573 days both lines and continuum have dropped by a factor $\sim 10$, although the Mg II $\wll 2796, 2804$ only  by a factor $\sim 3$.

\subsubsection{The X-shooter spectrum at 461 days}
\label{xshooter}

\begin{figure}
\begin{center}
\resizebox{85mm}{!}{\includegraphics[angle=0]{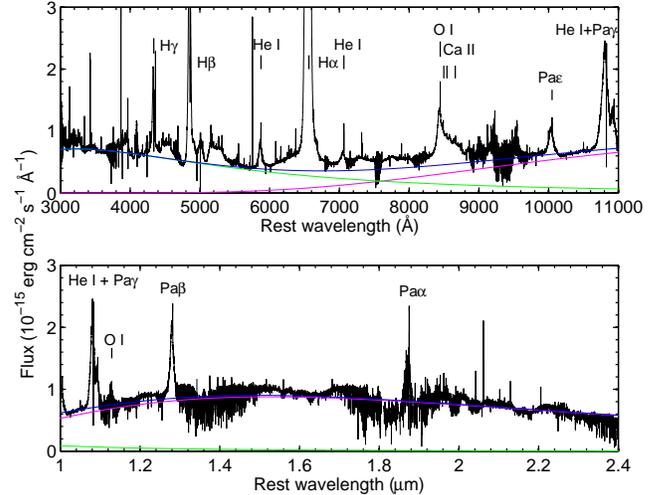}}
\caption{X-shooter spectrum from 2012 January 14   (day 461). The continuum has been fit with a two-component blackbody with temperatures of $9000$ K and $1870$ K, respectively. Note the many narrow lines from the circumstellar gas, as well as the strong, broad He I and H I lines in the NIR. 
}\label{fig_xshooterspec}
\end{center}
\end{figure}

We also obtained one medium/high resolution X-shooter spectrum from the VLT for 2012 Jan. 14  (day 461), shown in Figure \ref{fig_xshooterspec}. Because of the higher spectral resolution and NIR coverage we discuss this spectrum separately. 

Starting with the continuum, we note the excess in the NIR with a maximum at $\sim 1.5 \ \mu$m. This NIR excess agrees well with the rise in the J, H and K$_s$ bands we see in the photometry in Figure \ref{fig_photometry} later than $\sim 400$ days and is most likely caused by hot dust emission; its origin  will be discussed in Sect. \ref{sec_dust}. 
We fit this  well with a blackbody spectrum at a temperature of $\sim 1870$ K. The corresponding continuum in the optical is  fitted with a temperature of $\sim 9000$ K.  The value for the dust temperature differs by $\sim 8 \%$ from that obtained from the photometry in Table \ref{table_dust_param},  2045 K. This gives an estimate of the errors in this parameter from the errors in the photometry, spectral calibration and fitting errors. Because of the many lines in the optical region the error in the photospheric temperature is at least as large as this, although they here happen to be very close. 

In the NIR the most interesting lines are the strong broad and narrow components of He I $\wl 10,830$, as well as Pa$\alpha$, Pa$\beta$ and Pa$\gamma$. The He I $\wl 10,830$ confirms the identification of the He I $\wl 5876$ line in the optical. The $\wl 2.0581 \mu$m line is, however, only seen as a narrow absorption. 

We identify the broad, asymmetric  feature with a peak at $\sim 8435$ \AA \ with  O I $\wl 8446$. The extended red wing is a blend of  the Ca II triplet $\wll 8498.0, 8542.1, 8662.1$ and higher members of the Paschen series,  ($n = 12 - 15$). The identification with the O I $\wl 8446$ line is supported by the presence of the O I 1.129 $\mu$m line. These lines are likely to be a result of Ly$\beta$ fluorescence with O I, where the  upper level of the 1.129 $\mu$m line is pumped by Ly$\beta$, and then decays into the 1.129 $\mu$m and $\wl 8446$ lines \citep[][]{Grandi1980}. The relative flux of these lines  agrees well with that expected, although blending of the O I $\wl 8446$ line prevents a detailed comparison. 

O I $\wl 8446$ was also seen in the Type IIn SN 1995G \citep{Pastorello2002}, although in this case no NIR spectrum exists. 
There is also a line consistent with O I $\wl 8446$ in the SN 1995N spectrum in \cite{Fransson2002}, although this was identified as mainly Fe II $\wl 8451$. In the case of SN 2010jl there are, however, no lines corresponding to Fe II $\wll 9071, 9128, 9177$, which would be expected to accompany the Fe II $\wl 8451$ line.

\begin{figure*}
\begin{center}
\resizebox{150mm}{!}{\includegraphics[angle=0]{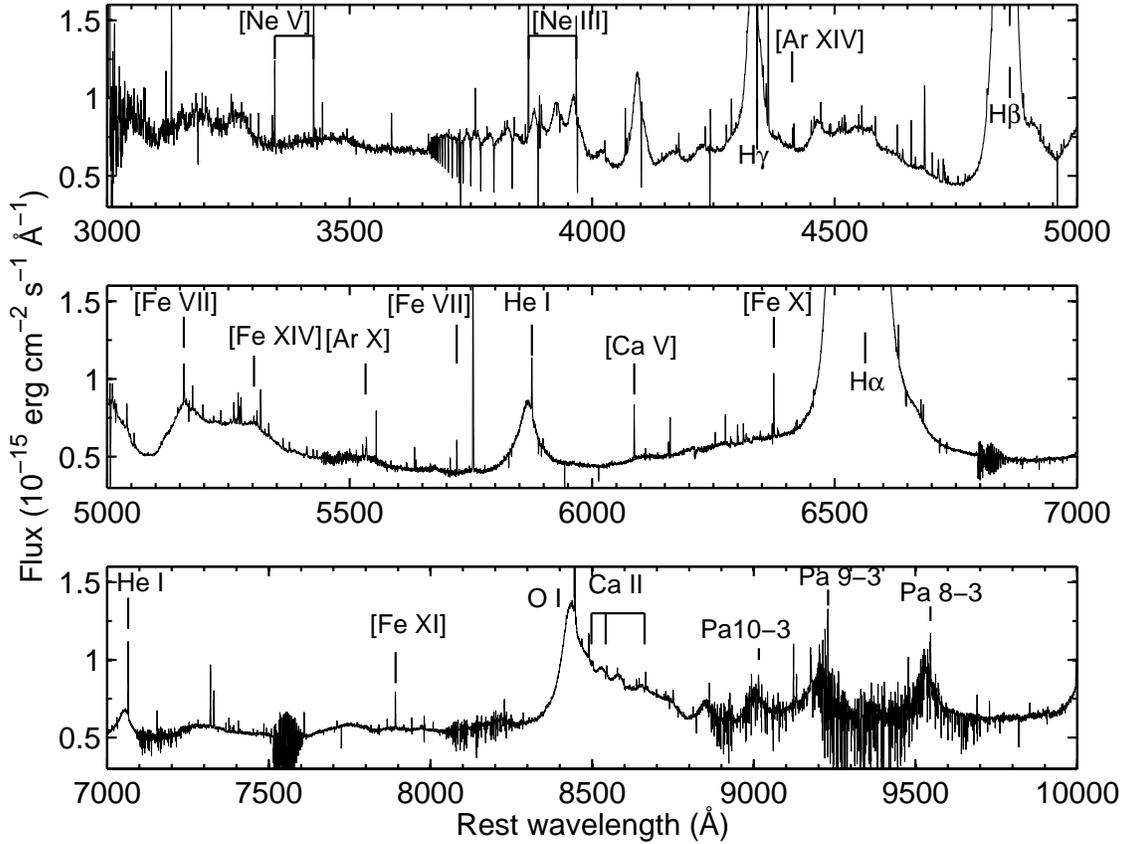}}
\caption{The optical part of the X-shooter spectrum from 2012 January 14  (day 461). A number of the most important lines from the CSM have been marked. 
}\label{fig_xshooterspec_opt}
\end{center}
\end{figure*}

The high resolution in combination with high S/N  allows us to identify a large number of weak, narrow emission lines not seen in the other  lower S/N spectra spectra which cast additional light on the CSM. In Figure \ref{fig_xshooterspec_opt} we show a close-up of the optical range with some of the most important lines marked. 
The most interesting of these lines are the large range of ionization stages of forbidden lines from [Fe V-XIV]. The [Fe VII] lines serve as diagnostics of the density and the excitation conditions in the CSM and are discussed in Sect. \ref{sec-fluxes}. 

Of the coronal lines [Fe X] $\wl 6374.5$ is strong, as well as  [Fe XI] $\wl 7891.8$, while [Fe XIV] $\wl 5302.9$ is only barely seen as a weak line. [Fe XIII] $\wl 10,746$ is not seen, but sits on the strong wing of He I $\wl$10,830.
Other high ionization lines include [Ne V] $\wll 3345.8, 3 425.9$, [Ca V] $\wl 6086.8$, [Ar X]  $\wl 5533.2$ and a weak  [Ar XIV]  $\wl 4412.6$. The presence of these lines  confirms the high state of ionization in the CSM of this SN. 

In these high S/N spectra the Balmer series can be traced  up to $n = 33$ ($\wl 3659.4$) with strong absorption components, but  only weak P-Cygni emission.
The Paschen series is seen from Pa$\alpha$ to $n=17$ ($\wl 8467.3$). In contrast to the Balmer lines, these  lines are all in pure emission. 
Br$\gamma$ is also seen in emission. 

\subsection{Density diagnostics of the narrow lines}
\label{sec-fluxes}

Using  fluxes from the previous section we use the forbidden and intercombination lines in the UV and optical as diagnostics of the density.

The ratios of the N III] $\wll 1746.8, 1748.7, 1749.7,1752.2, 1754.0$
  lines are sensitive to the electron density \citep{Keenan1994}. In
  particular, the $\wll 1752.2/1749.7$ ratio is $0.55-0.58$ in the range
  $10^5 - 10^8 \ccm$. Above this density the ratio decreases to become $0.20$ at
  $10^{10} \ccm$.   The $\wll 1754.0/1749.7$ ratio has a similar
  behavior. Unfortunately, the $\wl 1754.0$ line is at the very
  boundary of the wavelength range of the G160M grating, and its flux is
  uncertain. The $\wl$ 1748.7 and $\wl$ 1754.0 lines have the same upper level
  with transition rates within a few percent of each other, so  we
   use the $\wll$ 1748.7/1749.7 ratio instead of the $\wll 1754.0/1749.7$ ratio for the diagnostics. 
  
\begin{figure}
\begin{center}
\resizebox{85mm}{!}{\includegraphics[angle=0]{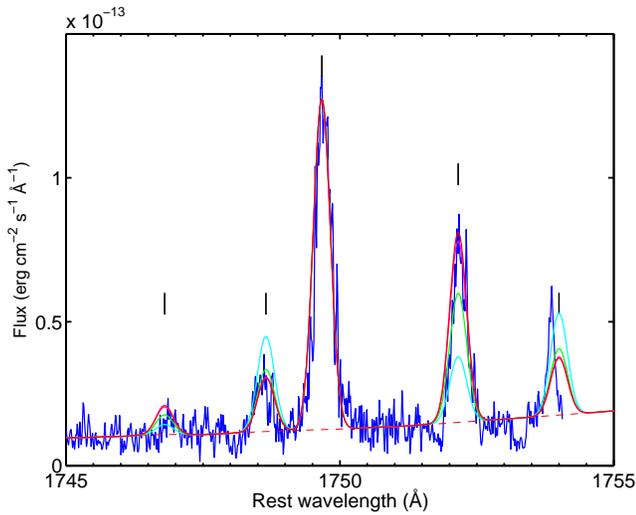}}
\caption{Fits to the N III] $\wll 1746.8, 1748.7, 1749.7,1752.2, 1754.0$ multiplet (marked by the vertical lines) from the COS spectrum from day 44. The red line is for $10^7 \ccm$, the magenta for $10^8 \ccm$, the green for  $10^9 \ccm$, and the cyan for $10^{10} \ccm$. The dashed red line marks the continuum. The dip at 1753.3 \AA \ is an interstellar absorption line. 
}
\label{fig_niii_fit_spec}
\end{center}
\end{figure}

  Figure \ref{fig_niii_fit_spec} shows a fit of these lines for
  different electron densities between $10^7 - 10^{10} \ccm$ for the day 44 COS spectrum. The fits show that (with the exception of the
  $\wl 1754.0$ line) all lines agree with an electron density  in the range $10^5 - 10^8
  \ccm$. Higher values give  too low a value for the
  $\wll 1752.2/1749.7$ ratio. We have also examined the day 107 and day 621 COS
  spectra and find essentially the same density constraint, although
  the S/N is somewhat lower.

\begin{figure}
\begin{center}
\resizebox{85mm}{!}{\includegraphics[angle=0]{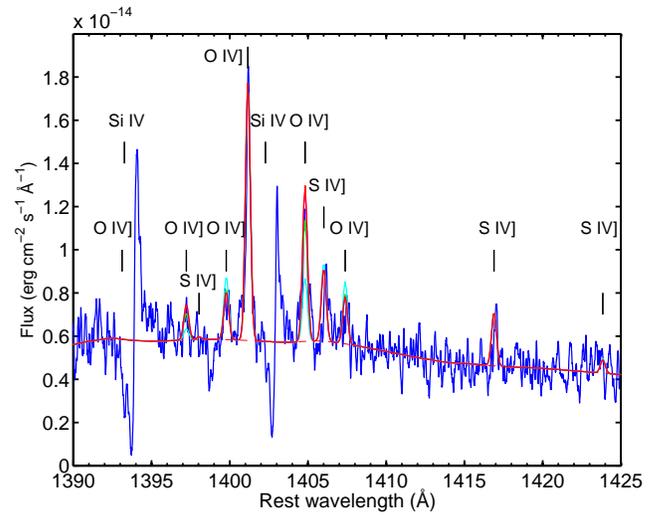}}
\caption{Same as Figure \ref{fig_niii_fit_spec}  for the O IV] $\wll 1393.1, 1397.2, 1399.8, 1401.2, 1404.8, 1407.4$ and S IV] $\wll 1398.0, 1404.8, 1406.0, 1416.9, 1423.8$ multiplets. The red line is for $10^7 \ccm$, the magenta for $10^8 \ccm$, the green for  $10^9 \ccm$, and the cyan for $10^{10} \ccm$. The solid red line marks the continuum. We have not attempted a fit of the Si IV $\wll 1393.3, 1402.3$ P-Cygni lines}.
\label{fig_oiv_fit_spec}
\end{center}
\end{figure}

Figure \ref{fig_oiv_fit_spec} shows a similar analysis of the O IV]
$\wll 1393.1, 1397.2, 1399.8, 1401.2, 1404.8, 1407.4$ and S
  IV] $\wll 1398.0, 1404.8, 1406.0, 1416.9, 1423.8$
    multiplets. As the red curve shows, the lines are fit with the
    same density interval, $10^5 - 10^8 \ccm$, as the N III]
      lines. This is not surprising since these ions are expected to
      arise in the same region.

The ratio of the N IV] lines at 1482.9 \AA \ and 1486.1
  \AA \ is sensitive to  density in the range $10^4 - 10^{8} \ccm$
  \citep{Keenan1995}, with a ratio of $\sim 1.6$ below $10^4 \ccm$ and
  decreasing above this density. From the June 20 2012 spectrum we estimate a $\wll 1482.9/1486.1$ flux ratio of $0.029 \pm 0.05$. From \citet[][their Figure 1]{Keenan1995} we find that this corresponds to an electron density $(3-6) \times 10^6 \ccm$ for temperatures in the range $10^4 - 2\times 10^4$ K. 
  
\begin{figure}
\begin{center}
\resizebox{85mm}{!}{\includegraphics[angle=0]{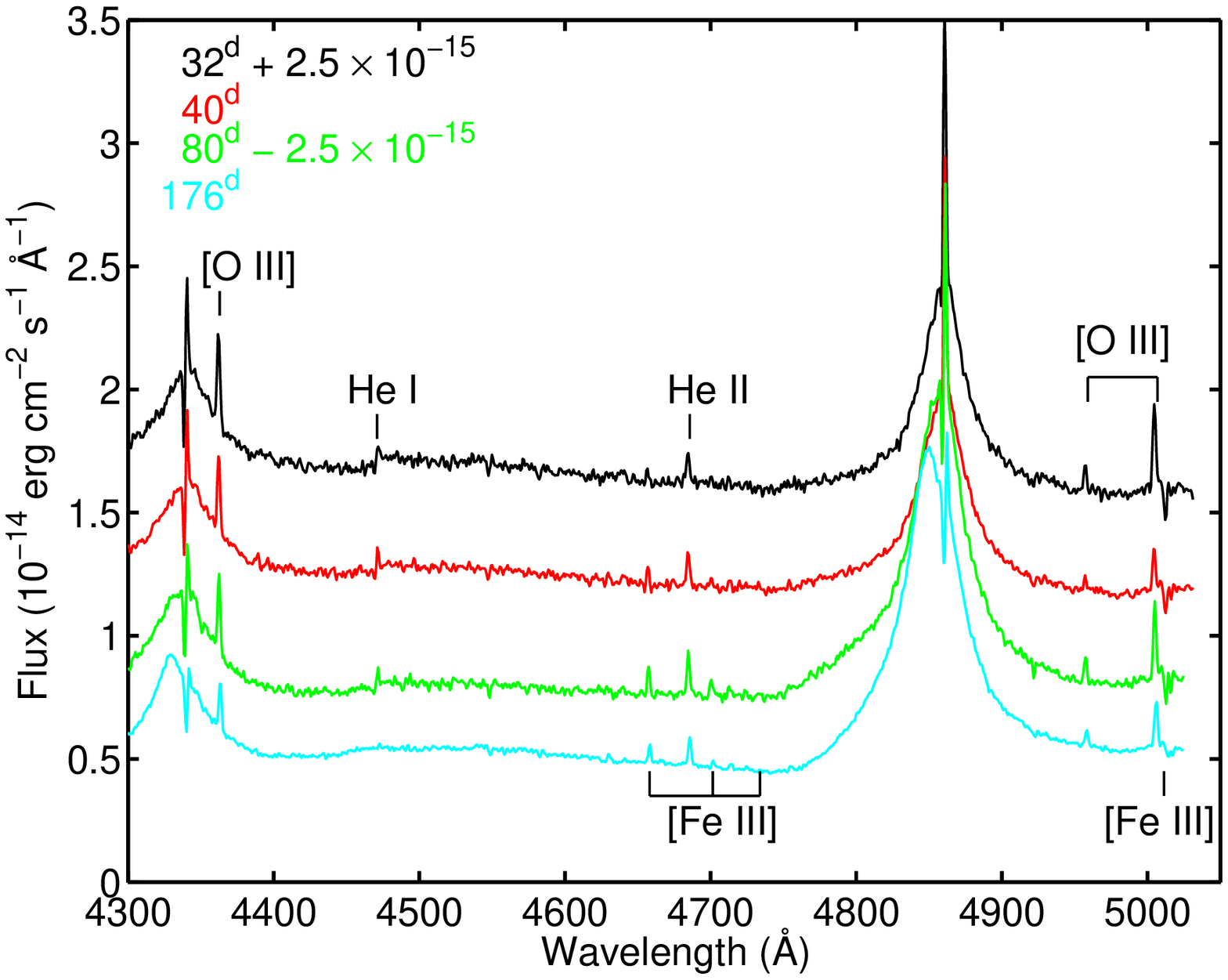}}
\caption{The [O III] and [Fe III] region from observations with NOT from day 32 to day 176.  Note the strong narrow [O III] $\wl 4363 $ line compared to the $\wl 5007$ line. For clarity the day 32 and day 80 spectra have been shifted in flux by the amount indicated in the figure. }
\label{fig_oiii_feiii}
\end{center}
\end{figure}

The O III] $\wll 1660.8, 1666.2$, and [O III] $\wll 4363.2, 4958.9, 5007.0$ lines are
especially useful as density and temperature diagnostics
\cite[][]{Crawford2000}. Figure \ref{fig_full_stis_spec} shows the 
     [O III] $\wll 4363.2, 4958.9, 5007.0$ region for the two first
     STIS spectra and all three [O III] lines are easily
     identified. This is even  clearer in the ground based spectra in Figure \ref{fig_oiii_feiii}. The first thing to note is the very strong $\wl 4363$ line,
     which is usually faint compared to the $\wll 4949, 5007$
     lines. Typical [O III] $\wll 5007/4363$ ratios for H II regions
     are $\ga 50$ \citep{Osterbrock2006}. 
  \cite{Pilyugin2007} determine [O III] $\wll 4959, 5007/4363$ ratios of 166 and 140 for UGC 5189 at different radii, which is much larger than the ratio observed for SN 2010jl and strongly argues for a circumstellar origin of these lines.    

\begin{figure}
\begin{center}
\resizebox{85mm}{!}{\includegraphics[angle=0]{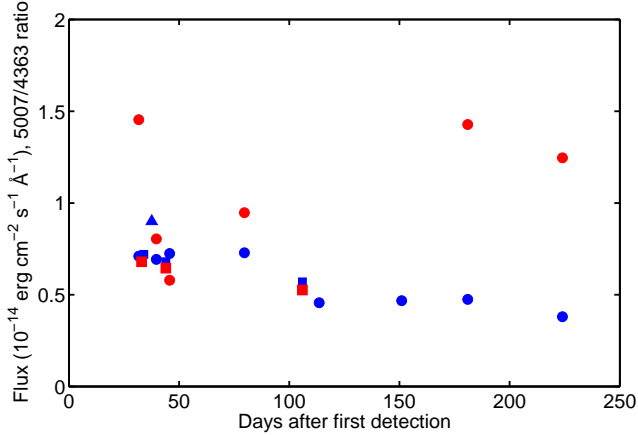}}
\caption{The flux of the narrow [O III] $\wl 4363$ line (blue) and the  [O III] $\wll 5007/4363$ ratio (red). Observations with STIS are marked with squares, with MMT with triangles and NOT/grism 16 with circles. The errors in the $\wl 4363$  fluxes are $\sim 25 \%$, while the ground based O III] $\wll 5007/4363$ ratios are systematically overestimated because of background contamination from nearby H II regions. The observations with the best seeing give the lowest line ratios, as is confirmed with the STIS observations, although the S/N of these are lower. }
\label{fig_oiii_ratio}
\end{center}
\end{figure}

       In Figure \ref{fig_oiii_ratio} we
       give the [O III] $\wll 5007/4363$ ratio and the [O III] $\wl 4363$ flux  for the different
       dates for selected  medium resolution spectra  with NOT and MMT (2010 Nov. 15), as well as the STIS observations.  The limited spectral resolution, as well as the
       S/N, make the fluxes of the STIS spectra uncertain. They do, however, have the advantage of minimizing the background contamination of especially the $\wl$ 5007 line. 

 We note that the flux of the $\wl 4363$ line shows a smooth decay by a factor of $\sim 2$ during the  period.  However,  the $\wll 5007/4363$ ratio varies by a large factor from observation to observation. The reason for this is that the $\wl 5007$ line is severely affected by nearby H~II regions in some of the ground based observations. Observations with poor seeing ($\ga 1\arcsec$) therefore have a much larger $\wl 5007$ flux than the ones with good seeing. This is also consistent with the STIS observations, which in spite of a low S/N, show a low $\wl 5007$ flux  and a low $\wll 5007/4363$ ratio. Because of the high $\wll 5007/4363$ ratio in the nearby H II regions (see above) the $\wl 4363$ line is only marginally affected by this. For the nebular analysis below we  only use the lowest $\wll 5007/4363$ ratios, which are in the range $0.5 - 0.8$. 

The region of O III]
       $\wll 1660.8, 1666.2$ is noisy in the STIS spectra, but
       does show a clear line above the noise (Figure \ref{fig_full_stis_spec}). The flux of
       the line is consequently uncertain. However, the fluxes in the COS and STIS spectra are consistent with each other for the dates of the two COS observations. The
observation from day 44 gives a total O III] $\wll 1661 + 1666$ flux
of $1.8 \times 10^{-14} \ergs$, close to that of the STIS
observation for the same date. We  conclude that the O III] $\wll 1661 +
1664$ flux is accurate to at least $\sim 50 \%$.
The expected line ratio of O III] $\wl 1666.2 / \wl 1660.8$ is 2.5, consistent with the measured ratio of $2.3$ from COS.
As a very conservative estimate we take O III]  $\wll (1661 + 1666)/5007 = 1.2 - 8$. 

\begin{figure}
\begin{center}
\resizebox{85mm}{!}{\includegraphics[angle=0]{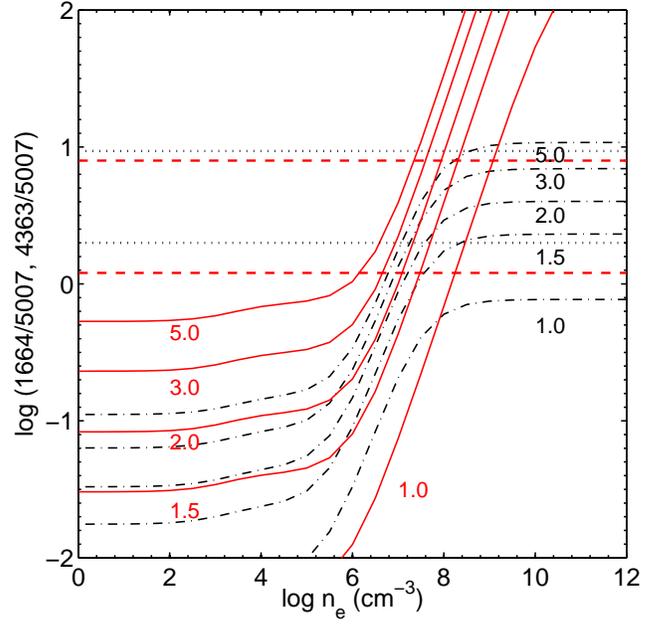}}
\caption{The [O III] $\wll 4363/5007$ (black, dash-dotted lines) and  $\wll 1666/5007$ (red, solid lines)  ratios as a function of density and temperature (in $10^4$ K). The corresponding observed ranges are plotted as horizontal dotted lines for the $\wll 4363/5007$ ratio and dashed lines for the $\wll 1666/5007$ ratio.}
\label{OIII_ratios}
\end{center}
\end{figure}

Figure \ref{OIII_ratios} shows the [O III] $\wll 4363/5007$ and $\wll
1666/5007$ ratios as a function of density and temperature. We have also plotted the observed ranges of these ratios as horizontal lines in the figure. The atomic data are taken from  \cite{Crawford2000} and  \cite{Palay2012}. From this figure we see that for
reasonable temperatures, i.e., $\la 50,000$ K, a $\wll 5007/4363$ ratio in the range $0.5 - 0.8$
requires an electron density $\ga 3 \times 10^6 \ccm$. Further, the $\wll 1664/5007$
ratio  results in a range $3 \times 10^6 - 10^9 \ccm$ for $T_{\rm e}
= (1-3) \times 10^4$ K. A higher reddening than we have assumed would only increase this range marginally. 

We identify faint, but clear narrow lines of [Fe III] at $4657\pm1$ \AA \ and $4701\pm1$ \AA \ in the high S/N NOT spectra (Figure \ref{fig_oiii_feiii}). These lines can be seen at all epochs with approximately the same fluxes. In addition, there may also be a weak line at $\sim 5010$ \AA, but this is  disturbed by the [O III] $\wl 5007$ line, as well as noise. 
These lines are interesting in that they can provide some independent diagnostics of the narrow line gas. \cite{Keenan1993} find that the [Fe III] $\wll 4702/4658$ ratio varies from $\sim 0.30$ for $n_{\rm e} \la 10^3 \ccm$ to $ 0.46-0.52$ for $n_{\rm e} \ga 10^5 \ccm$. The observed ratio is $\sim 0.5$, in agreement with the high density limit. The $\wll 4881/4658$ ratio increases from $\sim 0.2$ for $n_{\rm e} \approx 10^2 \ccm$ to $\sim 0.5$ for $n_{\rm e} \approx 10^5 \ccm$; at higher densities it rapidly decreases to $\la 0.1$. There is no sign of the $\wl 4881$ line in the spectra, consistent with a  density $n_{\rm e} \ga 10^5 \ccm$. The $\wl 5011.3$ line is, on the other hand, probably present. Together, these lines are consistent with $n_{\rm e} \ga 10^6 \ccm$.

The final diagnostics come from the [Fe VII] lines in the X-shooter spectrum from day 461 (Figure  \ref{fig_xshooterspec_opt}). The [Fe VII] $\wll 4988.6 / 5720.7$ ratio is 0.16. Using the diagnostic diagram from \cite{Keenan2001} we find that this corresponds to an electron density of $\sim 5 \times 10^5 \ccm$. We note, however, that the derived density is sensitive to the uncertain collision strengths, as is discussed in   \cite{Keenan2001}.

Summarizing these different density diagnostics, i.e., the N III], N IV], [O
    III], O IV], [Fe III] and [Fe VII] ratios, as well as the presence of P-Cygni absorptions in the Balmer lines, we find that these lines indicate
 a density $3 \times 10^6 - 10^8 \ccm$  for the region  emitting  the narrow lines
      in SN 2010jl.

\subsection{CNO abundances}
\label{sec-cno}

The strong [N II], N III], N IV] and N V lines in the STIS and COS spectra (Figs. \ref{fig_full_spec} and \ref{fig_full_stis_spec}) indicate a nitrogen enrichment in the narrow line region of the CSM of the SN. 
 
To estimate the relative CNO abundances we note that for a photoionized plasma the ionization zones of N III and C III coincide closely with each other, and similarly for the C IV, N IV and O III zones \citep[][their Figure 6a]{Kallman1982}. The N V and O IV zones overlap, but not to the same extent as the lower ionization ions. In addition, the temperature in these zones is nearly constant. Using the line fluxes for these ions allows us to derive the
relative ionic abundances and, with some assumptions, elemental
abundances. Because the excitation energies are similar, the line ratios are insensitive to
the temperature for reasonable ranges, $(1 - 3)\times 10^4$ K. Here we
use $T_{\rm e} =2\times10^4$  K. They do depend on the electron density, but this
is mainly important above $\sim 10^9 \ccm$, where the lines are suppressed, and densities this high were 
excluded through the analysis in the previous Section. The narrow range of wavelengths also makes the results insensitive to the reddening.

The most interesting lines for the CNO analysis analysis are N IV] $\wl 1486.5$, C IV $\wll 1548.2, 1550.8$,  O III] $\wll 1660.8, 1666.2$, N III] $\wll 1746.8-1754.0$, and C III] $\wl 1908.7$. With the exception of the C III] lines, all are within the range of our COS spectra and accurate relative fluxes can be derived (within a few percent). The C III] line, however, poses a special problem because of the lower S/N and low spectral resolution  of the STIS spectra (Figure \ref{fig_full_stis_spec}). An additional complication is that the line sits on top of a broad electron scattering feature centered on the  Si III] $\wl 1892.0$ line. This is also the case for the N III] $\wll 1746.8-1754.0$ lines, but in this case it is easy to determine the background flux from the high resolution COS spectra. An estimate of the systematic error in the N III] flux from STIS can consequently be obtained if one compares the COS and STIS fluxes for the same epoch. For the  day 44 observation we get for the total N III] flux $(11.5\pm 2.0) \EE{-14} \ergcms$ in the STIS observation, depending on where the 'continuum' level is set, compared to $ 7.9 \EE{-14}\ergcms$ for the COS spectrum. For the day 107 spectrum the corresponding numbers are $(5.57\pm 0.8) \EE{-14} \ergcms$ and $(5.33\pm 0.1) \EE{-14} \ergcms$, respectively. There is  a tendency to overestimate the flux from the STIS observations, mainly from the difficulty in determining the continuum level. The same is likely to be the case  for the C III] line. 

As a consequence of this discussion, to minimize the systematic error  we use the N III]/C III] ratio from the STIS observations, in spite of the lower S/N in the STIS observations for N III] compared to the COS flux. To estimate the systematic error in the C III] flux we have calculated the line flux for a range of assumptions about the background 'continuum'.  The error may still be up to $\sim 50 \%$ and dominates the error budget in the relative N III/C III ratio. For the other line ratios, N IV]/C IV and N IV]/O III] we use fluxes from the COS spectra. For the C IV resonance doublet we assume two limiting cases. In the first case we subtract the full absorption component from the emission component to correct for the scattering, as discussed in Sect. \ref{sec-narrow}. This may underestimate the net emission by up to $\sim 60 \%$. In the other case we neglect this correction and use the observed flux in the emission component. This is likely to overestimate the net emission.  These limits should bracket the thermal excitation emission and is the best we can do without knowledge about the radius of the emission region. The adopted line ratios with errors are given in Table  \ref{table_cno}. There is a fairly large scatter in both the N III]/C III] ratio and the N IV]/C IV ratio, reflecting the difficulties discussed above. 
\begin{deluxetable*}{lccc}
\tabletypesize{\footnotesize}
\tablecaption{Observed and derived ionic abundances. \label{table_cno}}
\tablewidth{0pt}
\tablehead{
\colhead{Date} & \colhead{N III/C III} & \colhead{N IV/C IV} & \colhead{N IV/O III}  \\ 
}
\startdata
Line fluxes used&&&\cr
34 days&$13.5\pm2.0~/~2.99\pm1.2$    &--&--\cr
44 days&$11.5\pm2.0~/~2.96\pm1.2$ &$5.47\pm0.18~/~1.62\pm0.28$\tablenotemark{b}&$5.47\pm0.18~/~1.76\pm0.05$\cr
 &&$5.47\pm0.18~/~2.48\pm0.22$\tablenotemark{c}&\cr
107 days &$5.57\pm0.8~/~2.24\pm1.0$&$5.17\pm0.15~/~0.85\pm0.13$\tablenotemark{b}&$5.17\pm0.15~/~1.25\pm0.05$ \cr
&&$5.17\pm0.15~/~1.39\pm0.11$\tablenotemark{c}&\cr
Observed line ratios&&&\cr
34 days&$4.51\pm1.93$    &--&--\cr
44 days&$3.89\pm1.71$ &$3.37\pm0.59$\tablenotemark{b}&$3.11\pm0.14$\cr
&&$2.21\pm0.21$\tablenotemark{c}&\cr
107 days &$2.49\pm1.16$&$6.08\pm0.95$\tablenotemark{b}&$4.14\pm0.20$ \cr
&&$3.71\pm0.31$\tablenotemark{c}&\cr
573 days&$2.49\pm1.16$&$6.08\pm0.95$\tablenotemark{b}&$4.14\pm0.20$ \cr
&&$3.71\pm0.31$\tablenotemark{c}&\cr
621 days&-&$6.08\pm0.95$\tablenotemark{b}&$4.14\pm0.20$ \cr
&&$3.71\pm0.31$\tablenotemark{c}&\cr
&&&\cr
Derived ionic ratios \tablenotemark{a}&&&\cr
&&&\cr
Nov 11 &$37.9\pm 16.2 - 30.3 \pm 13.0$&--&--\cr
Nov 22 & $32.7\pm 14.4 - 26.1 \pm11.5$ & $22.2\pm3.9  - 22.9 \pm4.0$\tablenotemark{b}& $0.74\pm 0.04 - 0.70 \pm 0.04$ \cr
&&$14.6\pm 1.4 - 15.0 \pm1.4$\tablenotemark{c} &\cr
Jan 23 & $20.9\pm \phantom{0}9.8 - 16.7 \pm\phantom{0}7.8$ & $40.1\pm 6.3 - 41.3 \pm6.5$\tablenotemark{b}  & $0.99\pm 0.05 - 0.93 \pm 0.05$  \cr
&&$24.5\pm 1.0 - 25.9 \pm1.2$\tablenotemark{c} &\cr
 \enddata
\tablenotetext{a}{~Assuming $n_{\rm e}$ in the range $10^6 - 10^9 \ccm$ and $T_{\rm e}=2\times10^4$ K and E$_{B-V} = 0.027$ mag.} 
\tablenotetext{b}{Corrected for scattering.} 
\tablenotetext{c}{No correction for scattering.} 
\end{deluxetable*}

Using these line ratios we can now derive  relative ionic abundances. For the collision strengths and radiative transition rates  we use data from the Chianti data base  \citep{Landi2012}. 
In Table \ref{table_cno} we show the derived ionic abundances for the different dates and $T_{\rm e} =2\times10^4$ K, typical of an X-ray heated plasma. As discussed above, the results are not very sensitive to the temperature. To show the dependence on the assumed density we give the abundance ratios for both   $n_{\rm e} = 10^6 \ccm$ and  for $n_{\rm e} = 10^9 \ccm$. 

There are several things to note here. The scatter and large errors in the STIS  fluxes are  reflected in the N III/C III ratio.  
Taking an average of the three epochs we get $n({\rm N III})/n({\rm C III) }=30.5 \pm 13.7$ for $10^6 \ccm$ and   $n({\rm N III})/n({\rm C III})=  24.3 \pm 11.0$ for $10^9 \ccm$.  Also, the N IV/C IV ratio varies greatly with both epoch and the assumption about the scattering contribution. The scattering introduces an uncertainty of $\sim 60 \%$, while the variation between the different epochs amounts to $70-80 \%$. These variations are probably related to each other. A decrease in the continuum flux means that the scattering contribution will decrease.  There may therefore be an evolution from a line dominated by scattering to a line where scattering is less important. The N IV/O III ratio does not have any of these complications since this ratio involves two intercombination lines. Consequently, it is relatively constant  between the two epochs, and has also small individual errors.

In addition to the observational errors and the uncertainty of the scattering correction, there is some uncertainty in the correspondence of the ionization zones. Although both the C III and N III zones, as well as the C IV, N IV and O III zones, nearly coincide in the X-ray photoionized models by \cite{Kallman1982} this is not obviously so for our case with a different ionizing spectrum and density, although it is likely that the narrow lines are indeed excited by X-rays (Sect. \ref{sec_csm}). 

In spite of these caveats, the need for a very high N/C ratio is
clear. From the N III/C III and N IV/C IV ratios we find a 'best value'  $n({\rm N})/n({\rm C})=25 \pm 15$. The N/O ratio is better constrained to be $n({\rm N})/n({\rm O})=0.85\pm 0.15$. For comparison, the corresponding solar ratios are $n({\rm N})/n({\rm C}) =0.25$ and $n({\rm N})/n({\rm O})=0.14$  \citep{Asplund2009}.

\section{Discussion}
\label{sec-discussion}

\subsection{The narrow circumstellar lines}
\label{sec_csm}

Our analysis in Sect. \ref{sec-results}  provides several pieces of important information about the SN and its CSM that provide clues to  the nature of the progenitor and its evolutionary status. In particular, we have evidence for a CSM of the  progenitor with a velocity of $\sim 100 \kms$ and a density $3\times10^6 - 10^8 \ccm$.
Further, the derived CNO abundances indicate a large nitrogen enrichment, $n({\rm N})/n({\rm C})=25 \pm 15$ and $n({\rm N})/n({\rm O})=0.85\pm 0.15$, typical of CNO processed gas.

A large  CNO enrichment has been observed for a number of SNe, either in the ejecta itself or in the CSM. This includes SN 1979C (IIL), SN 1995N (IIn), SN 1998S (IIn),  SN 1993J (IIb) and SN 1987A (IIpec) \citep[see][for a summary of these with references]{Fransson2005}. As for potential progenitors, CNO enrichment is observed for several types of evolved massive stars. In particular, this is consistent with an LBV scenario for the progenitor. The best studied cases  are for Eta Carinae and AG Carinae, which we discuss below. 

\cite{Hillier2001} make a detailed analysis of STIS observations of the central source in Eta Carinae. The mass loss rate is determined to $\sim 10^{-3}  \ml$ for a wind velocity of $\sim 500 \kms$. From the weakness of the electron scattering wings they conclude that the wind has to be clumpy, with a filling factor of $\sim 10 \%$.
Hillier et al. also find that the N abundance in the primary star is at least a factor of 10 higher than solar, while C and O are severely depleted. 
In the nebula of Eta Carinae the CNO abundances can be determined more reliably. In particular, \cite{Dufour1997} and \cite{Verner 2005} find that N is enhanced by a factor 10 -- 20, while C and O are depleted by factors 50 -- 100.
\cite{Smith2004} also find a strong radial gradient in the N/O ratio  in Eta Carinae, with a high N/O ratio close to the star, decreasing outwards. 

Most of the mass in the CSM of Eta Carinae is in a molecular shell surrounding the central star. \cite{Smith2006} estimates a total mass of $\sim 11  \Msun$. The distance from the star to this hour-glass shaped shell is between $3\EE{16}$ cm and $3\EE{17}$ cm. Especially interesting is that \cite{Smith2006} derives a  density $\ga 3\EE6 \ccm$ of the molecular shell around Eta Carinae.  This density bound is similar to that we find for the CSM of SN 2010jl.  In addition, gas with similar densities are seen even closer, at distances $\sim 10^{15}$ cm  from the star, as [Fe II] emission.

Besides Eta Carinae, AG Carinae is one of the most extensively studied LBV stars, showing a variation of the S-Dorados type. Over the two periods studied the mass loss rate has varied between $(1.5 - 3.7)\times 10^{-5} \ml $, while the wind velocity varied between $300 - 105 \kms$, in anti-phase with the mass loss \citep{Groh2009}.  It has clearly undergone a more intense mass loss period earlier, as is apparent from the fact that it has a circumstellar nebula with an ionized mass of $\sim 4.2 \Msun$  and a dynamical age of $\sim 8.5\times10^3$ years \citep{Nota1995}. An even higher mass of $\sim 25 \Msun$ in the neutral medium has been estimated based on IR imaging, assuming a standard dust to gas ratio of 1: 100  \citep{Voors2000}. From a spectral analysis of the central star \cite{Groh2009} find a N mass fraction of $11.5\pm3.4$ times solar, while  C is only $0.11\pm0.03$ times solar, and O is $0.04\pm0.2$ times solar, clearly indicating CNO processed material. The abundances in the nebula are more uncertain, with an N/O ratio of $6\pm 2$, compared to $39^{+28}_{-18}$ for the star. It is  likely that the surface of the star has  undergone more CNO processing than the nebula. 

The wind velocity we find for SN 2010jl, $\sim 100 \kms$, is not that of a red supergiant, whose velocities are in the range $10-40 \kms$ \citep{Jura1990}, so we rule these out as progenitors. Here we differ from \cite{Zhang}, who find a wind velocity of $28 \kms$, but this is a result of their use of the velocity of maximum absorption in the P-Cygni profile and not the maximum blue velocity. The maximum absorption velocity we observe is also considerably higher than theirs (Figure \ref{fig3}). Their conclusion that the progenitor was a red supergiant is based on a wind velocity that is too low. We find a velocity that is more typical of LBV stars, which have velocities in the range $100 - 1000 \kms$ \citep[][]{Smith2011b}. 
The velocity we find, $\sim 100  \kms$, is on the low end of this range, which may be a potential problem with the LBV interpretation.  However, 
the expansion velocity of the molecular shell in  Eta Carinae is highly anisotropic, ranging from $\sim 60 \kms$ at the equator to  $\sim 650 \kms$  at the poles \citep{Smith2006}, and is likely to vary substantially with time. As we noted above, the velocity in the high mass loss phase of AG Carinae is  very similar to the one we derive for SN 2010jl. 

Although it is clear that the narrow lines originate in fairly dense circumstellar gas, it is not obvious what excites them. The recombination time is $\sim 1/\alpha n_{\rm e} \sim  5.8 (n_{\rm e} / 10^7 \ccm)^{-1}$ days for H and $T_{\rm e} \sim 10^4$ K. The gas is therefore in near equilibrium with the ionizing radiation. An early, undetected UV burst is therefore not likely to be important for the excitation. The widths of the lines,  $\la 100 \kms$, also makes it unlikely that the gas  is shock ionized. 

Instead we believe that the lines are excited by the same X-rays as are observed with the Chandra satellite  \citep{Chandra2012}. As we discuss in Sect. \ref{sec_broad}, these are likely to come from fast shocks connected to an anisotropic high velocity ejection. We can estimate the state of ionization from the models in \cite{Kallman1982}. With an X-ray luminosity $\sim 3\EE{41} \ergs$, corresponding to the absorbed flux of $\sim 10^{-12} \ \rm erg \  cm^{-2} \ s^{-1}$  \citep{Chandra2012}, the ionization parameter, $\zeta = L/n_{\rm e} r^2$, becomes  $\zeta \sim  300  (n_{\rm e} / {10^7 \ccm})^{-1} (r / 10^{16} \rm \ cm)^{-2}$. From the optically thin  Model 1 in \cite{Kallman1982} we find that the presence of N III -- N V requires $\log \zeta \sim 0 - 1$. Optically thick models with a density $10^{11} \ccm$ have the same ions at a somewhat higher ionization parameter, $\log \zeta \sim 1.5$. The presence of [Fe XIV] (Sect. \ref{xshooter}) is also consistent with this ionization parameter. The observed X-ray luminosity indicates a distance of  $ \sim  (2-20) \EE{16}  (n_{\rm e} / 10^7 \ccm)^{-1/2}$ cm. The radius of the shock at 500 days after outburst is $\sim 1.3 \EE{16 }(V_{\rm s}/{3000 \kms}) (t/500  \ {\rm days})$ cm, where $V_{\rm s}$ is the shock velocity. There is therefore a rough consistency  between the estimated distance of the narrow line emitting region and the location of the SN shock wave. 

This estimate of the ionization parameter is approximate for several reasons. First, the density has a considerable uncertainty. Further, as we discuss below, the shock velocity, and therefore the shock radius, is also uncertain and may be in the range $500 - 3000 \kms$.  Finally, the X-ray models in \cite{Kallman1982} are not completely appropriate, since these assume a  10 keV bremsstrahlung spectrum, while the X-ray spectrum of SN 2010jl is heavily absorbed, having a deficit of photons below $\sim 1$ keV. 
The  \cite{Kallman1982} models were also calculated for a density of $10^{11} \ccm$, while the narrow lines in SN 2010jl  indicate a considerably lower density. Nevertheless, this estimate shows that it is  likely that the narrow lines originate in X-ray ionized gas at a distance and density comparable to the molecular shell of Eta Carinae discussed above. In Sect. \ref{sec-big} we discuss more detailed photoionization calculations that confirm these estimates.

The fact that we observe fairly 'normal' P-Cygni profiles with both a deep absorption and a red emission wing implies that the CSM at the distance where the narrow lines arise must be fairly symmetrically distributed, unlike the distribution for the broad component as described below.  This is discussed further in Sect. \ref{sec-big}.

\subsection{Origin of the dust emission}
\label{sec_dust}

The  NIR excess seen in the photometry in Figs. \ref{fig_photometry} and \ref{fig_sed}, as well as the X-shooter spectrum in Figure \ref{fig_xshooterspec}, is most likely a result of hot dust. From the photometric light curve in Figure \ref{fig_bollc} we see that by day 400 the flux in the NIR is higher than that in the optical. This NIR/optical ratio increases with time and at the last epochs the NIR luminosity dominates completely. 

 In Sect. \ref{sec_phot_results} we used  simple blackbody fits of the optical and NIR photometry. We have investigated this in more detail following \cite{Kotak2009}. We have here used the grain emissivities from \cite{Draine1984} and
\cite{Laor1993} and the escape probability formalism from \cite{Osterbrock2006}. For the grain sizes we use the MRN distribution with $dn/da \propto a^{-3.5}$ \citep{Mathis1977}  with a minimum grain size $0.005 \mu$m and a maximum size $0.05 \mu$m. We then fit the sum of the dust spectrum with a given optical depth in the V-band, $\tau_{\rm V}$, and a photospheric blackbody spectrum, minimizing the chi-square with respect to the photospheric effective temperature and radius  $T_{\rm eff}$ and $R_{\rm phot}$, and the corresponding parameters for the dust shell, $T_{\rm dust}$, $R_{\rm dust}$. A more detailed modeling, calculating the temperature of the dust as function of radius, as in \cite{Andrews2011}, requires assumptions about the geometry, and is out of the scope of this paper. 

\begin{figure}
\begin{center}
\resizebox{85mm}{!}{\includegraphics[angle=0]{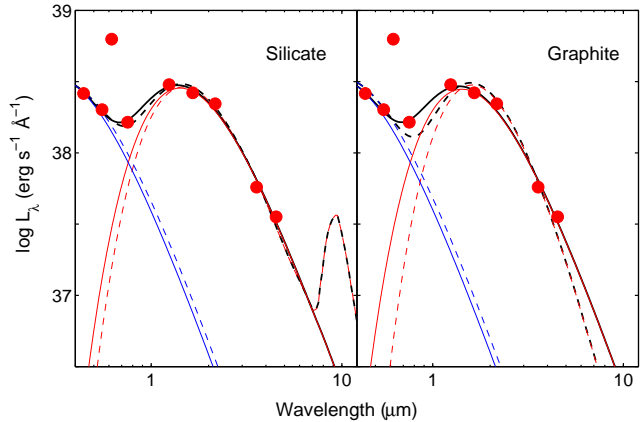}}
\caption{SED fits to the extinction corrected photometry at 465 days for two different dust compositions, silicates and graphite.  The blue curves show the photospheric contribution, while the red show the dust component and the black curves the total. The dashed lines show the optically thin models with $\tau_{\rm V}=0.04$, while the solid lines show the corresponding optically thick models. As discussed in the text, the R-band, which is dominated by the H$\alpha$ line, is not included in the fits. We note the $10 \mu$m silicate feature for the optically thin model, which is the main discriminator for the dust composition. . }
\label{fig_sed_dust}
\end{center}
\end{figure}

In Figure \ref{fig_sed_dust} we show an example of this kind of fit for day 465 for which we have observations at 3.6 $\mu$m and  4.5 $\mu$m. 
From this figure it is clear that without any 
 observations longward of  4.5 $\mu$m   it is impossible to discriminate between the different compositions and only marginally between different optical depths for the dust. This is in agreement with the modeling by \cite{Andrews2011}.

The main difference between the optically thin and thick models is the  larger radius of the former in order to produce the required luminosity. The dust temperature is  lower in the optically thin models. At 465 days the best fit silicate model in Figure \ref{fig_sed_dust} with an optical depth in the V-band of $\tau_{\rm V}=0.04$ had $T_{\rm dust}=1680$ K and $R_{\rm dust}=3.3\times 10^{17}$ cm. The photospheric values were less affected with $T_{\rm eff}=8900$ K and $R_{\rm phot}=6.0\times 10^{14}$ cm. The corresponding graphite model  had $T_{\rm dust}=1360$ K, $R_{\rm dust}=7.8\times 10^{17}$ cm, $T_{\rm eff}=8600$ K and $R_{\rm phot}=6.6\times 10^{14}$ cm. These values should be compared to $T_{\rm dust}=2040$ K, $R_{\rm dust}=2.2\times 10^{16}$ cm, $T_{\rm eff}=9200$ K and $R_{\rm phot}=5.6\times 10^{14}$ cm for the optically thick model (Table \ref{table_dust_param}).   We find similar results for the other epochs. As Figure  \ref{fig_sed_dust} shows, the separation between the dust and photospheric components are only marginally affected. The light curves of the photospheric and dust components based on  the blackbody fits in Sect. \ref{sec_phot_results} should therefore be accurate.

As an independent check we find from the X-shooter spectrum at 461 days a dust temperature $\sim 1870$ K and radius $R_{\rm dust} \sim 2.5 \times 10^{16}$ cm for the optically thick model.
To fit the NIR continuum  of the X-shooter spectrum in Figure \ref{fig_xshooterspec} we need a blackbody radius of  on day 461, in good agreement with that derived from the broad band photometry. For the dates with only broad band photometry we give $R_{\rm dust}$ in Figure \ref{fig_dust_tbb_rbb}. 

It should be noted that the blackbody radius only represents a lower limit to the NIR emitting region. A covering factor, $f$, less than unity will increase the radius by $R_{\rm dust} \propto f^{-1/2}$. As was discussed above, an optically thin model would increase this further. 

The blackbody radius can be compared to the shock radius, given by $R_{\rm s} \sim V_{\rm s} t \sim 1.0 \times 10^{16} (V_{\rm s}/3000 \kms) (t/400 \ {\rm days})$ cm, where we have scaled the velocity to what we believe is the maximum velocity of the shock (Sect. \ref{sec_broad}). With this velocity, or higher, the blackbody radius is comparable to the shock radius. In addition, a lower shock velocity is likely in the directions where the column density is larger than that inferred from the X-rays, which
would give a smaller shock radius (Sect. \ref{sec-big}). 

The  blackbody and shock radii can also be compared to  the evaporation radius, given by $
 R_{\rm evap} = [L_{\rm max} Q_{\rm abs} /  (16\pi \sigma T^{4}_{\rm evap} Q_{\rm emiss}(T_{\rm evap})) ]^{1/2} $ \citep{Draine1979}. Here $L_{\rm max}$ is the maximum bolometric luminosity of the SN, 
$Q_{\rm abs}$ is the wavelength averaged  dust absorption efficiency over the SN spectrum and $Q_{\rm emiss}$ is the  the Planck averaged dust emission efficiency, $\sigma$ is the Stefan-Boltzmann constant, and $T_{\rm evap}$ is the evaporation temperature. $Q_{\rm abs} \propto a$, where $a$ is the size of the dust grain. 
The parameter $Q_{\rm abs} /  Q_{\rm emiss}$ depends upon the evaporation temperature of 
the dust, the grain size of the dust, and the  effective temperature of the SN, $T_{\rm eff}$. For 
our calculations of 
 $R_{\rm evap}$ we adopted values of $T_{\rm evap}$ $=$ 1900 K for graphite and $T_{\rm evap}$ $=$ 1500 K for silicates and assumed $T_{\rm SN}$ $=$  6,000 K and 10,000 K.   For both graphite
and for silicates we use grain emissivities from \cite{Draine1984} and
\cite{Laor1993}. 
Because $R_{\rm evap} \propto L_{\rm max} ^{1/2}$ we give values for $L_{\rm max} = 10^{43} \ergs$ in Table \ref{table_evap}, which can then be scaled to different luminosities.
\begin{deluxetable*}{lccc}
\tabletypesize{\footnotesize}
\tablecaption{Dust evaporation radii for $L_{\rm bol}=10^{43} \ergs$. \label{table_evap}}
\tablewidth{0pt}
\tablehead{
\colhead{a}& \colhead{$T_{\rm eff}$}&\colhead{$R_{\rm 0 \ evap}$}&\colhead{$R_{\rm 0 \  evap}$}\\
$\mu$m&K&$10^{17}$ cm&$10^{17}$  cm\\
\colhead{}& \colhead{}&\colhead{Silicates$^a$}&\colhead{Graphite$^a$}
}
\startdata
0.001    & 10,000     &2.59	        &1.43 \\
1.0	 &	      &1.30        &0.34 \\
0.001	 &  \phantom{0}6,000     &0.86          &0.86 \\
1.0	 &            &0.72 		&0.35 \\
 \enddata
\tablenotetext{a}{Assuming $T_{\rm evap}=1500$ K for silicates and $T_{\rm evap}=1900$ K for graphite.}  
\end{deluxetable*}

  We note that $R_{\rm evap}$ is not very sensitive to $T_{\rm eff}$, except for very small dust grains. As a typical value at shock break-out we use $T_{\rm eff} = 10^4$ K for our estimates below. The peak luminosity of the SN was $L_{\rm max} \sim 3 \times 10^{43} \ergs$   (Figure \ref{fig_bollc}). Using a grain size of $\sim 0.001 \ \mu$m we get $R_{\rm evap} \sim 2.5 \times 10^{17}$ cm for graphite and $R_{\rm evap} \sim 4.5 \times 10^{17}$ cm for silicates. For a grain size of $\sim 1 \ \mu$m the corresponding values are $R_{\rm evap} \sim 6.0 \times 10^{16}$ cm and $R_{\rm evap} \sim 2.3 \times 10^{17}$ cm, respectively.

The dust temperatures we find at $\la 500$ days, $1600 - 2000$ K, are close to the evaporation (or formation) temperature for dust. It may therefore either be newly formed dust, close to the shock, or dust heated to close to the evaporation temperature by either the radiation from the SN or the shock wave. We now discuss different scenarios for the origin of the dust emission.

The cool dense shell formed behind a radiative shock has often been mentioned as a favorable place for dust formation. The formation and survival of dust in this environment is, however, difficult to explain at epochs earlier than $\sim 320$ days. Dust formation in the dense post shock gas requires temperatures $\la 1900$ K. At these epochs the shock is most likely in the optically thick region of the SN. The photospheric temperature at these epochs is $\ga 7000$ K (Table \ref{table_dust_param}). 
Even if the shock is radiative it is therefore unlikely to cool to less than this temperature, and may be higher.  A related argument comes from noticing that the shock radius is well within the evaporation radius even at $\sim 200$ days.  As Table \ref{table_evap} shows, $R_{\rm evap} \ga 3\times10^{16}$ cm for a luminosity of $\sim 10^{43} \ergs$ at 200 days, independent of dust size or composition for $T_{\rm eff} \ga 6000$ K. It is therefore  difficult to see how any dust could form close to the shock region for a realistic shock velocity.

It is also difficult to see how the large IR luminosity at late epochs can be produced by such dust. The photospheric luminosity is then low and the X-ray luminosity is likely to be less than that at the last observation by \cite{Chandra2012} on  day 373, $\sim 7\EE{41} \ergs$, which is a factor of $\sim 5$ lower than the NIR luminosity at 500 days. This agrees with the conclusions by \cite{Andrews2011}, although with somewhat different arguments.

 Reprocessing of radiation from the SN, possibly resulting in evaporation of pre-existing dust  or heating and evaporation of the dust by the shock, is more likely. As Andrews et al. conclude, there is  evidence for pre-existing dust around the progenitor star. The dust heating mechanism is, however, not clear. Previous treatments of dust in Type IIn SNe have discussed both heating by the radiation from the SN in combination with an echo and collisional heating by the hot gas behind the shock \citep[e.g.,][]{Gerardy2002,Fox2010}.  
 
 We first discuss the case where 
 the dust emission comes from dust which is collisionally evaporated by the shock of the SN. 
 As discussed in next section, there is also direct evidence for such a velocity component in other Type IIn SNe. Massive dust shells into which the shock is propagating are also natural if the progenitor is an LBV.
However,  if we  compare the shock radius with the evaporation radius at early epochs we find that even the lowest values of  the latter, $R_{\rm evap} \sim 7 \times 10^{16}$ cm, is considerably larger than the shock radius at the first epochs, $R_{\rm s} \sim 2.6 \times 10^{15} (V_{\rm s}/3,000 \kms) (t/100 \ {\rm days})$ cm, where we have scaled to the velocity derived from the temperature of the X-rays.  This radius should be seen as an upper limit to the shock radius. We therefore exclude this possibility. 

A more attractive alternative is based on an echo from radiatively heated dust, as has been discussed for other Type IIns.    
The  plateau in the IR light curve between $\sim 200$ to $\sim 450$ days and slow decline thereafter  (Figs. \ref{fig_photometry} and \ref{fig_bollc} ) would then be expected.  Also the energetics may be consistent with this. The integrated optical luminosity from the SN during the first $\sim 400$ days was $\sim 3.2 \times 10^{50}$ ergs. In addition, there is a considerable luminosity in the UV, as well as at shorter wavelengths. This can be compared to the integrated luminosity in the dust component, which we estimate from the blackbody fits in Figure \ref{fig_dust_tbb_rbb}; we find a total energy of $\sim 2.7 \times 10^{50}$ ergs. There is therefore  reasonable consistency between the energy emitted from the central source and that emitted by the dust. It does, however, require a large covering factor of the dust, as well as a high optical depth in these directions, which is consistent with the blackbody shape of the spectrum. As Andrews et al. discuss, this may indicate an anisotropic geometry; they  discuss torus models with different inclinations. 

To explain the plateau as an echo the inner radius would  have to be $\ga 450/2$ light days  or $\sim 6 \times 10^{17}$ cm, depending on the inclination of a disk or torus. At the same time the high temperature indicates that the inner radius is close to the evaporation radius. 
A radius of $\sim 6 \times 10^{17}$ cm  is, however, considerably larger that the estimates earlier for $R_{\rm evap}$,  even if the luminosity used for this estimate may be an underestimate, considering the fact that only the observed wavelengths were included, and also that the peak luminosity may have been before the first observations. A solution to this discrepancy may be that most of the emission comes from very small grains.  Some support for the presence of small grains in winds come from observations and modeling of the emission from some O-rich AGB stars in the LMC, where a grain distribution with a minimum size of $0.01 \mu$m to $0.1 \mu$m gave the best fit \citep{Sargent2010}. 
Although the most appealing model,  this scenario   has some difficulties to simultaneously explain both the light curve and the high dust temperature, unless one stretches the parameters. 

We have calculated light curves from an echo from a spherically symmetric dust shell of inner radius $6 \times 10^{17}$ cm. We, however, find that the rise of the light curve occurs too rapidly compared to the observed NIR light curve in Figure \ref{fig_bollc}  and the total dust luminosity in Figure  \ref{fig_dust_tbb_rbb}. 
\cite{Emmering1988} have studied echoes from asymmetric dust distributions. As shown in their Figs. 2 and 3, more disk like distributions give a considerably slower rise for low values of the inclination of the disk or torus. A low value of the inclination also agrees with the low reddening in the spectrum of the SN, while a high optical depth in the disk is needed for the high dust to SN luminosity, discussed above.  Even more asymmetric distributions with most of  the dust behind the SN relative to our line of sight may also be consistent with the light curve, although we have not studied this quantitatively.   The increase of the dust emitting area and simultaneous decrease of the dust temperature (Fig. \ref{fig_dust_tbb_rbb}) are, however, qualitatively consistent with an echo, as the light echo paraboloid expands to larger radii into dust with decreasing temperature.

In the discussion about different dust emission mechanisms for Type IIn SNe \citet[][see also \cite{Gerardy2002}]{Fox2011} proposed  a scenario where pre-existing dust was radiatively heated by the radiation from ongoing circumstellar interaction. Although the heating of the dust is radiative, echo effects are less important and the shell can  be at a smaller evaporation radius, without invoking small grains.  If the late luminosity, including all wavelengths, is of the same order as the maximum bolometric luminosity this may account for the high temperature observed. The main problem is that the luminosity of the late IR emission is high, $\ga 5\times 10^{42} \ergs$ at 500 days, while the optical flux from the SN is decreasing rapidly. This scenario would therefore require a very large EUV-X-ray luminosity. There may  be some indication of this from the H$\alpha$ luminosity, as discussed in Sect. \ref{sec-big}. 
 
\subsection{The broad emission lines}
\label{sec_broad}

The profiles of the broad lines give important information about  the structure of the envelope and the dynamics of the shock wave. 
As we discuss in the previous and next Sections, dust formation has severe problems in both the ejecta and in connection to the shock at early epochs due to the high luminosity and temperature of the radiation. 
Instead, the  blueshift of the lines is more easily explained as a result of the line formation and dynamics in the outer parts of the
extensive envelope/CSM of the progenitor star, as was also suggested as a possible alternative by Smith
et al. (2012).  As 
discussed in detail by e.g., \citet{Chevalier2011}, for a radiation dominated shock in an extended envelope the radiation starts to escape when $\tau_{\rm e} \la c/V_{\rm s}$. The region outside the shock is therefore optically thick and line radiation emitted here will be Compton scattered and can give rise to the broad wings observed, even if the emission is coming from regions which have not yet been accelerated by the shock radiation. Only when the region with $\tau_{\rm e} \la 1$ becomes accelerated  will the macroscopic velocity become dominant.

An emission line undergoing electron scattering in a hot medium at rest will give rise to a symmetric line profile, since the velocity broadening is provided by microscopic, thermal velocities of the electrons \citep{Munch1948}. However, if the medium is moving with a bulk velocity towards us, this can introduce  a blueshift of the line profile. For a spherically symmetric expansion one expects a redshift of the line peak \citep[][]{Auer1972,Fransson1989}. This is clearly not the case here. However, if the scattering medium is primarily moving coherently towards us one may instead get a blueshift of the peak, although still symmetric around the expansion velocity. We believe that this is the case here. 

\begin{figure}
\begin{center}
\resizebox{85mm}{!}{\includegraphics[angle=0]{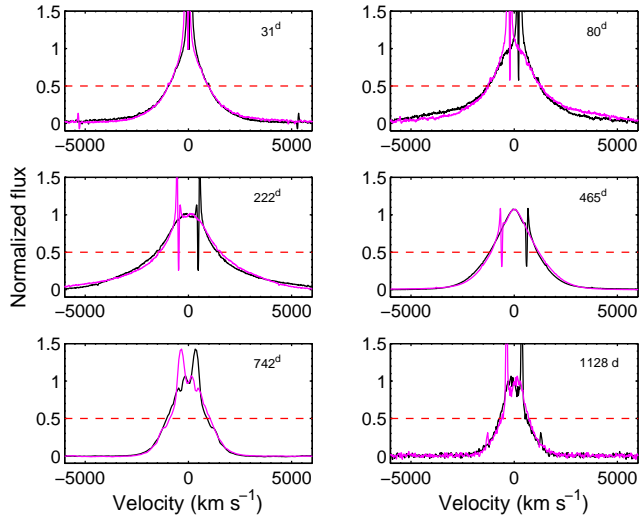}}
\caption{The broad H$\alpha$ line for selected dates after shifting the line to the red by  a velocity given in the upper panel of Figure \ref{Ha_HWHM} (black line). The magenta line gives the reflected line profile. Note the nearly perfect symmetry between the red and blue wings, characteristic of electron scattering. The 'horns' close to the center of the line are due to the P-Cygni absorption from the narrow component coming from the CSM which is close to the rest velocity of the host galaxy. The separation between the 'horns' is therefore a measure of the velocity shift of the broad line component. }
\label{Ha_shift}
\end{center}
\end{figure}

We can test this interpretation quantitatively  by first shifting each line in Figure \ref{fig3b} by a macroscopic velocity, $V_{\rm bulk}$, to the red. We adjust this so that a reflection about zero velocity gives a best $\chi^2$ fit for Doppler shifts between $\sim 1000 - 5000 \kms$,
$\chi^2 = \sum_i{ [F_\lambda(V-V_{\rm bulk}) - F_\lambda(-(V-V_{\rm bulk}))]^2]/\sigma^2}$, where $V$ is the velocity rest wavelength
and $\sigma$ is the r.m.s. of the flux. The lower velocity limit is chosen to avoid influence on the line profile by the narrow P-Cygni absorption and emission, and the scattering wings of this emission. We treat $V_{\rm bulk}$ as a free parameter, representing the macroscopic velocity of the emitting gas. 
The result of this procedure for a selection of days is shown in Figure \ref{Ha_shift}.   

From this figure we see that  this simple, linear transformation results in nearly perfectly symmetric line profiles between the red and blue wings for essentially all epochs. 
This therefore  argues  that the  blueshift in Figure  \ref{fig3b} is due to a macroscopic velocity. In addition, the symmetric line profiles are a natural result of electron scattering. Other processes, like dust or other absorption processes would result in asymmetric lines  about the peak in flux, with a time dependent line shape \citep[e.g.,][]{Lucy1989}. Electron scattering in an expanding medium which is optically thick, however, naturally results in symmetric profiles, as demonstrated explicitly below. 
\begin{figure}
\begin{center}
\resizebox{85mm}{!}{\includegraphics[angle=0]{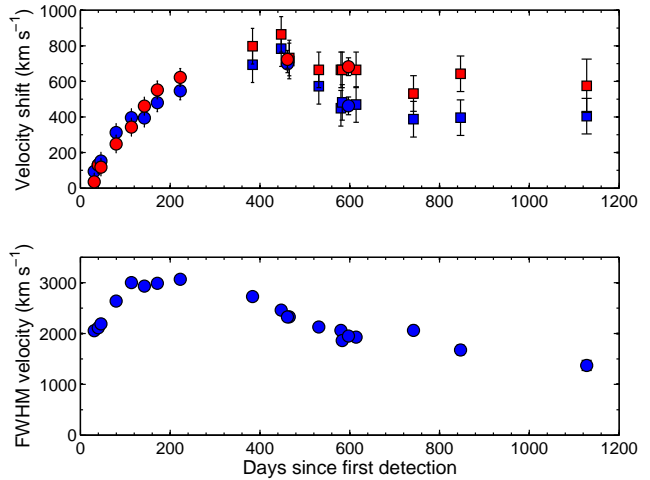}}
\caption{Upper panel: Velocity shift, representing a macroscopic velocity of the H$\alpha$ emitting region as function of time.   Blue markers represent $V_{\rm bulk}$ from measurements of the symmetry of the lines, as discussed in the text, and red markers the peak velocity of the broad component. 
Circles represents measurement with medium resolution spectra, while the squares are measurements from low resolution   spectra. Lower panel: Full width, half maximum (FWHM) width of H$\alpha$ as function of time. 
}
\label{Ha_HWHM}
\end{center}
\end{figure} 

 In the upper panel of Figure \ref{Ha_HWHM} we show with blue markers $V_{\rm bulk}$ as function of time. In this  panel we also show the velocity of the peak of the broad component as red markers. For the early epochs this agrees well with $V_{\rm bulk}$, but is somewhat higher for the later epochs. The broad agreement of these two determinations is another way of showing the symmetry of the line around the velocity shifted peak of the broad component. Because the peak velocity is more difficult to determine in the late spectra and is somewhat influenced by the narrow component, we prefer to use $V_{\rm bulk}$ as the macroscopic velocity for the rest of the paper.
To quantify the width of the line profiles we use the Full Width at Half Maximum (FWHM), as measured from the continuum subtracted line profiles. This is shown in the lower panel of Figure \ref{Ha_HWHM}.

During the first 300-400 days there is an increase in $V_{\rm bulk}$ to $\sim 700 \kms$. After this epoch the velocity decreases slightly and then becomes nearly constant at $\sim 400-500 \kms$. Considering the errors in the last observations the significance of this decrease is marginal. The FWHM width increases from $\sim 2000 \kms$ to $\sim 3000 \kms$ during the first $\sim 200$ days, and then decreases slowly to $\sim 1500 \kms$ on day 1128.  

Before discussing this result in more detail, we first test the electron scattering interpretation further by fitting the line profiles for a few different dates. This is done by a Monte Carlo code developed for this purpose, which is similar to that developed by \cite{Auer1972} and \cite{Chugai2001}.  The details of this is discussed at length in e.g., \cite{Pozdniakov1977} and \cite{Gorecki1984}. Except for the Compton recoil, which is negligible at these wavelengths, this calculates the line profile for an arbitrary temperature and spatial distribution of electrons.  In this calculation relativistic effects are unimportant and are ignored. Because the macroscopic bulk velocity can be transformed away in the  way described above, we only consider  scattering in a static medium.  

For photons injected at a specific optical depth, $\tau_{\rm e}$, the parameters determining the line shape are only $\tau_{\rm e}$ and temperature, $T_{\rm e}$.  The optical depth determines the average number of scatterings, $ N_{\rm sc} \approx \tau_{\rm e}^2$.
The thermal velocity of the electrons is $v_{\rm rms} = 674 (T_{\rm e}/10^4 \ {\rm K})^{1/2} \  \kms$. 
The FWHM width of the line is $\sim  N_{\rm sc}^{1/2} v_{\rm rms} \approx     674 \  \tau_{\rm e} (T_{\rm e}/10^4 \ {\rm K})^{1/2} \  \kms$. For a constant electron temperature medium with no internal emission our Monte-Carlo calculations show that the FWHM can be fitted more accurately with $\Delta v_{\rm FWHM} = 900 (T_{\rm e}/10^4 \ {\rm K})^{1/2} \tau_{\rm e} \  \kms$. A similar scaling also applies if the photons are internally generated.
This means that  there is some degeneracy in temperature and optical depth for a given $\Delta v_{\rm HWHM}$. There is, however, a constraint on $ \tau_{\rm e}$ from the luminosity of the unscattered fraction of photons relative to those scattered  is approximately $\exp(-\tau_{\rm e})$.

\begin{figure}[t!]
\begin{center}
\resizebox{85mm}{!}{\includegraphics[angle=0]{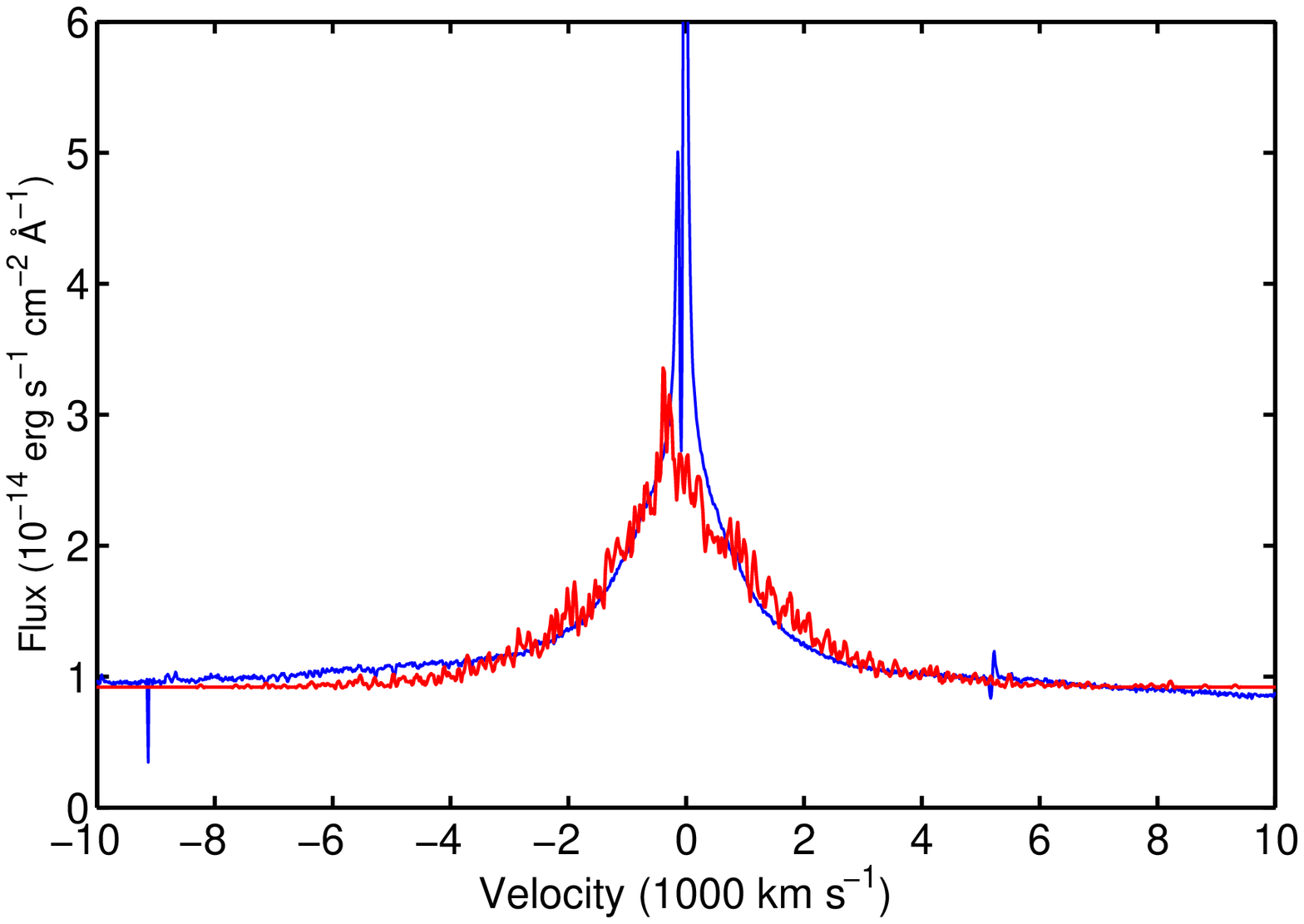}}
\resizebox{85mm}{!}{\includegraphics[angle=0]{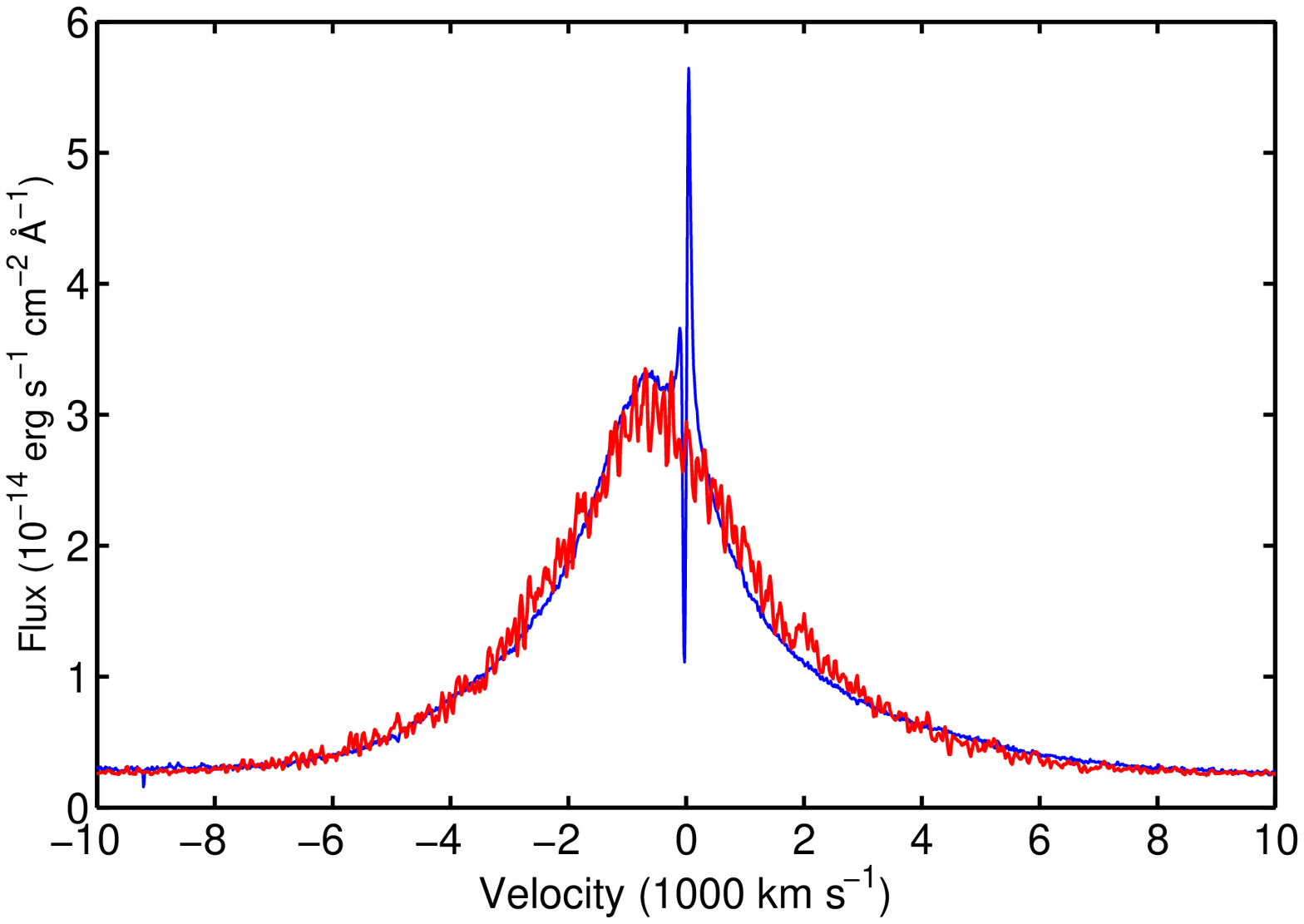}}
\caption{Electron scattering line profiles for day 31 (left) and day 222 (right).  The observed profiles (from NOT) are in blue and the calculated profiles from the Monte Carlo  simulations  in red.}
\label{el_scatt_fits}
\end{center}
\end{figure}

In Figure  \ref{el_scatt_fits} we show fits of the H$\alpha$  profile for two different dates, day 31 and day 222,  covering both the very early and late phases. In this model we have used an $n_{\rm e} \propto r^{-2}$ density and assumed that the photons are injected in the scattering region with an emissivity $\propto n_{\rm e} ^2$, as applies if recombination dominates the H$\alpha$ emission. Note that the narrow component of the H$\alpha$ line is at the rest velocity of the galaxy and does not shift in velocity as the broad line becomes blueshifted, showing that these photons are not the dominant source for the scattering. The injected photons span a velocity range of $0-700 \kms$, as expected for radiative acceleration (see Section \ref{sec_radacc}), which decreases the un-scattered narrow peak, in agreement with the observations. The temperature is taken to decrease from $2 \times 10^4$ K to $1 \times 10^4$ K,  similar to what is found in detailed simulations of shocks \citep[e.g., Figure 6 in ][]{Moriya2011}, as well as from our photoionization calculations in Sect. \ref{sec-big}. However, the profile is not sensitive to these assumptions. 

As can be seen from the figure, we get  excellent fits for both dates, except at zero velocity, where the P-Cygni profile from the CSM affects the profile. This confirms that electron scattering dominates the line formation.  There is  no need for two different velocity components in the broad lines as advocated by \cite{Zhang}. The same conclusion is reached by \cite{Borish2014} for the day 36 Pa$\beta$ line. 

The fact that the line profile can be explained by electron scattering plus a linear velocity transformation is  surprising. As was pointed out above, one expects a redshift rather than blueshifted line profile if the expansion of the ejecta/CSM is spherically symmetric \citep[see ][Figure 6]{Fransson1989}. Instead we propose that the  H$\alpha$ emission comes from a highly asymmetric, almost planar region expanding towards us (although possibly at some inclination relative to the LOS),  as discussed in Sect. \ref{sec-big}.  

\subsection{Origin of the line shifts}

\subsubsection{Dust absorption from the ejecta}
\label{sec_ejectadust}
In Figure  \ref{Ha_HWHM} we see a gradual shift of  the H$\alpha$ peak. We have already in Secs. \ref{sec_dust} and \ref{sec_broad} argued against  dust in the ejecta or at the reverse shock as a reason for this. Nevertheless, this has been discussed by several authors so we consider additional arguments on this topic. 

\cite{Smith2012} analyze the line profiles during the first 236 days and discuss two different scenarios for the observed blueshift.  From a comparison of the optical and near-IR hydrogen line profiles at $\sim 100$ days they claim to see a wavelength dependent asymmetry, which they interpret as a sign of wavelength dependent extinction, indicative of dust formation in the post-shock gas. However,  Smith et al., also discuss electron scattering as an explanation for the line profile. 

\cite{Maeda2013} argue for a dust origin based on the line shift, drop in luminosity and IR execss. However, their analysis has several limitations which we  discuss based on our observations. 
The line shifts discussed by Maeda et al. are based on a single spectrum on day 513 with a   limited S/N for lines other than H$\alpha$ (see their Figs. 4 and 5). The line shifts derived from these observations are therefore highly uncertain, and do not include the red wings of the lines. 
As we show in Figs. \ref{Ha_shift} and \ref{el_scatt_fits}, we can  get a satisfactory fit of both wings with an electron scattering profile. 

\begin{figure}[t!]
\begin{center}
\resizebox{85mm}{!}{\includegraphics[angle=0]{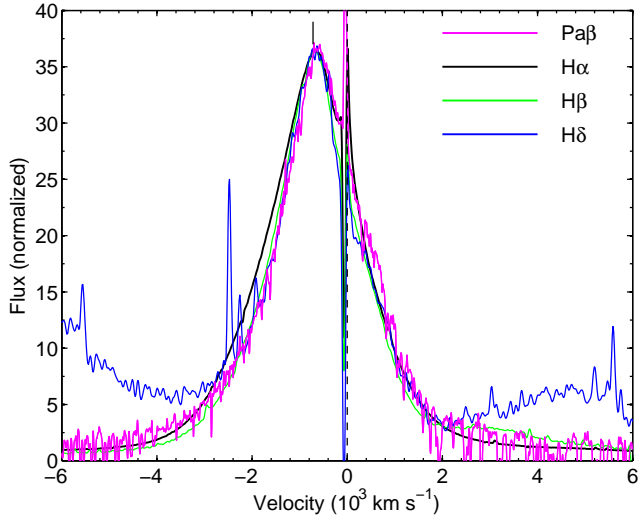}}
\caption{ Comparison of the scaled line profiles of H$\alpha$, H$\beta$, H$\delta$ and Pa$\beta$ from the X-shooter spectrum from day 461. The dashed line gives zero velocity,
while the solid line gives the -720 $\kms$ blueshift of the H$\alpha$ peak. 
Note the near coincidence of the peaks of all lines with that of H$\alpha$.}
\label{ha_hb_hei_pb}
\end{center}
\end{figure}

Maeda et al. also claim to observe a systematic shift of the velocity with wavelength, characteristic of dust extinction. 
In Figure \ref{ha_hb_hei_pb} we  compare the line profiles of our high S/N X-shooter spectrum for H$\alpha$, H$\beta$, H$\gamma$,  and Pa$\beta$ 
at  461 days, an epoch not very different from that of Maeda et al.  For these lines we have applied a smoothing of $\sim 50\kms$ to the spectra and normalized the  fluxes to the same peak flux for the broad component and the same 'continuum' level between $\pm (6000-7000) \kms$. This is somewhat lower than for the determination of the continuum level of the H$\alpha$ line in Figure \ref{fig3b}, but is necessary to avoid line contamination for H$\beta$ and Pa$\beta.$ As seen in Figure \ref{ha_hb_hei_pb}, this is even more serious for H$\gamma$ and we have therefore scaled this line to the level of the other lines at the peak and at the maximum velocity where the line dominates,  $\sim 2000 \kms$.
The high spectral resolution  makes it possible to isolate the narrow P-Cygni component of the lines, which contaminates the low resolution  line profiles of Maeda et al. Both the spectral resolution and the S/N are crucial for a reliable determination of the peak velocity as well as the general line profile. Although the day 461 spectrum is the best one in terms of S/N and coverage to the NIR, we do not see any significant difference in the wavelength shifts for any other epochs. This includes the UV lines (Figs. 14 and 15), which should be most sensitive to extinction.  

As can be seen from the figure, there is no notable difference in the velocity of the broad lines.  The peak of H$\alpha$ is shifted by $\sim 720 \kms$ for this date. The H$\delta$, H$\beta$, Pa$\beta$ and He I $\wl 10,830$ lines have their peaks at $640 \kms$, $670 \kms$, $700 \kms$ and $690 \kms$, respectively. From variations of the smoothing we estimate  an error of  $\pm 50 \kms$. For H$\alpha$, with a very high S/N, the error is $\la 30 \kms$. We  do  therefore not find any significant trend of the peak velocity or line shape with wavelength. 

Although distorted by the interstellar absorptions one can  also compare the Ly$\alpha$ line profile  to the H$\alpha$ and H$\beta$ lines in Figure  \ref{fig_lprof_lya} for days 44,  107 and 621. It is clear that there is no significant difference in these in spite of the large wavelength difference and large time span. 
 This is in contrast to the changing  line profiles of the Ly$\alpha$, Mg II and  H$\alpha$ in SN 1998S \citep[e.g., Figure 8 in][]{Fransson2005}, where such a difference was indeed seen, indicative of dust formed behind the reverse shock. The line profiles in SN 1998S were also very asymmetric.
Thus we find that the main argument for ejecta dust in Maeda et al. is doubtful. 

Another indicator of ejecta dust discussed by Maeda et al., the NIR excess, has already been discussed as the probable result of an echo. The third, the drop in the light curve, has a natural explanation in terms of the shock either exiting the circumstellar shell or transiting to a momentum driven phase, as proposed by \cite{Ofek2013}, and discussed in Sect. \ref{sec_energy}. 

After this paper was first submitted \citet[][in the following G14]{Gall2014} published an analysis of the line shift based on dust absorption.  
In contrast to \cite{Andrews2011} and this paper, G14 interpret the NIR dust emission as well as the blueshifts being caused by dust formed  behind the reverse shock at an initial radius of $\sim 2\EE{16}$ cm. There is, however, a number of problems with this interpretation. First, it requires the shock to have reached this radius at their first observation, 26 days after peak or 66 days after first detection. This requires a shock velocity of $\sim 35,000 \kms$. To support this G14 claim to see velocities up to $\sim 20,000 \kms$ from an assumed P-Cygni absorption in H$\beta$ extending to this velocity. If this would be real, there would in that case also be an emission component up to a similar velocity, because the photosphere, assumed to be at $\sim 7500 \kms$, will only occult a small fraction of the ejecta. Such a broad emission component is not seen. The only way around this is to assume that  the ejecta is extremely asymmetric. The 'P-Cygni absorption' is instead likely to be the result of contributions by weaker emission lines bluewards of H$\beta$. 
Further, there are  no indications from similar high velocities from any other lines. In particular, the Ly$\alpha$ line, being a resonance line, should show such absorption and emission if real. As seen in Fig. \ref{fig_lprof_lya} this is not the case. Nor do any other of the UV resonance lines show such a component (Fig. \ref{fig_lprof_niv_civ}). There are also other observations disfavoring  similar high velocities. The analysis of the X-rays imply a  velocity $\la 6000 \kms$  \citep{Ofek2013}. The only direct evidence for high expansion velocities comes from NIR spectra, where the He I $\wl 10,830$ line shows evidence for expansion velocities of $\sim 5500$ between 100-200 days towards us \citep{Borish2014}. 

There are additional problems with the scenario in G14. If the emission lines would be coming from the cool dense shell and absorbed by the dust formed in this the lines would be expected to have a boxy profile, not a symmetric profile peaked at low velocities. Boxy profiles are indeed seen in objects like SN 1993J \citep{Matheson2000} and SN 1995N \citep{Fransson2002}, where the cool dense shell is believed to dominate the emission. Also in objects where there indeed are strong indications for dust formation, like SN 1998S and SN 2006jc, the line profiles are  less centrally peaked and more irregular  \citep{Leonard2000,Pozzo2004, Fransson2005,Smith2008a}.  As we have shown, the line profiles are instead well characterized by that resulting from electron scattering. 
The mechanism behind the blueshift is not discussed in the G14 paper. It would, however, be surprising if this gave an intrinsically symmetric profile a simple shift in velocity, as we find, rather than a more irregular shape, given the expected clumping in a cold dense shell \citep{Chevalier1995}. 

Taken these results together, we
therefore conclude that a cold dense shell at $\sim 35,000 \kms$ is  unlikely and that the dust emission is instead coming from preexisting dust, as concluded by \cite{Andrews2011}.
If the shock velocity is much lower than  $\sim 35,000 \kms$ the shock will be well inside the evaporation radius even for the large grains G14 propose, as Table \ref{table_evap} and also the estimates in G14 show. Formation of the dust is then very difficult, as discussed before. 

We are also surprised that G14 exclude the H$\alpha$ line in their analysis, given that this is by far the highest S/N line, and therefore best suited for this kind of analysis. In addition, it should be noted that the asymmetry of H$\alpha$ and other lines is sensitive to the assumed continuum level. A continuum subtraction has apparently not been done, as can be seen in  Fig. 5 of G14. The claims for a strong asymmetry in H$\alpha$  is therefore doubtful. Instead, it can be seen from our Fig. \ref{fig_full_stis_spec} that H$\beta$ is considerably more complicated to analyze, with a  number of interfering lines. 

G14 further argue against preexisting dust based of the high dust temperature observed. It is not clear what this is based on, and  on the contrary this is a natural consequence of dust evaporation by the strong flux from the SN.  The same conclusion is reached by  \cite{Borish2014}.

Taken together we therefore think the results regarding the dust formation and derived properties of this in G14 are highly questionable. 

Based on this and our earlier discussion we  exclude the dust alternative, and instead we believe that the line shifts are due to a bulk velocity  of the scattering material. There are at least two  possibilities for this.

\subsubsection{Gradual contribution from post-shock gas}
\label{sec_viscshock}
One alternative is that the velocity shift is caused by an increasing contribution of emission from  the cooling gas behind the outgoing viscous shock. A possible scenario would  be that at the first epochs the scattering optical depth in front of the shock is large enough that most of the line photons are emitted from un-shocked gas with low velocities. As the optical depth decreases an increasing contribution of the emitted photons will come from shocked gas  behind the radiative forward shock. That this shock is indeed cooling is shown in Sect. \ref{sec_energy}. These photons will  be scattered by electrons both in front and behind the shock, producing the symmetric wings of the lines. There will therefore be a gradual shift from zero velocity to the velocity of the shock of the line profiles. 

The main problem with this model is the high density in the post-shock gas. The forward shock is expected to be radiative. The temperature immediately being the shock will be $\sim 10^8$ K, but will cool down to $(1-2)\times 10^4$ K. The compression behind the shock will therefore be a factor of $10^3-10^4$ in density, implying a density of  $\ga 10^{13} \ccm$ where the gas is cool enough to emit H$\alpha$ and other low ionization lines. Calculations discussed in Sect. \ref{sec-big} show that at densities $\ga 10^{11} \ccm$ the emission is in LTE and thermalized, with a very low efficiency of conversion into line emission. Most radiation will emerge as a blackbody continuum. The large H$\alpha$ luminosity is  very difficult to explain in this scenario. 

In principle, the H$\alpha$ emission could come from the unshocked ejecta, where the density is lower. From the swept up mass derived from the light curve (Eq. \ref{eq_mass} below) the column density of the cool shell is $\sim 10^{25} \  \rm cm^{-2}$, meaning that all X-rays below $\sim 15$ keV will be absorbed in the shell. Only the reverse shock will therefore be able to ionize this region. The luminosity of the reverse shock is only $\sim 10$ \% of the forward shock, so there will be a serious problem with the energy. Finally, it is not clear that a superposition of the emission from the pre-shock and post-shock gas will result in the observed symmetric profiles. 

Taken together, we believe that this explanation for the line shifts is unlikely.

\subsubsection{Radiative acceleration}

\label{sec_radacc}

A different  alternative is that the line shift is the result of acceleration of the pre-shock gas by the extremely energetic radiation field from the shock and thermalized optical radiation from the ejecta and cool shell.  This has been discussed earlier in different contexts \citep[e.g.,][]{Chevalier1976,Fransson1982}, but no conclusive observational evidence has been found. As a simple model we  consider  an optically thin shell at a radius, $r(t)$, illuminated from below by the radiation field of the SN, having a luminosity $L(t)$. 
Assuming 
that we can consider the radiation as radially free streaming, which should be a reasonable approximation at small optical depths close to the surface, the velocity is given by
\begin{equation}
V(t) = {\kappa  \over 4 \pi c} \int _0^{t} {L(t')  \over r(t')^2 } dt'  \ .
\label{eq_vel_acc}
\end{equation}
For $\kappa$ we use the electron scattering opacity, $\sim 0.35$, applicable for an ionized medium with a He/H ratio of 0.1.  

\begin{figure}[t!]
\begin{center}
\resizebox{85mm}{!}{\includegraphics[angle=0]{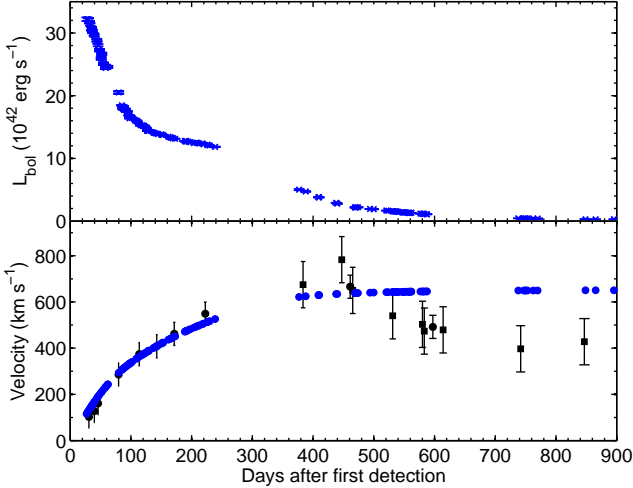}}
\caption{Upper panel: Bolometric light curve from the SN, excluding the IR dust echo, but including the UV and IR contributions from the direct SN emission. Note the linear luminosity scale in this figure,  most relevant for the radiative acceleration, compared to the logarithmic one in Figure \ref{fig_bollc}. Lower panel: The velocity shift found from H$\alpha$ together with the velocity predicted from the acceleration by the SN radiation from Eq. (\ref{eq_vel_acc}). }
\label{vel_acc}
\end{center}
\end{figure}

For  the bolometric luminosity from the SN we use the result from Figure \ref{bol_log_log}, shown on a linear scale in the upper panel of Figure \ref{vel_acc}. The total radiated energy from the SN (excluding the echo) was $\sim 6.5 \times 10^{50}$ ergs. 
Using this energy  and assuming a radius constant with time, we can estimate of the final velocity as $V \approx 670 (r/3\times 10^{15} \rm cm)^{-2} \kms$. This estimate is, however, only approximate. The reason is that the hydrodynamic time scale of this gas is comparable to the length over which most of the energy is emitted, $t_{\rm hydro} =  580 ~ (r/ 3 \times 10^{15} \  {\rm cm}) / (V/700 \kms) $ days. The increase of the radius of the shell must therefore be taken into account when calculating the velocity. This is really a hydrodynamic problem, which includes the interaction of the shell with the surrounding gas. 
We ignore the latter effects and treat the expansion as ballistic,  calculating the radius as $r = \int_{t_0}^t v(t') dt' + r_0$.
 The  only free parameter is the initial radius, $r_0 $, of the emitting shell, which gives the normalization of the velocity. 
 
 For  the best fit we find $r_0=2.6 \times 10^{15}$ cm, resulting in a velocity as a function of time shown by the blue dots in the lower panel of Figure \ref{vel_acc}. 
Given that there are uncertainties in the estimate of the bolometric luminosity,  we find a  good agreement with the observed velocity shift from Figure \ref{Ha_HWHM} and shown as black dots in Figure \ref{vel_acc}.  At  epochs later than $\sim 500$ days there is a considerable scatter in the observed line shift, and it is difficult to judge if there is a slight decrease of the velocity, as may happen if there is  a braking effect by the swept up CSM.  
 Apart from the general agreement with the evolution of the velocity, it is also interesting that the initial radius which gives the best fit, $r_0=2.6\times 10^{15}$ cm, is  close to that which gives the best fit to the photospheric radius from the blackbody fitting of the SED at the early epochs, $R_{\rm phot}\approx 3.2 \times 10^{15}$ cm (Table \ref{table_dust_param}). The final radius after $\sim 900$ days is $7  \times 10^{15}$ cm and the velocity $650 \kms$. 

As we discussed in Section \ref{sec_broad}, an important constraint on the source of the H$\alpha$ emission is the relation of the narrow unscattered line and the scattered wings, with the fraction of the former approximately given by  $\propto \exp(-\tau_{\rm e})$ for a planar geometry. Because the typical optical depth required for the wings is $2-5$, depending on the temperature, this would in general give a strong narrow line. This is indeed observed at early time, but disappears at later epochs at the same time as the blueshift starts (Figure \ref{fig3b}). In this scenario this may be explained by the fact that the source of the H$\alpha$ photons become increasingly spread out in velocity, reflecting the local, accelerated velocity, but now spanning an increasing velocity range. The peak flux will  decrease with the increasing velocity of this material, as observed. One therefore obtains a consistency check between the flux of the unscattered emission and the velocity shift. Note that the narrow CSM line most likely comes from a different, more distant component than the broad lines, based on its constant velocity and different flux evolution (Fig. \ref{fig_narrowHa}). 

The velocity of this radiatively accelerated material is considerably lower than the $\sim 3000 \kms$ one infers from the X-rays. After a time $t \approx r_0/(V_s - V_{\rm acc}) \approx 1.0 (r_0/2.6\times10^{15} {\rm cm})/(V_s/650 \kms -1)$ years the shock will sweep up the accelerated shell. For this not to happen too early, the shock velocity interior to the H$\alpha$ emitting shell has to be below $\sim 1500 \kms$, depending on the velocity of the accelerated shell, corrected for the LOS angle. As discussed earlier, a reason for a lower shock velocity may  be that the shell is anisotropic with the highest column density where the broad lines arise. The shock velocity will then be lower than that inferred from the X-rays, which are likely to come from directions of lower column density (see next section). 

The mass of the accelerated gas does not need to be large and is likely to be small. The scattering layer only has to have a $\tau_{\rm e} \ga 1$,  corresponding to $\sim 0.14 (r/3 \times 10^{15} {\rm~ cm})^2 (\Omega/4 \pi) \  \msun$, where $r$ is the radius of the shell and $\Omega$ the solid angle of the shell. To accelerate it to $\sim 1000 \kms$ only takes  $\sim 1.5 \times 10^{48} (\Omega/4 \pi)$ ergs, which is small compared to the total radiated energy.  Most of the mass in the CSM is instead in the optically very thick region inside the accelerated layer.

\subsection{Shock breakout, circumstellar interaction and energy source}
\label{sec_energy}

The high luminosity, $\sim 3\times 10^{43} \ergs$ at maximum, and large total radiated energy,  $\ga 6.5 \times 10^{50} \ \rm ergs$ (Section \ref{sec_phot_results}), require an efficient conversion of kinetic to radiative energy. 
In Section \ref{sec_dust} we argued that the flux from the optical and NIR up to $\sim 350$ days is dominated by the direct flux from the SN, while most of the NIR flux is from the echo after this epoch. 
We now discuss the energetics in terms of the shock propagation through a dense envelope. 

The shock breakout for extended SNe has been discussed by several authors \citep{Falk1977,Grasberg1987,Moriya2011,Moriya2013,Chevalier2011,Ginzburg2012}. 
For large optical depths  the shock will be radiation dominated  with a width $\tau_{\rm e} \sim c/ v$ \citep{Weaver1976}. 
 \cite{Chevalier2011} estimate the diffusion time scale of the CSM, assumed to have $\rho \propto r^{-2}$, as 
\begin{eqnarray}
 t_{\rm diff} &\approx& 6.6 \left({\dot M  \over 0.1 \mll}\right) \left({u_{\rm w} \over 100 \kms}\right)^{-1}  \nonumber \\
 &&\left({\kappa   \over 0.34 \  \rm cm^2 \ g^{-1}}\right)  \rm days.
 \label{eqdiff}
\end{eqnarray}
Here $\dot M$ is the mass loss rate and $u_{\rm w}$ is the wind velocity, and $\kappa$ the opacity. For the mass loss rates we derive below this is likely to be short for the epochs of interest here. 
As the shock approaches the surface at $\tau_{\rm e}  \la c/ v$  a viscous shock will form \cite[for a discussion see][]{Chevalier2011}.  If the circumstellar density is high the viscous shock will be radiative, implying an efficient conversion of kinetic to radiative energy. 

The cooling time for the forward shock is given by $t_{\rm cool} = 3 kT/n_{\rm e} \Lambda$, where the cooling rate can be approximated by $\Lambda = 2.4\times 10^{-23} T_8^{0.5} \ergs \ cm^3$, for temperatures above $\sim 2\times 10^7$ K. 
For close to solar abundances, the shock temperature is given by $T_{\rm e} = 1.2 \times 10^8 (V_{\rm s}/3000 \kms)^2$ K. 
For an ejecta profile given by $\rho_{\rm ejecta} \propto r^{-n}$ the shock radius is 
\begin{eqnarray}
R_{\rm s} &=& 9.47\times 10^{15} {(n-2)\over (n-3)} \left({V_{\rm s,320} \over 3000 \kms}\right) \left( {t \over {\rm years}}\right)^{(n-3)/(n-2)} \\
&=& 1.23\times 10^{16}   \left({V_{\rm s,320}\over 3000 \kms}\right) \left( {t \over {\rm years}}\right)^{0.82} \ \rm cm,
\end{eqnarray}
 for $n=7.6$ as we find below. 
Based on the shock velocity derived by \cite{Ofek2013}  we scale this and the following estimates to  a shock velocity at 320 days, $V_{\rm s,320}$, of  $3000 \kms$.  Note that this velocity may vary with the direction if the mass loss rate is anisotropic (see below), as we argue. This gives a density  immediately behind the forward shock of 
\begin{eqnarray}
n_{\rm s, e} &=& 8.1\times 10^{8} \left({\dot M  \over 0.1 \mll}\right) \left({u_{\rm w} \over 100 \kms}\right)^{-1} \nonumber\\ 
&&\left({V_{\rm s,320} \over 3000 \kms}\right) \left( {t \over {\rm years}}\right)^{-1.64} \ccm,
\end{eqnarray}
which leads to 
\begin{eqnarray}
t_{\rm cool}& =& 26.6   \left({\dot M \over 0.1 \mll}\right)^{-1} \left({u_{\rm w} \over 100 \kms}\right)  \nonumber\\ 
&& \left({V_{\rm s,320} \over 3000 \kms}\right)^{3}   \left( {t \over {\rm years}}\right)^{1.46} \rm days.
\end{eqnarray}
 Therefore, the forward shock should be radiative for several years as long as the shock is in the high density shell. 
 Further, assuming that the ingoing X-ray flux from the shock is thermalized in the ejecta and cool shell behind the shock, the blackbody luminosity is given by $(1/4) \dot M  V_{\rm s}^3/u_{\rm w}$. An equal amount will partly be absorbed and ionize the pre-shock gas and partly escape as the observed X-ray flux. Using this luminosity for the radiation density the ratio of Compton to free-free cooling is then
\begin{eqnarray}
{P_{\rm  Compton}\over P_{\rm free-free}} = 7.4\tau_e  \left({V_{\rm s} \over 10^4 \kms}\right)^4  
\end{eqnarray}
\citep{Chevalier2012}. Therefore, for shocks with velocity below $\sim 5000 \kms$ free-free cooling dominates, unless the optical depth is very high.

For a general circumstellar density given by $\rho = \dot M /(4 \pi u_{\rm w} r_0^2) (r_0/r)^s$  the ejecta velocity at the shock depends on the mass loss rate and time as 
\begin{equation}
V_{\rm s} \propto \left( { \dot M  \over v_{\rm w}} \right)^{-1/(n-s)} t^{-(3-s)/(n-s)}  
\label{eq_vshock}
\end{equation}
\citep[e.g.,][]{FLC1996},    assuming the interaction can be described by a similarity solution.   If the mass loss rate is a function of polar angle, $V_{\rm s}$ will therefore depend on angle  as well. For a radiative shock we  find for the luminosity 
 \begin{eqnarray}
L &=& 2 \pi  \rho V_{\rm s}^3 r^2 
=   {1 \over 2}   {\dot M \over u_{\rm w}}  \left(r_0 \over r\right)^{s-2} V_{\rm s}^3 \nonumber\\
 &\propto&  \left(\dot M \over u_{\rm w}\right)^{(n-5)/(n-s)}  t^{-[15+s(n-6)-2n]/(n-s)}  \ .
  \label{eq_lum}
 \end{eqnarray}
  For $s=2$
we obtain $L \propto t^{-3/(n-2)} $.

The fit of Eq. (\ref{eq_lum}) to the data in Figure \ref{bol_log_log} implies a value of $n=7.6$ for $s=2$, as shown by the dashed line  in this figure. The break at $\sim 320$ days in the light curve can either be caused by the break-out of the shock through the dense shell, or as a result of a transition to a momentum conserving phase, occurring when the swept up mass is comparable to the ejecta mass \citep{Ofek2013}.  The latter explanation has, however, been contested by \cite{Moriya2014}, who find that the transition to the momentum driven phase gives  too smooth and slow a decrease of the luminosity.  As we discuss in Section \ref{sec-big}, the breakout in a bipolar shell may be compatible with this.

 For  $s=2$ (consistent with the X-ray light curve, see below) we  get
 \begin{eqnarray}
L &=&
8.51~\times~10^{42}  \left({\dot M \over 0.1 \mll}\right) \left({u_{\rm w} \over 100 \kms}\right)^{-1}  \nonumber\\ 
&&\left({V_{\rm s,320} \over 3000 \kms}\right)^3 \left({t \over 320 \ \rm days}\right)^{-0.537}\ergs
\label{eq_lum_mdot}
\end{eqnarray}
For a luminosity at the break at $\sim 320$ days of $9.33 \times 10^{42} \ergs$ we then find
 \begin{equation}
\dot M =
0.11 \left({u_{\rm w} \over 100 \kms}\right)  \left({V_{\rm s,320} \over 3000 \kms}\right)^{-3} \ \mll \ .
\label{eq_massloss}
\end{equation}
The total mass swept up is then 
 \begin{equation}
\Delta M = \dot M  {V_{\rm s} \over u_{\rm w}} t =
2.89 \left({V_{\rm s,320} \over 3000 \kms}\right)^{-2} \ \msun \  .
\label{eq_mass}
\end{equation}
Both the mass loss and total mass are sensitive to the expansion velocity  of the ejecta. 
In this estimate we assume that the ejecta with a velocity $\sim 3000 \kms$ dominates the energy input to the optical light curve.
In reality there  should be a range in velocities and there is therefore a considerable uncertainty in this estimate. 
The fact that we observe both hard X-rays with a temperature  corresponding to a shock velocity $\sim 3000 \kms$   and with a column density corresponding to an electron scattering depth $\tau_{\rm e} \la 1$, as well as high column density gas with  $\tau_{\rm e} \ga 3$ argues for a highly anisotropic CSM.   This is further strengthened by the blue shoulder seen in the He I $\wl 10,830$ line between $\sim 100 - 200$ days after explosion by \cite{Borish2014}.  The velocity of this decreases during this interval from $\sim 6000 \kms$ to  $\sim 5000 \kms$, indicating deceleration of this material. The shock velocity in the X-ray obscured regions may, however, be considerably lower  (Eq. \ref{eq_vshock}). 
The mass loss rate above, as well as the total mass lost,  should  therefore be considered as lower limits. Note, however, that while the mass loss rate scales with the velocity of the CSM, the total mass lost is independent of this parameter.

\cite{Ofek2013} find a considerably higher mass loss rate of $\sim 0.8 \mll$. The main reason for this is that they are scaling to a higher wind velocity of $u_w = 300 \kms$, while we find $u_w \approx 100 \kms$. They also assume an efficiency of $\sim 0.25$ in converting the shock energy into the observed luminosity.  The reason for this is not  clear. Another important difference is that they assume an ejecta profile $n=10$, close to that found by  \cite{Matzner1999} for the ejecta of a radiative envelope, while we derive $n\approx 7.6$ from our bolometric light curve.  Their larger value of $n$ is a result of the flatter light curve they derive from the R-band only, $L\propto t^{-0.38}$ compared to our $L\propto t^{-0.54}$ and using $L\propto t^{-3/(n-2)}$ (Eq.   \ref{eq_lum}).

Even considering the above uncertainties, the mass loss rate and the total mass lost are very large, but as discussed in Sect. \ref{sec_csm}, these are of the same magnitude as that inferred from the CSM of local LBVs, like Eta and AG Carinae. We discuss the implications of this further in Section \ref{sec-big}, but we first place it in relation to other Type IIn SNe.

\subsection{Comparison to other Type IIn supernovae}

\label{sec-comp}

  Although interesting as a case by itself, it is at least as interesting to put SN 2010jl into the context of other Type IIn SNe. There have been several Type IIn SNe which show  similarities to SN 2010jl to various degrees, although in most cases these events are less extreme in terms of luminosity or total radiated energy. As we show in this section, most of these, SNe 1995N, 1998S, 2005ip, 2006jd, although observationally quite different, are related to SN 2010jl, while two others recently discussed, SN 2009ip and SN 2010mc, while having some properties in common, show major differences from SN 2010jl. 

SN 1998S is one of the best observed Type IIn SNe, and is interesting as a less extreme case of a Type IIn SN. 
Initially it showed typical Type IIn signatures with narrow symmetric lines \citep{Leonard2000,Fassia2001} dominated by  electron scattering of H and He I lines \citep{Chugai2001}
The symmetric lines 
disappeared  after about a week and instead broad P-Cygni profiles with an expansion velocity of $\sim 7000 \kms$ appeared.
Later optical spectra at $\ga 70$ days showed increasingly box-like profiles  typical of circumstellar interaction, with H$\alpha$ by far the strongest and a steep Balmer decrement.  Spectra later than a year showed a strong suppression of the red wing of H$\alpha$, as well as Ly$\alpha$, indicating dust formation in the ejecta or reverse shock \citep{Leonard2000,Pozzo2004, Fransson2005}.   As was remarked earlier, these line profiles were very different from thise in SN 2010jl. 

High velocity ejecta profiles of H$\alpha$ were also seen for SN 1995N \citep{Fransson2002}, SN 2005ip and 2006jd \citep{Smith2009,Stritzinger2012}. For the latter two maximum ejecta velocities of $16,000 - 18,000 \kms$ were seen at early times \citep{Smith2009,Stritzinger2012}. For SN 2006jd a high velocity wing  at $\sim 7000 \kms$ could be seen even at 1540 days. In addition, an intermediate velocity component with a FWHM of $\sim 1800 \kms$ was seen in both SNe. There were no strong indications of electron scattering wings in these SNe, although this may have contributed to the intermediate component or faded away before their discovery, as for SN 1998S. 

In the last HST spectra of SN 1998S broad features from O I $\wll 1302, 1356$ were seen, indicating processed material \citep{Fransson2005}. 
Recently, \cite{Mauerhan2012} found that at 14 years the spectrum of SN 1998S was
 dominated by strong lines of [O I-III]. This shows that the reverse shock has now propagated close to the core of the SN and that newly processed gas is dominating the spectrum. It is therefore no doubt that a core collapse has taken place. For SN 2010jl this phase has not yet occurred. 

 Hard X-rays with $kT \sim 10$ keV and a luminosity of $\sim (5-8) \times 10^{39} \ergs$ were observed for SN 1998S 2-3 years after explosion  \citep{Pooley2002}. 
  The X-ray light curve of SN 2006jd stayed nearly flat with an unabsorbed luminosity of $(3-4)\times 10^{41} \ergs$ between 400 - 1600 days \citep{Chandra2012b}, lower than for SN 2010jl. Also the column density of these SNe, $\la 1.5 \times 10^{21}$ cm$^{-2}$, were  considerably lower than for SN 2010jl and did not show any strong time evolution as for SN 2010jl. 

High resolution spectra of SN 1998S by \cite{Fassia2001} 
exhibited a low velocity P-Cygni line with a velocity of $40-50 \kms$ and a higher velocity extension with velocity $350 \kms$. The low velocity component argues for a red supergiant progenitor. The origin of the higher velocity component is not clear. 
Variations of the wind velocity is also seen in LBVs and S Doradus stars and such a progenitor is probably not excluded.  
In the case of SN 2005ip the narrow component was only marginally resolved with FWHM $\sim 120 \kms$ \citep{Smith2009}. 

The  very  different  optical spectra of  SN 1998S and the other mentioned SNe compared to SN 2010jl can naturally be explained as a result of the different CSM densities and shock velocities.  
The mass loss rate of SN 1998S was from late observations estimated to be quite modest by Type IIn standards, $\sim 2\times 10^{-5} \ml$ \citep{Fassia2001}. The time of optical depth unity to electron scattering is for a steady wind (for simplicity assumed to extend to infinity) given by
\begin{eqnarray}
t(\tau_{\rm e}=1)&=&{\kappa_{\rm T}\dot M \over 4 \pi u_{\rm w} V_{\rm s}} = \nonumber\\
&&680 \left({\dot M \over 0.1 \ml}\right)  
 \left({u_{\rm w}  \over 100 \kms}\right)^{-1}  \nonumber\\
 && \left({V_{\rm s}\over 3000 \kms}\right)^{-1} \ \rm days, 
\label{eq_tau_el}
\end{eqnarray}
where we have scale the parameters to SN 2010jl. Note, however, that we in Sect. \ref{sec_energy} argue that  $V_{\rm s}$ is considerably lower in the directions we observe.   For SN 1998S with $\dot M \approx  2\times 10^{-5} \ml$, $u_{\rm w} \approx 50 \kms$ and $V_{\rm s} \approx7000 \kms$  we find that $t(\tau_{\rm e}=1) \approx 0.1$ days. The fact that electron scattering wings were observed for several days argues for a denser shell at $\sim 10^{15}$ cm from the progenitor with $\dot M \approx 3\times 10^{-3} (u_{\rm w} /10 \kms) \ml$ \citep{Chugai2001}. 

Although there are uncertainties  in these quantities one can therefore conclude that {\it  the fast disappearance of the electron scattering wings and the transparency to the processed core in SN 1998S (and the other discussed medium luminosity Type IIns)  compared to SN 2010jl, is consistent with the much lower mass loss rate of the former. Because the X-ray column density is proportional to the electron scattering optical depth, this also explains  the much lower X-ray column densities for these SNe.} This argument also applies to the even higher luminosity 'Superluminous' Type II SNe \citep[e.g.,][]{Gal-Yam2012}.


The UVOIR luminosity of SN 1998S was at the peak $\sim 2\times 10^{43} \ergs$, while a blackbody fit gave a considerably higher luminosity of $\sim 6\times 10^{43} \ergs$ \citep{Fassia2000}. This is similar to the peak luminosity of SN 2010jl. The  initially much higher  effective temperature, $\sim 18,000$ K, compared to $\sim 7300$ K for SN 2010jl \citep{Zhang}, however, had the consequence that the R-band magnitude at maximum was only $\sim -19.1$, compared to $\sim -20.0$ for SN 2010jl. A main difference compared to SN 2010jl was that the decay was nearly exponential up to $\sim 100$ days with a fast decay rate of $\sim 25$ days. Including only the UVOIR luminosity we estimate an uncertain total energy of $\sim 1.1 \times 10^{50}$ ergs. Using instead the bolometric luminosity from the blackbody fits by Fassia et al., one gets a factor of $2 - 3$ higher radiated energy. 

As \cite{Pozzo2004} argue, the early light curve is likely to be dominated by the released shock energy, and the main reason for the high initial luminosity was a large extent of the progenitor, as indicated by the dense CSM during the first days, discussed above. The fast decay as well as the lower radiated total energy, however, indicates a  lower total mass of the CSM compared to SN 2010jl.   \cite{Chugai2001} estimate a total mass of $\sim 0.1 \Msun$ in the CSM of SN 1998S, $\sim 2$ orders of magnitude smaller than for SN 2010jl. 

SN 2005ip had a moderate peak luminosity for a Type IIn SN, $\sim 4\times 10^{42} \ergs$ \citep{Smith2009,Stritzinger2012}. It then decayed on roughly the ${}^{56}$Co time scale to a nearly constant level at $\sim 200$ days,   dominated by the NIR. The optical luminosity during this plateau phase was  $\sim 2.5\times 10^{41} \ergs$, decaying slowly as $t^{-0.3}$ from $\sim 200$ to $\ga 1600$ days. From blackbody fits Stritzinger et al. find a luminosity of $\sim 8\times 10^{41} \ergs$ for the NIR warm component. Stritzinger et al. also discuss optical and IR observations of another interesting Type IIn, SN  2006jd. This SN had a similar early luminosity, but considerably higher optical luminosity in the plateau phase, $\sim 8\times 10^{41} \ergs$, and nearly constant from $\sim 200$ to $\ga 1600$ days.

From \cite{Stritzinger2012} we estimate the total radiated energy of SN 2005ip to $\sim 4 \times 10^{49}$ ergs from the photospheric emission and $\sim 7 \times 10^{49}$ ergs in the dust component, and for SN 2006jd to $\sim 10^{50}$ ergs from the photospheric emission and $\sim 2 \times 10^{50}$ ergs in the dust component. Although the luminosity and energy are high, both these SNe are therefore considerably less extreme in terms of mass loss and degree of interaction compared to SN 2010jl. 
 
An independent indicator of strong mass loss is the large observed N/C ratio, characteristic of CNO processed material in the CSM.  This is constistent with the fact that high N/C ratios have been observed for all Type IIn, IIb and IIL SNe observed with HST, SN 1979C, 1993J, 1995N, 1998S and SN 2010jl  (Sect.  \ref{sec_csm}).


A common feature of all these SNe is also the presence of high ionization lines from the CSM. 
UV spectra of SN 1998S from day 28 to day 485  showed increasingly strong lines  from C III-IV and N III-V originating in the CSM. 
Also SN 2005ip and 2006jd revealed a number of narrow high ionization lines from the CSM. For SN 2006jd this included lines from [Fe X-XI] and [Ar X], while SN 2005ip showed even higher ionization lines, including [Fe XIV] $\wl 5302.9$  and [Ar XIV] $\wl 4412.3$ \citep{Smith2009}. 
High ionization lines were also seen in SN 1995N, which showed a very similar spectrum to SN 2006jd \citep{Fransson2002}. We note that \cite{Hoffman2008} observed similar high ionization coronal lines  in the Type IIn SN 1997eg. 
Nearly all of these SNe  had a strong X-ray flux. The connection between this and the presence of  circumstellar lines therefore indicates that the latter are excited by the X-rays in most of these cases. Conversely, the presence of circumstellar lines may be used as a diagnostic of a strong X-ray flux.  


NIR spectra and photometry of SN 1998S by  \cite{Fassia2000} revealed a dust excess already at 136 days. Later observations by \cite{Pozzo2004} showed a comparatively hot dust spectrum in the first observations, with a dust temperature $\sim 1000-1250$ K at $\sim 1$ year, decreasing to  $\sim 750-850$ K at 1198 days. 
Pozzo et al. argue for a dust echo for the first epochs, while the later emission may have an origin in condensed  the dust in the cool dense shell. In contrast to SN 2010jl,  there is no problem with dust evaporation at the shock radius for epochs later than $\sim 1$ year because the bolometric luminosity from the SN was then $\la 10^{41} \ergs$ \citep{Pozzo2004}. The dust evaporation radius is then $(1-2)\times 10^{16}$ cm (Table  \ref{table_evap} ), while the shock radius is $\sim 3\times 10^{16}$ cm for a shock velocity of $10^4 \kms$. 

Also the Type Ibn SN 2006jc showed evidence for dust formation behind the reverse shock \citep{Smith2008a,Mattila2008}. Even at maximum the luminosity of this SN was $\la 10^{42} \ergs$, declining to $\la 2\times 10^{41} \ergs$ later than 100 days \citep{Mattila2008}. The expansion velocity was estimated to be $\ga 8000 \kms$, indicating a shock radius of $\ga 7 \times 10^{15}$ cm at 100 days. The high dust temperature, $\sim 1800$ K at maximum, argues for carbon dust. From  Table \ref{table_evap} we then find a dust evaporation radius of $\sim 5.6 \times 10^{15}$ cm at 100  days for carbon dust with a size of 1 $\mu$m. Because of the much lower luminosity compared to SN 2010jl dust formation behind the reverse shock is therefore  possible.

\cite{Gerardy2002} discuss NIR photometry and spectra for several Type IIn SNe. The best observed of these is SN 1995N with NIR observations from 730 to 2435 days after explosion. During the first observations the NIR luminosity was $\sim (7-10)\times 10^{41} \ergs$, depending on the assumed dust emissivity. It then slowly decayed by a factor of $\sim 10$ at the time of the last observations. The dust temperature was $700 - 900$ K during most of the evolution. Also the other SNe observed showed similar high temperatures.  
From the fading of the red wing of the H$\alpha$ line in SN 1995N late dust formation  at an age of $\sim 1000$ days was indicated \citep{Fransson2002}. 

Evidence for dust formation  for SN 2005ip was first presented for this SN based on NIR photometry by \cite{Fox2009}.  Further observations in the  NIR and with Spitzer showed evidence for two dust components with different temperatures \citep{Fox2010}. Fox et al. argue that the hot component comes from newly formed dust in the ejecta or reverse shock, while the cool emission comes from pre-existing dust heated by the late circumstellar interaction.  Independent evidence for dust came from line profile asymmetries \citep{Smith2010}.

Both  SN 2005ip and 2006jd showed strong IR excesses already shortly after the explosion. The fit to the warm components reveal for both SNe a maximum dust temperature of $\sim 1600$ K at 100-200 days, decreasing to $\sim 1000$ K at 1000 days.   The luminosity of the dust component of SN 2006jd increased to $\sim 3\times 10^{42} \ergs$ at $\sim 500$ days and then decayed slowly to $\sim 5\times 10^{41} \ergs$ at $\sim 1700$ days. The fraction of the bolometric luminosity in this warm component increased already at $\sim 100$ days to $\sim 80 \%$, and to even higher values at later times. SN 2005ip showed a similar behavior, although the luminosity of the warm component was only $\sim 5\times 10^{41} \ergs$ and at a more constant level.


While we believe that an echo from pre-existing dust may dominate the NIR emission in SN 2010jl, the other discussed mechanisms may well be important for other Type IIn SNe. All these are physically plausible and their relative importance depends on the specific case in terms of progenitor mass loss rate, dust shell geometries,  SN luminosity, including the X-rays, and viewing direction. 

Besides the discussed SNe, two recent Type IIn SNe  SN 2009ip \citep{Smith2010a,Foley2011,Pastorello2012,Mauerhan2013} and SN 2010mc \citep{Ofek2013b} have received considerable attention. Both SNe showed minor outbursts with $M_{\rm R} \sim -14 - -15$ before the last major eruption. The main outburst had an absolute magnitude of $M_R \sim -18$, corresponding to a luminosity of $\sim (5-8) \times 10^{42} \ergs$, a factor $\sim 4-6$ lower than SN 2010jl. Another difference is that the flux decayed considerably faster during the first $\sim 60$ days, with an e-folding decay time scale of $\sim 20$ days. From the bolometric light curve we estimate a total radiated energy of $\sim 2 \times 10^{49}$ ergs for SN 2009ip and a similar energy for SN 2010mc. This is more than an order of magnitude lower than what we estimate for SN 2010jl. 


Both narrow lines with broad electron scattering wings, as well as broad ejecta lines, were seen.  \cite{Pastorello2012} stress the important observation that for SN 2009ip already in the minor outbursts a year before the most recent large outburst expanding material at a velocity of $\sim 12,500 \kms$ was seen. It is therefore clear that high velocities are not a unique signature of a core collapse, but can also occur in stages before this. In connection to the September 2012 eruption both an electron scattering profile with FWHM $\sim 550-800 \kms$ and a broad absorption extending to $14,000-15,000 \kms$ were seen. 

%
Based on the large velocities  and other arguments \cite{Mauerhan2013} propose a core collapse scenario, which is, however, challenged by  \cite{Pastorello2012}.  From the fact, mentioned above, that also the previous outbreaks showed similar high velocities,   high luminosity and  long term variability, they propose that SN 2009ip is instead a pulsational pair-instability event, and that the star may have survived the September 2012 outburst. 


From these comparisons we conclude that there are  similarities between SN 2010jl and these two objects. 
All three objects could be classified as Type IIn SNe, based on the narrow lines with profiles typical of electron scattering. It is  clear that they all have very dense CSM into which the shock waves are propagating.  
At least for SN 2009ip there is evidence for pre-existing dust  which may have been formed a few years before the last eruption \citep{Foley2011,Smith2013}, like SN 2010jl.

The main differences between these objects and SN 2010jl is that the latter showed a considerably higher peak luminosity,  slower decay and an order of magnitude higher total radiated energy compared to SNe 2009ip and 2010mc. We therefore believe that the basic mechanism for these objects is different from that in SN 2010jl. The fact that no very broad lines with velocities typical of core collapse SNe were seen for SN 2010jl is less important. 
As we have already argued, this is mainly an effect of different mass loss rates and total mass lost. 

\subsection{Putting it all together}
\label{sec-big}

In this section we summarize the main points of the previous discussion and discuss how this information can be put together in a coherent scenario for SN 2010jl. 

Most of the bolometric luminosity is produced by the radiation from the radiative shock with a velocity of up to $\sim 3000 \kms$, which propagates through the dense CSM resulting from a mass loss rate of $\ga 0.1 \ml$ and a velocity of $\sim 100 \kms$, which may be the result of a previous LBV-like eruption. The total mass lost is $\ga 3 \Msun$. The ingoing X-rays from the shock will be thermalized at early epochs in the dense shell behind the shock and later in the ejecta. This will there be converted into UV and optical continuum radiation with a spectrum close to a blackbody. Most of the outgoing X-rays will be absorbed by the pre-shock wind and will there give rise to UV and optical emission lines. 

An important issue is the relative location of the source of the UV and optical line emission and the electron scattering region. The fact that the strong high ionization UV lines, like the N IV], C IV, N III] lines, at early epochs had strong electron scattering wings besides the narrow component, shows that at least a large fraction of these arise in or interior to the electron scattering region. The same is true for the Balmer lines. 
As we have already discussed in Sect.  \ref{sec_csm}, the narrow lines  most likely arise in a more extended region outside the denser part of the CSM. Because the  velocity to which the gas is accelerated is $\propto r^{-2}$ (Eq. \ref{eq_vel_acc}) the absence of line shifts of the narrow component is consistent with a distance $\ga 2 \times 10^{16}$ cm of this  (Sect. \ref{sec_csm}).

To explain the broad wings of the lines the electron scattering optical depth of this CSM has to be $\ga 3$. The mass loss rates we find correspond to a total optical depth to electron scattering of a wind  with outer radius $R_{\rm shell} $
\begin{eqnarray}
\tau_{\rm tot} &=&  1.56    \left({\dot M \over 0.1 \mll}\right) \left({u_{\rm w} \over 100 \kms}\right)^{-1} \nonumber\\
&&\left({V_{\rm s,320}  \over 3000 \kms}\right)^{-1}   \left( {t \over {\rm years}}\right)^{-0.82} \left(1 - {R_s \over R_{\rm shell}   } \right) \ ,
\end{eqnarray}
if  completely ionized.  From this it is clear that to get large enough $\tau_{\rm tot}$ either the shock velocity has to be lower than that corresponding to the X-ray temperature, $\sim 3000 \kms$, or that the mass loss rate in these directions is higher than the average, or most likely both, as we discuss in the end of this section. 

One can estimate the extent of the ionized region, assuming that  most of the X-ray and EUV emission from the cooling shock is absorbed by the CSM in front of the shock, and will there give rise to a Str\"omgren zone (see e.g., \cite{Fransson1984} for a similar situation). We also assume that the X-rays emitted inwards are thermalized by the cool shell and ejecta, resulting in optical and UV emission, but only contributing marginally to the ionization. Balancing the number of ionizing photons, $\epsilon_i L / h\nu_{\rm ion} $ with the number of H II recombinations, $4 \pi \alpha_B \int n_e(r)^2 r^2 dr$, where $\epsilon_i$ is the fraction of the energy going into ionizations, and with $\tau_e = \sigma_T \int n_e dr$ one obtains  
\begin{equation}
\tau_e =  { \sigma_T \over \alpha_B } {m_p u_w \over  \dot M} {\epsilon_i L\over h\nu_{\rm ion}}.
\end{equation}
With $\alpha_B  = 2\times 10^{-13} T_4^{-0.7} {\rm cm^{3} \ s^{-1}}$ one finds
\begin{eqnarray}
\tau_e &=& 4.05 \epsilon_i \left({L\over 10^{43} \ergs}\right) T_4^{0.7}   \left({\dot M \over 0.1 \mll} \right)^{-1}   \nonumber\\
&&\left({u_{\rm w}\over100 \kms} \right).
\end{eqnarray}
Assuming that the X-ray/EUV luminosity is produced by the shock  and given by Eq. (\ref{eq_lum_mdot})
(for $n=7.6$), we get 
\begin{equation}
\tau_e = 
1.9 \epsilon_i T_4^{0.7}  \left({V_{\rm s,320}  \over 3000 \kms}\right)^3 \left( {t \over 320 \ {\rm days}}\right)^{-0.54} \ .
\label{eq_taue_s}
\end{equation}
Therefore, one obtains naturally a width of the ionized zone in front of the shock of the order of unity  {\it independent of the mass loss rate, wind velocity or radius. } This assumes that the total optical depth of the wind exceeds this value.

To estimate the conversion efficiency from X-rays to H$\alpha$, as well as to determine the general UV and optical spectrum expected, we have made some exploratory calculations  using CLOUDY \citep{Ferland2013}. For this we have taken a simple free-free X-ray spectrum and a density profile given by a $\rho \propto r^{-2}$ wind. The X-ray temperature was 10 keV and we studied a range of shock luminosities and mass loss rates. A problem is here that for the most interesting parameters the electron scattering optical depth is larger than unity (Eq. \ref{eq_taue_s}), which can not be handled properly by CLOUDY. As pointed out by \cite{Chevalier2012}, one effect of this is that the state of ionization is underestimated since the relevant quantity for this is the ratio of the radiation to matter density. This is given by $\zeta = \tau_e L/n r^2$, increasing the usual ionization parameter by a factor $\tau_e$. Nevertheless, the qualitative results are interesting. 

With regard to the efficiency of X-ray to H$\alpha$ conversion, we find that it is very difficult to obtain an efficiency larger than $\sim 5-7 \%$, and then only for large column densities and for densities less than $\sim 10^{10} \ccm$. Above this density the efficiency decreases rapidly due to thermalization of the H$\alpha$ line. The electron density we infer from the bolometric luminosity in Sect. \ref{sec_energy} is for  $n({\rm He)}/n({\rm H)}=0.1$
\begin{eqnarray}
n_{\rm e} &=& 2.86\times 10^{9}  \left({\dot M \over 0.1 \mll} \right)  \left({u_{\rm w}\over100 \kms} \right)^{-1}   \nonumber\\
&&\left({r \over  3\times10^{15} \ \rm cm} \right)^{-2}  \  \rm cm^{-3}, 
\end{eqnarray}
which is in the range where we  expect  a high H$\alpha$ efficiency. Further, the observed maximum H$\alpha$ luminosity was $\sim 1\times 10^{42}\ergs$ (Figure \ref{fig_haflux}). Although we find that there is  some contribution from the photospheric  emission from ionizations in the Balmer continuum, the observed H$\alpha$ luminosity implies a total X-ray luminosity of $\ga 10^{43} \ergs$. There is therefore a rough agreement between the bolometric luminosity, H$\alpha$ luminosity and a high H$\alpha$ efficiency. 
 
Compared to this the observed level of X-rays is $\sim 10^{42} \ergs$ \citep{Chandra2012}. This is by itself no problem, because most of the X-rays may be absorbed by the pre-shock gas, and is in fact needed to explain the UV emission lines. A major problem is, however, that, as discussed in the Introduction, the X-rays  at 59 days indicated a column density of $\sim 10^{24} \ {\rm cm}^{-2}$, decreasing to $3\times 10^{23} \ {\rm cm}^{-2}$ at 373 days\footnote{ Chandra et al. used a metallicity of 0.3 times solar in agreement with that derived for the host galaxy. The connection between the hydrogen column density and the X-ray absorption, which is dominated by metals, depends on the metallicity. The main indication of synthesized material in the outer layers is the CNO processing discussed in Sect. \ref{sec-cno}. This only results in a redistribution between the CNO elements, which have similar cross sections and ionization thresholds. As discussed later, we do not find any other indication of processing. It is therefore reasonable to assume that, except for a different host metallicity, the standard conversion between X-ray absorption column density and N$_H$, and therefore electron scattering depth,  applies. }. This column density   corresponds to an electron scattering depth of $\tau_{\rm e} \la1$, which is too low to explain the line profiles in the optical. 

There are several possible explanations to this discrepancy. \cite{Ofek2013} argue that the column density could be decreased by a higher ionization due to a large electron scattering optical depth. As we argue above, the optical depth of the {\it ionized} gas is limited and the effect will  not be dramatic. Instead we believe that the different column densities are due to large scale  asymmetries in the outflow, giving rise to different column densities in different directions. Alternatively, it could be due to small scale inhomogeneities resulting in a leakage of X-rays in the low density regions. This is also supported by the low column density \cite{Ofek2013} found from their NuSTAR + XMM observations at $728-754$ days. We discuss further evidence for these possibilities below. 

For a wind, most of  H$\alpha$  originates close to the shock, as can be seen from  $dL/dr = 4 \pi \alpha_B r^2 n_e^2 \propto 1/r^2$. This is also the region of largest optical depth to  electron scattering. Emission and scattering therefore take place in the same region, although some of the H$\alpha$ emission also occurs at low optical optical depths, resulting in a narrow line core as long as the bulk velocity of the emitting material is low (Sect. \ref{sec_radacc}). From our CLOUDY calculations we find that high ionization lines, like N III] $\lambda 1750$,  N IV] $\lambda 1486$ and C IV $\lambda \lambda 1548, 1551$, arise interior to the H$\alpha$ emitting region and therefore experience the same electron scattering as H$\alpha$. We also find that the luminosities of these lines are comparable to H$\alpha$, as is observed. For constant X-ray luminosity the H$\alpha$/H$\beta$ ratio increases as the density decreases, which may explain the trend seen in Figure \ref{fig_haflux}. The temperature in the H$\alpha$ emitting zone is $\sim (1.5-2)\times 10^4$ K. 

A related model is discussed by \cite{Dessart2009} who calculate the line formation in Type IIn SNe with application to SN 1994W. Although the authors point out that this is only a preliminary attempt  and related to a specific scenario, this is the only self-consistent modeling of the line formation we are aware of. The density of the envelope in the model by  \cite{Dessart2009} is $3\times 10^9 \ccm$, which is similar to what we find for the density in front of the shock. 
Dessart et al.  also point out a number of useful signatures, depending on the relative location of  the region where the Balmer emission originates and the electron scattering region. In particular they find that if the photons are internally generated,   H$\alpha$,  H$\beta$ and  H$\gamma$  form at increasing optical depth. The electron scattering depth  also increases for these lines, leading to increasingly strong wings for H$\beta$ and H$\gamma$ compared to H$\alpha$. If the lines, on the other hand,  are created outside the scattering region they are  expected to have similar line profiles. 

In principle, these possibilities can be tested with our observations. The limited S/N and relatively steep Balmer decrement makes it difficult to see any differences between these line profiles. The best set of line profiles are from our spectrum from X-shooter in Figure \ref{ha_hb_hei_pb}. This does show some minor differences in the width of the lines, but these are sensitive to the exact way the continuum is subtracted as well as to blending by other lines. The latter is especially important for the He I $\wl 10,830$ line, which is blended with Pa$\gamma$;   the red wing of H$\beta$ also shows indications of blending. It is therefore premature to draw any firm conclusions. 

As argued in Sect. \ref{sec_radacc}, we believe that the bulk velocity of $\sim 600-700 \kms$ of the Balmer emitting region is a result of the pre-acceleration of the circumstellar material in front of the main shock by the intense radiation from the thermalized radiation from the shock. This is supported by the agreement between the observed velocity evolution and that predicted from the evolution of the bolometric luminosity. 
A surprising result is  that to produce the line shift of the H$\alpha$ line we required that most of the emitting material was moving with similar velocity relative to us. This in turn requires the flow to subtend a fairly small solid angle. The expansion does, however, not necessarily have to be in our LOS, which would increase the true expansion velocity by $1/\cos \theta$, where $ \theta$ is the angle between these directions.

Both the line shifts and the X-ray observations therefore give evidence for a highly asymmetric ejecta interaction in SN 2010jl. On a larger scale we find evidence for an asymmetric dust distribution (Sect. \ref{sec_dust}). 
In addition, the early polarization measurements by \cite{Patat2011} at $14.7$ days after discovery  showed a large, constant continuum polarization of $1.2 - 2 \%$, indicating an asphericity with axial ratio $\sim 0.7$. Close to the center of H$\alpha$ the polarization decreased to a low level. The wings of the line have, however, a polarization close to the continuum level. As suggested by the authors, this is probably an indication of another component dominated by recombination. Our high resolution observations do show that especially at early phases the narrow component from the CSM is strong, and will therefore dominate in the line center. Most likely this  is produced mainly by recombination and is probably also coming from a more symmetric CSM, as  indicated by the strong absorption component of the  P-Cygni lines (Sect. \ref{sec_csm}).

Asymmetries have been discussed from time to time for Type IIn SNe. This includes the Type IIn SN 1988Z \citep{Chugai1994}, SN 1995N \citep{Fransson2002}, SN 1998S  \citep{Leonard2000,Fransson2005}, and  SN 2006jd  \citep{Stritzinger2012}. In these cases it has mainly been disk like or clumpy distributions that have been considered. 
A clumpy model does, however, not have the large scale asymmetry needed to match the observed bulk velocity of hydrogen lines. A disk like geometry also has problems because one would then expect the expansion into a disk to be cylindrically symmetric, which would not give the coherent velocity in the line of sight indicated by the broad line profiles above. 

Instead we believe that a bipolar outflow may be more compatible with the observations. As was noted in Sect. \ref{sec_csm}, the expansion velocity of the molecular shell in  Eta Carinae is highly anisotropic with a velocity of up to  $\sim 650 \kms$  at the poles \citep{Smith2006}.   Even more interesting is that most of the mass is ejected between $45\degr$ and the pole. 
Higher velocities up to $\sim 6000 \kms$ are also seen, although little mass is involved \citep{Smith2008}. 

The initially small shift of the lines in SN 2010jl, however, requires this high density shell to have a low velocity, $\la 100 \kms$, before the explosion,  similar to what we see from the narrow lines. The increasing velocity shift in H$\alpha$  may then be a direct representation of the gradual acceleration of this shell. The column density of the shell in Eta Carinae is $\sim 10^{24} \ \rm cm^{-2}$ located at a radius between $3\EE{16}$ cm and $3\EE{17}$ cm  \citep{Smith2006}. This scales as $N_{\rm H} \propto r^{-2}$, so if one would observe it at an earlier stage with a radius of a few $\times 10^{15}$ cm, as in SN 2010jl, the column density would be $2-3$ orders of magnitude higher, more than sufficient for explaining the large electron scattering optical depths we have evidence for in the broad lines. 

 The planar geometry of the H$\alpha$ emitting gas may resemble that seen in Eta Carinae at high latitudes. \cite{Smith2006} find above $\sim 60\degr$ latitude an almost perfectly planar shape of the molecular shell, containing most of the gas  \cite[see e.g., Fig. 4 in ][]{Smith2006}. If one would observe an explosion in this direction (or higher latitude) the expansion would clearly be planar. The large column density of the shell would also  prevent the observation of the rear parts of the CSM.

 The scenario we infer from the observations has many qualitative similarities to the one calculated by \cite{vanMarle2010}. Using 2-D hydrodynamical simulations they have calculated the interaction of a core collapse SN with a dense circumstellar shell ejected 2 years before explosion. The total shell mass was in most cases $ 10 \Msun$, while the ejecta mass for this shell mass was varied between $ 10 - 60 \Msun$. Models  with both spherical geometry and a density and mass distribution similar to the bipolar structure observed for Eta Carinae were calculated. Optically thin cooling, but no radiative transfer, was included.  
 At the time the shock collides with the shell the shock velocity is $\sim 5500 \kms$, for the chosen parameters, but decreases to 1000-2000 $\kms$ in the shell, depending on mass of ejecta, shell mass and SN energy. Because of the large density and low velocity, implying a shock temperature of $\la 10^7$ K,   the shock becomes radiative. After the shock has penetrated the shell the velocity again increases and the shock then in most cases becomes adiabatic. This evolution illustrates what happens when a SN with 'normal' energy interacts with a dense CSM. The low shock velocity resulting from this is  in line of what is observed for SN 2010jl. 

 In their bipolar simulations (models D01-D03) the ejecta first interact with the equatorial region and then later with the polar region. 
 Because of the smaller column density in the equatorial direction the shock penetrates this region first at which time the shock temperature increases rapidly. In the polar direction it takes a considerably longer time for the shock to reach the outer boundary of the shell and the shock velocity is also lower. This may illustrate the range of velocities which may be present in an anisotropic CSM, and which may explain the different column densities inferred from the X-rays and the electron scattering wings, discussed above.  These anisotropic models demonstrate nicely how one can have both high shock velocities (in the equatorial directions), giving rise to hard X-rays,  and slow shocks still in the optically thick circumstellar shell (in the polar directions) at the same time. The drop in luminosity during the breakout phase is in the bipolar models  $L \propto t^{-\alpha}$, where $\alpha = 3.6 - 4.4$ \citep[Figure 16 in ][]{vanMarle2010}. This depends on the parameters of the models, but is at least in qualitative agreement with the observations, which show that $L(t) \propto t^{-3.4}$ after the break (Sect. \ref{sec_phot_results}). 
 
 For the spherical models in \cite{vanMarle2010} the light curves in general consist of a rapid rise when the shock encounters the shell and a slowly decreasing plateau. As the shock has passed the outer shell, there is a rapid drop as the shock becomes adiabatic. Because of the gradual emergence of the shock with latitude the bipolar case results in a smoother evolution of the light curve without the same sharp break as the spherical shell. This result depends on the angular dependence of the column density. A stronger concentration to the poles would give a sharper drop of the light curve, in line of what we see in SN 2010jl.  
The efficiency in converting kinetic energy into radiation was found to be in the range $10-40 \%$, decreasing with increasing ejecta mass and expansion velocity of the shell (determining the distance of the shell from the SN). With an efficiency in this range the radiated energy of $\sim 6.5 \times 10^{50}$ ergs would correspond to an explosion energy of $\sim (4\pm 3)\times 10^{51}$ ergs. 

The fact that we do not see evidence for expansion in the opposite direction is a natural result of the obscuration of this region by the optically thick material in the core of the SN and from the ionized material exterior to this moving towards us.  The decline in the flux at late epochs is a result of the  decrease in the photospheric radius  (see Table \ref{table_dust_param}), which in principle might indicate that the extent of the photosphere is not large enough to obscure the receding side of the SN. However, the photospheric radius is the thermalization radius which must be deep inside the $\tau_{\rm e} = 1$ radius. In our last spectra at $\sim 1100$ days the optical depth to electron scattering is likely to be $\ga 1$.  The obscuration of the back-side of the SN by the scattering region  even at late epochs is therefore not a problem. As we have already discussed, at the late epochs  the shock should have penetrated the dense circumstellar shell in the low column density regions close to the equator, while it is still inside the optically thick region in our line of sight, although outside the photosphere.

With regard to the narrow lines we note that a steady wind with a mass loss rate of $10^{-3} \ml$, as in the quiescent state in Eta Carinae, and wind velocity of $100 \kms$ corresponds to a density of $\sim 3 \times 10^{6} (\dot M/10^{-3} \ml) (u_w/100 \kms)^{-1} (r/10^{16}$ $\rm cm)^{-2} \ccm$, which is similar to what we find from the nebular diagnostics of the narrow lines for SN 2010jl, but below that of the Balmer lines. A larger radial extent of the 'narrow line region', as indicated from the ionization parameter (Sect. \ref{sec_csm}) is natural in this context. 

The  deep absorption component  in the P-Cygni profile of the narrow H$\alpha$ line (Figure \ref{fig3b}) means that this material must have an appreciable covering factor of the underlying emission from the SN. The emission component of this, as well as the depth of the absorption, tends to decrease with time. This may be a result of the expansion of the SN relative to the slow CSM. An example of this kind of evolution can be seen in the analytical models in Figure 5 of \cite{Fransson1984}, where the relative size of the scattering layer to the 'photospheric' emission is varied. 

The dust shell, giving rise to the echo,  is at a considerable larger distance, $\ga 10^{17}$ cm. The high temperature and large distance argues for small dust grains. To reproduce the NIR light curve the dust  should have an anisotropic distribution. The dust is likely to have been formed in previous mass loss phases of the progenitor, as seen in Eta Carinae.

Finally, we remark that even in our last full spectrum at $\sim 850$ days there is no clear transition to a nebular stage, although the spectrum is dominated by emission lines,  with H$\alpha$ by far the strongest line, rather than continuum emission (Figure \ref{fig_fast_spec}).  Electron scattering is apparently still important in the line emitting region, and there are no signs of any processed material in even the last spectrum.   This may in principle  speak in favor of the pure LBV scenario without core collapse. Because of the extreme mass loss rate the hydrogen envelope is, however, likely to still be opaque to the emission from the core (Sect. \ref{sec-comp}). We will therefore have to wait until the optical depth is low enough for any firm conclusions.

\subsection{Progenitor scenarios}
\label{sec_prog}

As discussed by \cite{Miller2010} \citep[see also][]{Smith2008b,Smith2010}, the extremely luminous Type IIn SNe, with absolute magnitude brighter than $M_V = -20$, may arise from ejections where the ejected shell has not expanded to more than a few $\times 10^{15}$ cm, resulting in a dense, optically thick shell. This would correspond to SN 2010jl. Ejected shells which have expanded to larger radii, and therefore have lower density and optical depth, would give rise to more `normal' Type IIns, like SNe 1988Z and 2008iy, with absolute magnitudes fainter than $ -20$. These are also expected to be strong X-ray and radio sources, in contrast to the more luminous Type IIns. As we discuss in Sect. \ref{sec-comp}, the mass loss rate and its duration may be the most important parameters distinguishing these properties. 

Although we have in this paper mainly discussed our results in terms of conditions around LBVs, the standard LBV scenario has a number of problems in explaining the Type IIn SNe. In particular, the high frequency of Type IIns may be incompatible with very massive progenitors and also the evolutionary state of the LBVs may be a problem.  Recently there have been different suggestions, discussed below, addressing these problems.  One should note that the LBV interpretation is mainly phenomenological, based on observational characteristics like the very dense CSM, the CNO processing and the typical velocities for the CSM.    It has also been suggested that a  large fraction of Type IIn SNe may arise as a result of explosions in red supergiants with enhanced mass loss  \citep{Fransson2002,Smith2009}, which would alleviate these problems.

A single star version has recently been put forward by \cite{Groh2013} based on a moderately fast rotating star with a mass $20 - 25 \Msun$. The average rotational velocity on the main sequence was $\sim 200 \kms$ for these models. The rotation in combination with mixing results in a more massive He-core than for a non-rotating star. This has the consequence that the star ends up as a hot supergiant, with effective temperature $\sim 2\times10^4$ K, and with a thin hydrogen envelope. The heavy mass loss in the red supergiant and final hot phase results in a dense CSM. 

The most  interesting result in the paper by Groh et al. is  that a full NLTE atmospheric calculation of  the hydrostatic and wind regions in the final phase gives  a spectrum similar to that of an LBV, with a large number of emission lines. Because of heavy mass loss and mixing, CNO products are transported to the surface, with essentially complete CNO burning.  Their 20 $\Msun$ model has N/C = 128 and N/O = 16 at the end of C-burning. These are considerably higher than our determination for SN 2010jl, including the uncertainty in these. As discussed by Groh et al., this could perhaps be explained by He-burning products mixed to the surface. It may, however, also indicate less complete CNO burning, as is e.g., seen in the $15 \Msun$ models by \cite{Ekstrom2012}. These lower mass stars, however, end their life as red supergiants, not LBV-like objects. A dense CSM may nevertheless be present, but the wind velocity would probably be  lower than we observe. 

The mass loss rate Groh et al. find for the $20 - 25 \Msun$ models in the pre-SN phase is $(1-4)\times 10^{-4} \ml$ and wind velocity $270 - 320 \kms$. The wind velocity is considerably higher that what we find for SN 2010jl. Scaling to a lower wind velocity this mass loss rate corresponds to an electron density of  $ \sim 2.3 \times 10^5 (\dot M / 10^{-4} \ml) (u_{\rm w}/300 \kms)^{-1} (r/10^{16} \rm cm)^{-2} \ccm$, where $r$ is the radius of the line emitting material. We have assumed a He/H ratio of 0.8 by mass \citep{Groh2013}. Depending on the mass loss rate and radius of the line emitting gas, this may be compatible with what we observe for the narrow lines from the CSM in SN 2010jl. 
However, there are large uncertainties in this estimate.
First, as discussed in Sect. \ref{sec-fluxes} and \ref{sec_broad}, there is considerable uncertainty in the radius of the line emitting gas. Also, the mass loss rate of the models is highly uncertain since, as \cite{Groh2013} remark, these stars are close to the Eddington luminosity. An increased mass loss is therefore not unexpected, as discussed by \citet[][and references therein]{Grafner2012}. The mass loss rate could also have been higher in the previous red supergiant phase, further boosting the density of the CSM at large radii. Rotational effects could also make the mass loss  highly anisotropic. Overall, we find that given these uncertainties there is at least some rough agreement between the models and our observations. This scenario may therefore solve the evolutionary and rate problems, while at the same time at least qualitatively explain the observational LBV characteristics. A major issue is to explain the mechanism behind the outbursts needed to explain the dense inner shell.  

As a second alternative, based now on a binary model, \cite{Chevalier2012b} has proposed a merger scenario, where a compact companion in the form of either a neutron star or a black hole may merge with a normal companion star. The energy released by the in-spiraling will release a large energy into the envelope of the companion star, resulting in heavy mass loss. A dense CSM will therefore result, mainly concentrated to the orbital plane of the binary. This may have implications for the line profiles, but a more definite comparison is outside the scope of this paper. We also remark that an LBV-like eruption may also be the result of a merger or possibly only a tight binary encounter in a lower mass system. This would then be in better agreement with the relatively high frequency of Type IIn SNe. The high luminosity would, however, be difficult to understand.

A third scenario is discussed by \cite{Quataert 2012} based on convective motions connected to the carbon burning and later stages when the core luminosity is super Eddington. This turbulence in turn excites internal gravity waves which may convert a fraction of their energy into sound waves. These finally  dissipate their energy and cause mass loss from the stellar envelope. The advantage with this model is that it directly connects the stellar explosion and strong mass loss of the progenitor.  
\cite{Shiode2013} have developed this further and made predictions for the duration of the wave-driven mass loss, as well as the total mass lost. Most of the mass loss occurs in the Ne and O burning stages, which limits the duration to $\la 10$ years before explosion. Only for stars with He-core masses below $\sim 10 \Msun$, or ZAMS masses below $\sim 20 \Msun$ do they find substantial  mass loss, with a total mass lost between $0.1 -1 \Msun$. While the wind velocity we find,  $\sim 100 \kms$, is close to the escape velocities of their models, the total mass lost in the wave-driven phase is substantially lower than what we find for SN 2010jl. With a duration of $\la 10$ years also the extent of the dense shell, $\la 3\times 10^{15}$ cm, is on the low side.  
Although interesting in that this scenario directly connects the heavy mass loss phase with the explosion, many details connected to the conversion efficiency, effects of binary interaction etc. are uncertain and require numerical work along the lines of \cite{Meakin2006} to test this scenario. \cite{Smith2014} have recently discussed the consequences of instabilities from turbulent convection in the latest burning phase, which may trigger strong mass loss just before explosion.

\section{Conclusions}
\label{sec-conculsions}

SN 2010jl represents the best observed Type IIn supernova to date, and in addition one of the brightest. In this paper we have combined our optical, NIR and UV observations with  X-ray observations  to get a full view of this SN in both time and in wavelength. Our most important conclusions are:
\begin{itemize}
\item We have presented one of the most complete data set of any Type IIn SN, covering the UV, optical and IR range. It also represents a nearly complete coverage in time from the explosion to $\sim 1100$ days.  
\item We find a large number of narrow UV, optical and NIR lines from the CSM from both low and very high ionization stages, including coronal lines, most likely excited by the X-ray emission from the SN shock wave. The UV lines provide strong evidence for CNO processed gas in the CSM, with N/C$=25\pm15$ and N/O=$0.85\pm0.15$. 
The density of the CSM and distance to the narrow line emission are consistent with observations of LBVs in our Galaxy. 
\item The expansion velocity of the CSM makes red supergiants unlikely as progenitors, but is consistent with LBVs.  
\item The profiles of the broad lines are symmetrical  in the co-moving frame up to $\sim 1100$ days, with a shape typical of electron scattering, but are shifted by an increasing velocity to $\sim 600-700 \kms$  along the line of sight. 
The  profiles of the broad lines show no strong evidence of wavelength dependence, as would be expected if dust was responsible for the   shift of the broad lines. 
Instead, the  shift of the broad lines which develop at late stages is explained as a result of the bulk velocity of the gas, rather than as a result of dust in the ejecta.
\item We find that the most likely explanation for the velocity shift is radiative acceleration by the  flux from the SN.  This is a natural consequence of the large radiated energy of this SN in combination with the gradual release of this. 
\item The optical depth needed to produce the electron scattering wings, the low and coherent bulk velocity inferred from the velocity  shifts together with the comparatively low X-ray column densities and the high  X-ray temperature provide strong evidence for an asymmetric shock wave, with both varying velocities and column density, consistent with what is inferred from polarization measurements at early epochs. A bipolar outflow from an LBV a few years before the explosion would be compatible with these observations, although we can not exclude a one-sided asymmetry. 
\item The NIR dust excess is likely to originate from an echo at a distance of $\sim 6\times 10^{17}$ cm in the CSM of the SN.  The dust is heated by the radiation from the SN to nearly the evaporation temperature. The high dust temperature and large distance may require a small grain size. 
\item The expansion velocity of the CSM is $\sim 100 \kms$ and the mass loss rate has to be $\ga 0.1 \mll$ to explain the bolometric light curve by circumstellar interaction.  The total mass lost is $\ga 3 \Msun$.  These numbers are likely lower limits, depending strongly on the uncertain shock velocity and anisotropies in this.
\item There are no indications of a nebular stage, or any processed material even at $\sim 850$ days. The last spectrum is dominated by a strong H$\alpha$ line. The core region is, however, likely to still be opaque as a result of the dense CSM. 
\item When compared to other Type IIn SNe we find strong similarities with other well observed objects, like SNe 1995N, 1998S, 2005ip and 2006jd in terms of early electron scattering line profiles, total radiated energy, CSM properties and X-ray emission. The main difference is the several order of magnitude higher mass loss rate in SN 2010jl, which slows down the evolution to the nebular stage. 
\item The UV spectrum will add to the small set of nearby UV bright objects, which may dominate future high redshift SN surveys. 
 \end{itemize}

\acknowledgments

We are grateful to Eran Ofek and Avishay Gal-Yam for useful comments on the draft. This research was supported by the Swedish Research Council and
National Space Board. The Oskar Klein Centre is funded by the Swedish
Research Council. The 
CfA Supernova Program is supported by NSF grant AST-1211196 to the Harvard College Observatory and has also been supported by PHY-1125915 to the Kavli Institute of Theoretical Physics. F.B. acknowledges support from FONDECYT through Postdoctoral grant 3120227 and
from the Millennium Center for Supernova Science through grant P10-064-F (funded by �Programa Bicentenario de Ciencia y Tecnolog�a de CONICYT� and �Programa Iniciativa Cient\'�fica Milenio de MIDEPLAN�). The research of RAC is supported by NASA grant NNX12AF90G. SB is partially supported by the PRIN-INAF 2011 with the project �Transient Universe: from ESO Large to PESSTO� . Support for Program GO-12242 was provided by NASA through a grant from the Space Telescope Science Institute, which is operated by the Association of Universities for Research in Astronomy, Incorporated, under NASA contract NAS5-26555. Based on observations made with the Nordic Optical Telescope, operated by the Nordic Optical Telescope Scientific Association at the Observatorio del Roque de los Muchachos, La Palma, Spain, of the Instituto de Astrofisica de Canarias under Period P42, P43 and P46  and P47 (P.I. Sollerman) .The data presented here were obtained in part with ALFOSC, which is provided by the Instituto de Astrofisica de Andalucia (IAA) under a joint agreement with the University of Copenhagen and NOTSA. Based on observations collected at the European Organisation for Astronomical Research in the Southern Hemisphere, Chile (Program 088.D-0195, P.I. Sollerman)

{\it Facilities:} \facility{HST (COS,STIS)}, \facility{VLT (X-shooter)}, \facility{FLWO:1.5m (FAST)}, \facility{MMT}, \\
 \facility{NOT(ALFOSC,NOTCAM)}, \facility{Magellan:Baade}


\end{document}